\documentclass[longauth]{aa}

\usepackage{graphicx}
\usepackage{natbib}
\usepackage{scalerel}
\usepackage{comment}
\usepackage{threeparttable}
\usepackage{array}
\usepackage{multirow,tabularx}
\usepackage[normalem]{ulem} 

\usepackage[table]{xcolor}

\usepackage{txfonts}

\newcommand{\phosphoros}{\texttt{Phosphoros}}
\newcommand{\nnpz}{\texttt{NNPZ}}
\newcommand{\unif}{\texttt{UNIF}}

\makeatletter
\renewcommand*\aa@pageof{, page \thepage{} of \pageref*{LastPage}}
\makeatother

\usepackage[utf8]{inputenc}

\usepackage[switch, modulo]{lineno}
              
\renewcommand{\linenumbers}[0]{}

\usepackage{euclid}
\DeclareSIUnit{\jansky}{Jy}
\DeclareSIUnit{\uJy}{\micro\jansky}

\usepackage{glossaries}
\newacronym{sfr}{SFR}{star-formation rates}
\newacronym{snr}{S/N}{signal-to-noise ratio}
\newacronym{sed}{SED}{spectral energy distribution}
\newacronym{psf}{PSF}{point-spread function}
\newacronym{zpdf}{$z$PDF}{redshift posterior distribution function}
\newacronym{NIR}{NIR}{near-infrared}
\newacronym{EWS}{EWS}{Euclid Wide Survey}
\newacronym{EDF}{EDF}{Euclid Deep Field}
\newacronym{EAF}{EAF}{Euclid Auxiliary Field}

\defcitealias{Q1-TP005}{EC:T26}
\defcitealias{DR1-TP012}{EC:E26}

\usepackage{xspace}
\newcommand{\threebytwo}{3$\times$2pt\xspace}

\usepackage[pdfencoding=auto,psdextra]{hyperref}
\hypersetup{
    colorlinks=true,
    linkcolor=blue,
    filecolor=magenta,      
    urlcolor=blue,
    citecolor=blue
}
\begin{document}
%
%
\title{Euclid Data Release 1 (DR1)}
\subtitle{Photometric-redshift pipeline overview and validation}

\newcommand{\orcid}[1]{} 
\author{Euclid Collaboration: W.~G.~Hartley\orcid{0000-0002-4563-5101}\thanks{\email{william.hartley@unige.ch}}\inst{\ref{aff1}}
\and S.~Paltani\orcid{0000-0002-8108-9179}\inst{\ref{aff1}}
\and F.~Dubath\orcid{0000-0002-6533-2810}\inst{\ref{aff1}}
\and M.~Tucci\inst{\ref{aff1}}
\and A.~Enia\orcid{0000-0002-0200-2857}\inst{\ref{aff2},\ref{aff3}}
\and M.~Bolzonella\orcid{0000-0003-3278-4607}\inst{\ref{aff3}}
\and N.~Morisset\inst{\ref{aff1}}
\and H.~Degaudenzi\orcid{0000-0002-5887-6799}\inst{\ref{aff1}}
\and S.~Fotopoulou\orcid{0000-0002-9686-254X}\inst{\ref{aff4}}
\and C.~Saulder\orcid{0000-0002-0408-5633}\inst{\ref{aff5},\ref{aff6}}
\and Y.~Kang\orcid{0009-0000-8588-7250}\inst{\ref{aff7}}
\and C.~Benoist\orcid{0009-0003-0880-6295}\inst{\ref{aff8}}
\and F.~J.~Castander\orcid{0000-0001-7316-4573}\inst{\ref{aff9},\ref{aff10}}
\and D.~Delley\orcid{0000-0002-4958-7469}\inst{\ref{aff5}}
\and G.~Desprez\orcid{0000-0001-8325-1742}\inst{\ref{aff11}}
\and H.~Hildebrandt\orcid{0000-0002-9814-3338}\inst{\ref{aff12}}
\and O.~Ilbert\orcid{0000-0002-7303-4397}\inst{\ref{aff13}}
\and M.~Manera\orcid{0000-0003-4962-8934}\inst{\ref{aff14},\ref{aff15}}
\and D.~C.~Masters\orcid{0000-0001-5382-6138}\inst{\ref{aff16}}
\and R.~Pello\orcid{0000-0003-0858-6109}\inst{\ref{aff13}}
\and L.~Pozzetti\orcid{0000-0001-7085-0412}\inst{\ref{aff3}}
\and W.~Roster\orcid{0000-0002-9149-6528}\inst{\ref{aff5}}
\and R.~Saglia\orcid{0000-0003-0378-7032}\inst{\ref{aff6},\ref{aff5}}
\and M.~Salvato\orcid{0000-0001-7116-9303}\inst{\ref{aff5}}
\and M.~Schefer\inst{\ref{aff1}}
\and S.~A.~Stanford\orcid{0000-0003-0122-0841}\inst{\ref{aff17}}
\and F.~Tarsitano\orcid{0000-0002-5919-0238}\inst{\ref{aff18},\ref{aff19}}
\and A.~Viitanen\orcid{0000-0001-9383-786X}\inst{\ref{aff1},\ref{aff20},\ref{aff21}}
\and A.~H.~Wright\orcid{0000-0001-7363-7932}\inst{\ref{aff12}}
\and B.~Altieri\orcid{0000-0003-3936-0284}\inst{\ref{aff22}}
\and S.~Andreon\orcid{0000-0002-2041-8784}\inst{\ref{aff23}}
\and N.~Auricchio\orcid{0000-0003-4444-8651}\inst{\ref{aff3}}
\and C.~Baccigalupi\orcid{0000-0002-8211-1630}\inst{\ref{aff24},\ref{aff25},\ref{aff26},\ref{aff27}}
\and M.~Baldi\orcid{0000-0003-4145-1943}\inst{\ref{aff28},\ref{aff3},\ref{aff29}}
\and A.~Balestra\orcid{0000-0002-6967-261X}\inst{\ref{aff30}}
\and S.~Bardelli\orcid{0000-0002-8900-0298}\inst{\ref{aff3}}
\and P.~Battaglia\orcid{0000-0002-7337-5909}\inst{\ref{aff3}}
\and A.~N.~Belikov\inst{\ref{aff11},\ref{aff31}}
\and A.~Biviano\orcid{0000-0002-0857-0732}\inst{\ref{aff25},\ref{aff24}}
\and E.~Branchini\orcid{0000-0002-0808-6908}\inst{\ref{aff32},\ref{aff33},\ref{aff23}}
\and M.~Brescia\orcid{0000-0001-9506-5680}\inst{\ref{aff34},\ref{aff35}}
\and J.~Brinchmann\orcid{0000-0003-4359-8797}\inst{\ref{aff36},\ref{aff37},\ref{aff38}}
\and S.~Camera\orcid{0000-0003-3399-3574}\inst{\ref{aff39},\ref{aff40},\ref{aff41}}
\and G.~Ca\~nas-Herrera\orcid{0000-0003-2796-2149}\inst{\ref{aff42}}
\and V.~Capobianco\orcid{0000-0002-3309-7692}\inst{\ref{aff41}}
\and C.~Carbone\orcid{0000-0003-0125-3563}\inst{\ref{aff43}}
\and V.~F.~Cardone\inst{\ref{aff21},\ref{aff44}}
\and J.~Carretero\orcid{0000-0002-3130-0204}\inst{\ref{aff45},\ref{aff46}}
\and M.~Castellano\orcid{0000-0001-9875-8263}\inst{\ref{aff21}}
\and G.~Castignani\orcid{0000-0001-6831-0687}\inst{\ref{aff3}}
\and S.~Cavuoti\orcid{0000-0002-3787-4196}\inst{\ref{aff35},\ref{aff47}}
\and K.~C.~Chambers\orcid{0000-0001-6965-7789}\inst{\ref{aff48}}
\and A.~Cimatti\inst{\ref{aff49}}
\and C.~Colodro-Conde\inst{\ref{aff50}}
\and G.~Congedo\orcid{0000-0003-2508-0046}\inst{\ref{aff51}}
\and C.~J.~Conselice\orcid{0000-0003-1949-7638}\inst{\ref{aff52}}
\and L.~Conversi\orcid{0000-0002-6710-8476}\inst{\ref{aff53},\ref{aff22}}
\and Y.~Copin\orcid{0000-0002-5317-7518}\inst{\ref{aff54}}
\and A.~Costille\inst{\ref{aff13}}
\and F.~Courbin\orcid{0000-0003-0758-6510}\inst{\ref{aff55},\ref{aff56},\ref{aff57}}
\and H.~M.~Courtois\orcid{0000-0003-0509-1776}\inst{\ref{aff58}}
\and M.~Cropper\orcid{0000-0003-4571-9468}\inst{\ref{aff59}}
\and A.~Da~Silva\orcid{0000-0002-6385-1609}\inst{\ref{aff60},\ref{aff61}}
\and G.~De~Lucia\orcid{0000-0002-6220-9104}\inst{\ref{aff25}}
\and H.~Dole\orcid{0000-0002-9767-3839}\inst{\ref{aff62},\ref{aff63}}
\and X.~Dupac\inst{\ref{aff22}}
\and S.~Dusini\orcid{0000-0002-1128-0664}\inst{\ref{aff64}}
\and S.~Escoffier\orcid{0000-0002-2847-7498}\inst{\ref{aff65}}
\and M.~Fabricius\orcid{0000-0002-7025-6058}\inst{\ref{aff5},\ref{aff6}}
\and M.~Farina\orcid{0000-0002-3089-7846}\inst{\ref{aff66}}
\and S.~Farrens\orcid{0000-0002-9594-9387}\inst{\ref{aff67}}
\and F.~Faustini\orcid{0000-0001-6274-5145}\inst{\ref{aff21}}
\and S.~Ferriol\inst{\ref{aff54}}
\and F.~Finelli\orcid{0000-0002-6694-3269}\inst{\ref{aff3},\ref{aff68}}
\and N.~Fourmanoit\orcid{0009-0005-6816-6925}\inst{\ref{aff65}}
\and M.~Frailis\orcid{0000-0002-7400-2135}\inst{\ref{aff25}}
\and E.~Franceschi\orcid{0000-0002-0585-6591}\inst{\ref{aff3}}
\and M.~Fumana\orcid{0000-0001-6787-5950}\inst{\ref{aff43}}
\and L.~Gabarra\orcid{0000-0002-8486-8856}\inst{\ref{aff69}}
\and S.~Galeotta\orcid{0000-0002-3748-5115}\inst{\ref{aff25}}
\and K.~George\orcid{0000-0002-1734-8455}\inst{\ref{aff70}}
\and W.~Gillard\orcid{0000-0003-4744-9748}\inst{\ref{aff65}}
\and C.~Giocoli\orcid{0000-0002-9590-7961}\inst{\ref{aff3},\ref{aff29}}
\and J.~Gracia-Carpio\orcid{0000-0003-4689-3134}\inst{\ref{aff5}}
\and A.~Grazian\orcid{0000-0002-5688-0663}\inst{\ref{aff30}}
\and F.~Grupp\inst{\ref{aff5},\ref{aff6}}
\and L.~Guzzo\orcid{0000-0001-8264-5192}\inst{\ref{aff71},\ref{aff23},\ref{aff72}}
\and S.~Gwyn\orcid{0000-0001-8221-8406}\inst{\ref{aff73}}
\and S.~V.~H.~Haugan\orcid{0000-0001-9648-7260}\inst{\ref{aff74}}
\and S.~Hemmati\orcid{0000-0003-2226-5395}\inst{\ref{aff75}}
\and H.~Hoekstra\orcid{0000-0002-0641-3231}\inst{\ref{aff42}}
\and W.~Holmes\orcid{0009-0007-8554-4646}\inst{\ref{aff76}}
\and I.~M.~Hook\orcid{0000-0002-2960-978X}\inst{\ref{aff77}}
\and F.~Hormuth\inst{\ref{aff78}}
\and A.~Hornstrup\orcid{0000-0002-3363-0936}\inst{\ref{aff79},\ref{aff80}}
\and M.~Huertas-Company\orcid{0000-0002-1416-8483}\inst{\ref{aff50},\ref{aff81},\ref{aff82}}
\and K.~Jahnke\orcid{0000-0003-3804-2137}\inst{\ref{aff83}}
\and M.~Jhabvala\inst{\ref{aff84}}
\and S.~Kermiche\orcid{0000-0002-0302-5735}\inst{\ref{aff65}}
\and A.~Kiessling\orcid{0000-0002-2590-1273}\inst{\ref{aff76}}
\and B.~Kubik\orcid{0009-0006-5823-4880}\inst{\ref{aff54}}
\and M.~K\"ummel\orcid{0000-0003-2791-2117}\inst{\ref{aff6}}
\and M.~Kunz\orcid{0000-0002-3052-7394}\inst{\ref{aff85}}
\and H.~Kurki-Suonio\orcid{0000-0002-4618-3063}\inst{\ref{aff86},\ref{aff87}}
\and A.~M.~C.~Le~Brun\orcid{0000-0002-0936-4594}\inst{\ref{aff88}}
\and P.~Liebing\orcid{0009-0005-8043-7150}\inst{\ref{aff59}}
\and S.~Ligori\orcid{0000-0003-4172-4606}\inst{\ref{aff41}}
\and P.~B.~Lilje\orcid{0000-0003-4324-7794}\inst{\ref{aff74}}
\and V.~Lindholm\orcid{0000-0003-2317-5471}\inst{\ref{aff86},\ref{aff87}}
\and I.~Lloro\orcid{0000-0001-5966-1434}\inst{\ref{aff89}}
\and M.~Magliocchetti\orcid{0000-0001-9158-4838}\inst{\ref{aff66}}
\and G.~Mainetti\orcid{0000-0003-2384-2377}\inst{\ref{aff90}}
\and O.~Mansutti\orcid{0000-0001-5758-4658}\inst{\ref{aff25}}
\and S.~Marcin\inst{\ref{aff91}}
\and O.~Marggraf\orcid{0000-0001-7242-3852}\inst{\ref{aff92}}
\and M.~Martinelli\orcid{0000-0002-6943-7732}\inst{\ref{aff21},\ref{aff44}}
\and N.~Martinet\orcid{0000-0003-2786-7790}\inst{\ref{aff13}}
\and F.~Marulli\orcid{0000-0002-8850-0303}\inst{\ref{aff93},\ref{aff3},\ref{aff29}}
\and R.~J.~Massey\orcid{0000-0002-6085-3780}\inst{\ref{aff94}}
\and N.~Mauri\orcid{0000-0001-8196-1548}\inst{\ref{aff49},\ref{aff29}}
\and C.~J.~R.~McPartland\orcid{0000-0003-0639-025X}\inst{\ref{aff80},\ref{aff95}}
\and E.~Medinaceli\orcid{0000-0002-4040-7783}\inst{\ref{aff3}}
\and S.~Mei\orcid{0000-0002-2849-559X}\inst{\ref{aff96},\ref{aff97}}
\and M.~Meneghetti\orcid{0000-0003-1225-7084}\inst{\ref{aff3},\ref{aff29}}
\and E.~Merlin\orcid{0000-0001-6870-8900}\inst{\ref{aff30}}
\and G.~Meylan\orcid{0000-0001-6503-0209}\inst{\ref{aff98}}
\and P.~Monaco\orcid{0000-0003-2083-7564}\inst{\ref{aff99},\ref{aff25},\ref{aff26},\ref{aff24}}
\and M.~Moresco\orcid{0000-0002-7616-7136}\inst{\ref{aff93},\ref{aff3}}
\and C.~Moretti\orcid{0000-0003-3314-8936}\inst{\ref{aff25},\ref{aff24},\ref{aff26}}
\and L.~Moscardini\orcid{0000-0002-3473-6716}\inst{\ref{aff93},\ref{aff3},\ref{aff29}}
\and R.~Nakajima\orcid{0009-0009-1213-7040}\inst{\ref{aff92}}
\and C.~Neissner\orcid{0000-0001-8524-4968}\inst{\ref{aff15},\ref{aff46}}
\and R.~C.~Nichol\orcid{0000-0003-0939-6518}\inst{\ref{aff100}}
\and S.-M.~Niemi\orcid{0009-0005-0247-0086}\inst{\ref{aff101}}
\and C.~Padilla\orcid{0000-0001-7951-0166}\inst{\ref{aff15}}
\and F.~Pasian\orcid{0000-0002-4869-3227}\inst{\ref{aff25}}
\and W.~J.~Percival\orcid{0000-0002-0644-5727}\inst{\ref{aff102},\ref{aff103},\ref{aff104}}
\and V.~Pettorino\orcid{0000-0002-4203-9320}\inst{\ref{aff101}}
\and S.~Pires\orcid{0000-0002-0249-2104}\inst{\ref{aff67}}
\and G.~Polenta\orcid{0000-0003-4067-9196}\inst{\ref{aff105}}
\and M.~Poncet\inst{\ref{aff106}}
\and L.~A.~Popa\inst{\ref{aff107}}
\and F.~Raison\orcid{0000-0002-7819-6918}\inst{\ref{aff5}}
\and A.~Refregier\inst{\ref{aff19}}
\and A.~Renzi\orcid{0000-0001-9856-1970}\inst{\ref{aff108},\ref{aff64},\ref{aff3}}
\and J.~Rhodes\orcid{0000-0002-4485-8549}\inst{\ref{aff76}}
\and G.~Riccio\inst{\ref{aff35}}
\and I.~Risso\orcid{0000-0003-2525-7761}\inst{\ref{aff32},\ref{aff33},\ref{aff23}}
\and E.~Romelli\orcid{0000-0003-3069-9222}\inst{\ref{aff25}}
\and M.~Roncarelli\orcid{0000-0001-9587-7822}\inst{\ref{aff3}}
\and C.~Rosset\orcid{0000-0003-0286-2192}\inst{\ref{aff96}}
\and B.~Rusholme\orcid{0000-0001-7648-4142}\inst{\ref{aff75}}
\and Z.~Sakr\orcid{0000-0002-4823-3757}\inst{\ref{aff109},\ref{aff110},\ref{aff111}}
\and D.~Sapone\orcid{0000-0001-7089-4503}\inst{\ref{aff112}}
\and B.~Sartoris\orcid{0000-0003-1337-5269}\inst{\ref{aff8}}
\and M.~Sauvage\orcid{0000-0002-0809-2574}\inst{\ref{aff67}}
\and M.~Schirmer\orcid{0000-0003-2568-9994}\inst{\ref{aff83}}
\and P.~Schneider\orcid{0000-0001-8561-2679}\inst{\ref{aff92}}
\and T.~Schrabback\orcid{0000-0002-6987-7834}\inst{\ref{aff7}}
\and M.~Scodeggio\inst{\ref{aff43}}
\and A.~Secroun\orcid{0000-0003-0505-3710}\inst{\ref{aff65}}
\and E.~Sefusatti\orcid{0000-0003-0473-1567}\inst{\ref{aff25},\ref{aff24},\ref{aff26}}
\and E.~Sihvola\orcid{0000-0003-1804-7715}\inst{\ref{aff86},\ref{aff87}}
\and C.~Sirignano\orcid{0000-0002-0995-7146}\inst{\ref{aff108},\ref{aff64}}
\and G.~Sirri\orcid{0000-0003-2626-2853}\inst{\ref{aff29}}
\and J.~Skottfelt\orcid{0000-0003-1310-8283}\inst{\ref{aff113}}
\and A.~Spurio~Mancini\orcid{0000-0001-5698-0990}\inst{\ref{aff114}}
\and L.~Stanco\orcid{0000-0002-9706-5104}\inst{\ref{aff64}}
\and P.~Tallada-Cresp\'{i}\orcid{0000-0002-1336-8328}\inst{\ref{aff45},\ref{aff46}}
\and A.~N.~Taylor\inst{\ref{aff51}}
\and H.~I.~Teplitz\orcid{0000-0002-7064-5424}\inst{\ref{aff16}}
\and I.~Tereno\orcid{0000-0002-4537-6218}\inst{\ref{aff60},\ref{aff61}}
\and N.~Tessore\orcid{0000-0002-9696-7931}\inst{\ref{aff59}}
\and S.~Toft\orcid{0000-0003-3631-7176}\inst{\ref{aff115},\ref{aff95}}
\and R.~Toledo-Moreo\orcid{0000-0002-2997-4859}\inst{\ref{aff116},\ref{aff117}}
\and F.~Torradeflot\orcid{0000-0003-1160-1517}\inst{\ref{aff46},\ref{aff45}}
\and A.~Tsyganov\inst{\ref{aff118}}
\and I.~Tutusaus\orcid{0000-0002-3199-0399}\inst{\ref{aff9},\ref{aff10},\ref{aff110}}
\and J.~Valiviita\orcid{0000-0001-6225-3693}\inst{\ref{aff86},\ref{aff87}}
\and T.~Vassallo\orcid{0000-0001-6512-6358}\inst{\ref{aff25},\ref{aff70}}
\and G.~Verdoes~Kleijn\orcid{0000-0001-5803-2580}\inst{\ref{aff11}}
\and A.~Veropalumbo\orcid{0000-0003-2387-1194}\inst{\ref{aff23},\ref{aff33},\ref{aff32}}
\and Y.~Wang\orcid{0000-0002-4749-2984}\inst{\ref{aff75}}
\and J.~Weller\orcid{0000-0002-8282-2010}\inst{\ref{aff6},\ref{aff5}}
\and A.~Zacchei\orcid{0000-0003-0396-1192}\inst{\ref{aff25},\ref{aff24}}
\and G.~Zamorani\orcid{0000-0002-2318-301X}\inst{\ref{aff3}}
\and F.~M.~Zerbi\orcid{0000-0002-9996-973X}\inst{\ref{aff23}}
\and E.~Zucca\orcid{0000-0002-5845-8132}\inst{\ref{aff3}}
\and V.~Allevato\orcid{0000-0001-7232-5152}\inst{\ref{aff35}}
\and I.~T.~Andika\orcid{0000-0001-6102-9526}\inst{\ref{aff6}}
\and M.~Ballardini\orcid{0000-0003-4481-3559}\inst{\ref{aff119},\ref{aff120},\ref{aff3}}
\and P.~Bergamini\orcid{0000-0003-1383-9414}\inst{\ref{aff3}}
\and S.~Borgani\orcid{0000-0001-6151-6439}\inst{\ref{aff99},\ref{aff24},\ref{aff25},\ref{aff26},\ref{aff121}}
\and E.~Bozzo\orcid{0000-0002-8201-1525}\inst{\ref{aff1}}
\and C.~Burigana\orcid{0000-0002-3005-5796}\inst{\ref{aff122},\ref{aff68}}
\and M.~Calabrese\orcid{0000-0002-2637-2422}\inst{\ref{aff123},\ref{aff43}}
\and A.~Cappi\inst{\ref{aff8},\ref{aff3}}
\and F.~Caro\orcid{0009-0003-1053-0507}\inst{\ref{aff21}}
\and T.~Castro\orcid{0000-0002-6292-3228}\inst{\ref{aff124},\ref{aff25},\ref{aff24}}
\and S.~Conseil\orcid{0000-0002-3657-4191}\inst{\ref{aff125},\ref{aff54}}
\and D.~Di~Ferdinando\inst{\ref{aff29}}
\and J.~A.~Escartin~Vigo\inst{\ref{aff5}}
\and G.~Fabbian\orcid{0000-0002-3255-4695}\inst{\ref{aff62}}
\and J.~Garc\'ia-Bellido\orcid{0000-0002-9370-8360}\inst{\ref{aff109}}
\and T.~Gasparetto\orcid{0000-0002-7913-4866}\inst{\ref{aff21}}
\and A.~Gregorio\orcid{0000-0003-4028-8785}\inst{\ref{aff99},\ref{aff25},\ref{aff26}}
\and C.~M.~Gutierrez\orcid{0000-0001-7854-783X}\inst{\ref{aff50},\ref{aff126}}
\and A.~Hall\orcid{0000-0002-3139-8651}\inst{\ref{aff51}}
\and A.~Loureiro\orcid{0000-0002-4371-0876}\inst{\ref{aff127},\ref{aff128}}
\and J.~Macias-Perez\orcid{0000-0002-5385-2763}\inst{\ref{aff129}}
\and C.~Mancini\orcid{0000-0002-4297-0561}\inst{\ref{aff43}}
\and R.~Maoli\orcid{0000-0002-6065-3025}\inst{\ref{aff130},\ref{aff21}}
\and R.~B.~Metcalf\orcid{0000-0003-3167-2574}\inst{\ref{aff93},\ref{aff3}}
\and S.~Nadathur\orcid{0000-0001-9070-3102}\inst{\ref{aff131}}
\and E.~Sarpa\orcid{0000-0002-1256-655X}\inst{\ref{aff25}}
\and V.~Scottez\orcid{0009-0008-3864-940X}\inst{\ref{aff132},\ref{aff133}}
\and M.~Sereno\orcid{0000-0003-0302-0325}\inst{\ref{aff3},\ref{aff29}}
\and K.~Tanidis\orcid{0000-0001-9843-5130}\inst{\ref{aff134}}
\and M.~Tenti\orcid{0000-0002-4254-5901}\inst{\ref{aff29}}
\and A.~Troja\orcid{0000-0003-0239-4595}\inst{\ref{aff25}}
\and M.~Viel\orcid{0000-0002-2642-5707}\inst{\ref{aff24},\ref{aff25},\ref{aff27},\ref{aff26},\ref{aff121}}
\and J.~R.~Weaver\orcid{0000-0003-1614-196X}\inst{\ref{aff135}}
\and M.~Wiesmann\orcid{0009-0000-8199-5860}\inst{\ref{aff74}}
\and Abdurro'uf\orcid{0000-0002-5258-8761}\inst{\ref{aff136}}
\and J.~A.~Acevedo~Barroso\orcid{0000-0002-9654-1711}\inst{\ref{aff76}}
\and R.~Adam\orcid{0009-0000-5380-1109}\inst{\ref{aff8}}
\and Y.~Akrami\orcid{0000-0002-2407-7956}\inst{\ref{aff109},\ref{aff137}}
\and G.~Angora\orcid{0000-0002-0316-6562}\inst{\ref{aff35},\ref{aff119}}
\and M.~Archidiacono\orcid{0000-0003-4952-9012}\inst{\ref{aff71},\ref{aff72}}
\and G.~Aric\`o\orcid{0000-0002-2802-2928}\inst{\ref{aff29}}
\and F.~Atrio-Barandela\orcid{0000-0002-2130-2513}\inst{\ref{aff138}}
\and M.~Aubert\orcid{0009-0002-7667-8814}\inst{\ref{aff54}}
\and E.~Aubourg\orcid{0000-0002-5592-023X}\inst{\ref{aff96},\ref{aff139}}
\and M.~Baes\orcid{0000-0002-3930-2757}\inst{\ref{aff140}}
\and N.~Bartolo\inst{\ref{aff108},\ref{aff64},\ref{aff30}}
\and L.~Bazzanini\orcid{0000-0003-0727-0137}\inst{\ref{aff119},\ref{aff3}}
\and J.~R.~Bermejo-Climent\inst{\ref{aff141},\ref{aff142}}
\and D.~Bertacca\orcid{0000-0002-2490-7139}\inst{\ref{aff108},\ref{aff30},\ref{aff64}}
\and M.~Bethermin\orcid{0000-0002-3915-2015}\inst{\ref{aff143}}
\and F.~Beutler\orcid{0000-0003-0467-5438}\inst{\ref{aff51}}
\and A.~Blanchard\orcid{0000-0001-8555-9003}\inst{\ref{aff110}}
\and H.~B\"ohringer\orcid{0000-0001-8241-4204}\inst{\ref{aff5},\ref{aff70},\ref{aff144}}
\and M.~Bonici\orcid{0000-0002-8430-126X}\inst{\ref{aff102},\ref{aff43}}
\and F.~Bouch\`e\orcid{0000-0002-4663-1786}\inst{\ref{aff145},\ref{aff47}}
\and M.~L.~Brown\orcid{0000-0002-0370-8077}\inst{\ref{aff52}}
\and S.~Bruton\orcid{0000-0002-6503-5218}\inst{\ref{aff146}}
\and F.~Buitrago\orcid{0000-0002-2861-9812}\inst{\ref{aff147},\ref{aff148},\ref{aff149}}
\and A.~Calabro\orcid{0000-0003-2536-1614}\inst{\ref{aff21}}
\and P.~Carrilho\orcid{0000-0003-1339-0194}\inst{\ref{aff150}}
\and C.~S.~Carvalho\inst{\ref{aff149}}
\and N.~Chartab\orcid{0000-0003-3691-937X}\inst{\ref{aff75}}
\and F.~Cogato\orcid{0000-0003-4632-6113}\inst{\ref{aff3}}
\and T.~E.~Collett\orcid{0000-0001-5564-3140}\inst{\ref{aff131}}
\and M.~Correnti\orcid{0000-0001-6464-3257}\inst{\ref{aff21},\ref{aff105}}
\and B.~Csizi\orcid{0000-0003-3227-6581}\inst{\ref{aff7}}
\and O.~Cucciati\orcid{0000-0002-9336-7551}\inst{\ref{aff3}}
\and H.~Dannerbauer\orcid{0000-0001-7147-3575}\inst{\ref{aff50},\ref{aff126}}
\and T.~de~Boer\orcid{0000-0001-5486-2747}\inst{\ref{aff48}}
\and A.~D\'iaz-S\'anchez\orcid{0000-0003-0748-4768}\inst{\ref{aff151}}
\and S.~Di~Domizio\orcid{0000-0003-2863-5895}\inst{\ref{aff32},\ref{aff33}}
\and J.~M.~Diego\orcid{0000-0001-9065-3926}\inst{\ref{aff152}}
\and P.-A.~Duc\orcid{0000-0003-3343-6284}\inst{\ref{aff143}}
\and V.~Duret\orcid{0009-0009-0383-4960}\inst{\ref{aff153}}
\and Y.~Fang\orcid{0000-0002-0334-6950}\inst{\ref{aff6}}
\and A.~Farina\orcid{0009-0000-3420-929X}\inst{\ref{aff23},\ref{aff33}}
\and A.~M.~N.~Ferguson\orcid{0000-0001-7934-1278}\inst{\ref{aff51}}
\and A.~Finoguenov\orcid{0000-0002-4606-5403}\inst{\ref{aff86}}
\and A.~Fontana\orcid{0000-0003-3820-2823}\inst{\ref{aff21}}
\and A.~Franco\orcid{0000-0002-4761-366X}\inst{\ref{aff154},\ref{aff155},\ref{aff156}}
\and N.~Frusciante\orcid{0000-0002-7375-1230}\inst{\ref{aff157},\ref{aff158}}
\and A.~Fumagalli\orcid{0009-0004-0300-2535}\inst{\ref{aff25}}
\and Y.~Fu\orcid{0000-0002-0759-0504}\inst{\ref{aff42},\ref{aff11}}
\and K.~Ganga\orcid{0000-0001-8159-8208}\inst{\ref{aff96}}
\and R.~Gavazzi\orcid{0000-0002-5540-6935}\inst{\ref{aff13},\ref{aff159}}
\and E.~Gaztanaga\orcid{0000-0001-9632-0815}\inst{\ref{aff9},\ref{aff10},\ref{aff131}}
\and Z.~Ghaffari\orcid{0000-0002-6467-8078}\inst{\ref{aff25},\ref{aff24}}
\and F.~Giacomini\orcid{0000-0002-3129-2814}\inst{\ref{aff29}}
\and F.~Gianotti\orcid{0000-0003-4666-119X}\inst{\ref{aff3}}
\and G.~Gozaliasl\orcid{0000-0002-0236-919X}\inst{\ref{aff160},\ref{aff86}}
\and A.~Gruppuso\orcid{0000-0001-9272-5292}\inst{\ref{aff3},\ref{aff29}}
\and M.~Guidi\orcid{0000-0001-9408-1101}\inst{\ref{aff28},\ref{aff3}}
\and N.~A.~Hatch\orcid{0000-0001-5600-0534}\inst{\ref{aff161}}
\and C.~Hern\'andez-Monteagudo\orcid{0000-0001-5471-9166}\inst{\ref{aff126},\ref{aff50}}
\and T.~Heuberger\orcid{0009-0009-3474-8467}\inst{\ref{aff91}}
\and J.~Hjorth\orcid{0000-0002-4571-2306}\inst{\ref{aff162}}
\and M.~Jauzac\orcid{0000-0003-1974-8732}\inst{\ref{aff163},\ref{aff94},\ref{aff164},\ref{aff165},\ref{aff110}}
\and S.~Joudaki\orcid{0000-0001-8820-673X}\inst{\ref{aff45},\ref{aff131}}
\and J.~J.~E.~Kajava\orcid{0000-0002-3010-8333}\inst{\ref{aff166},\ref{aff167},\ref{aff168}}
\and D.~Karagiannis\orcid{0000-0002-4927-0816}\inst{\ref{aff169},\ref{aff119},\ref{aff170}}
\and J.~Kim\orcid{0000-0003-2776-2761}\inst{\ref{aff69}}
\and C.~C.~Kirkpatrick\inst{\ref{aff20}}
\and A.~Kov\'acs\orcid{0000-0002-5825-579X}\inst{\ref{aff142},\ref{aff141}}
\and I.~Kova{\v{c}}i{\'{c}}\orcid{0000-0001-6751-3263}\inst{\ref{aff140}}
\and K.~Koyama\orcid{0000-0001-6727-6915}\inst{\ref{aff131}}
\and S.~Kruk\orcid{0000-0001-8010-8879}\inst{\ref{aff22}}
\and M.~C.~Lam\orcid{0000-0002-9347-2298}\inst{\ref{aff51}}
\and M.~Lattanzi\orcid{0000-0003-1059-2532}\inst{\ref{aff120}}
\and S.~Lee\orcid{0000-0002-8289-740X}\inst{\ref{aff171},\ref{aff76}}
\and J.~Le~Graet\orcid{0000-0001-6523-7971}\inst{\ref{aff65}}
\and M.~Lembo\orcid{0000-0002-5271-5070}\inst{\ref{aff159}}
\and M.~D.~Lepinzan\orcid{0000-0003-1287-9801}\inst{\ref{aff25}}
\and J.~Lesgourgues\orcid{0000-0001-7627-353X}\inst{\ref{aff172}}
\and T.~I.~Liaudat\orcid{0000-0002-9104-314X}\inst{\ref{aff139}}
\and L.~Linke\orcid{0000-0002-2622-8113}\inst{\ref{aff7}}
\and E.~A.~Magnier\orcid{0000-0002-7965-2815}\inst{\ref{aff48}}
\and F.~Mannucci\orcid{0000-0002-4803-2381}\inst{\ref{aff173}}
\and M.~Marinucci\orcid{0000-0003-1159-3756}\inst{\ref{aff174}}
\and F.~R.~Marleau\orcid{0000-0002-1442-2947}\inst{\ref{aff7}}
\and C.~J.~A.~P.~Martins\orcid{0000-0002-4886-9261}\inst{\ref{aff175},\ref{aff37}}
\and M.~Migliaccio\inst{\ref{aff176},\ref{aff177}}
\and M.~Miluzio\inst{\ref{aff22},\ref{aff178}}
\and G.~Morgante\inst{\ref{aff3}}
\and C.~Murray\orcid{0000-0002-4668-1273}\inst{\ref{aff67}}
\and K.~Naidoo\orcid{0000-0002-9182-1802}\inst{\ref{aff131},\ref{aff83}}
\and P.~Natoli\orcid{0000-0003-0126-9100}\inst{\ref{aff119},\ref{aff120}}
\and A.~Navarro-Alsina\orcid{0000-0002-3173-2592}\inst{\ref{aff92}}
\and A.~Nersesian\orcid{0000-0001-6843-409X}\inst{\ref{aff179},\ref{aff140}}
\and S.~Nesseris\orcid{0000-0002-0567-0324}\inst{\ref{aff109}}
\and L.~Nicastro\orcid{0000-0001-8534-6788}\inst{\ref{aff3}}
\and M.~Oguri\orcid{0000-0003-3484-399X}\inst{\ref{aff180},\ref{aff181}}
\and F.~Oppizzi\orcid{0000-0003-3904-8370}\inst{\ref{aff33},\ref{aff64},\ref{aff108}}
\and F.~Pace\orcid{0000-0001-8039-0480}\inst{\ref{aff39},\ref{aff40},\ref{aff41},\ref{aff61}}
\and M.~J.~Page\orcid{0000-0002-6689-6271}\inst{\ref{aff59}}
\and D.~Paoletti\orcid{0000-0003-4761-6147}\inst{\ref{aff3},\ref{aff68}}
\and G.~Papini\orcid{0000-0002-5810-318X}\inst{\ref{aff3},\ref{aff93}}
\and F.~Passalacqua\orcid{0000-0002-8606-4093}\inst{\ref{aff64}}
\and K.~Paterson\orcid{0000-0001-8340-3486}\inst{\ref{aff83}}
\and C.~Pattison\orcid{0000-0003-3272-2617}\inst{\ref{aff131}}
\and R.~Paviot\orcid{0009-0002-8108-3460}\inst{\ref{aff67},\ref{aff106}}
\and A.~Pisani\orcid{0000-0002-6146-4437}\inst{\ref{aff65}}
\and L.~Pizzuti\orcid{0000-0001-5654-7580}\inst{\ref{aff182}}
\and J.~J.~Popp\orcid{0000-0002-3724-1727}\inst{\ref{aff183}}
\and D.~Potter\orcid{0000-0002-0757-5195}\inst{\ref{aff184}}
\and M.~Poulain\orcid{0000-0002-7664-4510}\inst{\ref{aff185}}
\and G.~W.~Pratt\inst{\ref{aff67}}
\and S.~Quai\orcid{0000-0002-0449-8163}\inst{\ref{aff93},\ref{aff3}}
\and M.~Radovich\orcid{0000-0002-3585-866X}\inst{\ref{aff30}}
\and F.~Ricci\orcid{0000-0001-5742-5980}\inst{\ref{aff186},\ref{aff21}}
\and G.~Rodighiero\orcid{0000-0002-9415-2296}\inst{\ref{aff108},\ref{aff30}}
\and M.~Sahl\'en\orcid{0000-0003-0973-4804}\inst{\ref{aff187}}
\and D.~B.~Sanders\orcid{0000-0002-1233-9998}\inst{\ref{aff48}}
\and C.~Scarlata\orcid{0000-0002-9136-8876}\inst{\ref{aff188}}
\and J.~Schaye\orcid{0000-0002-0668-5560}\inst{\ref{aff42}}
\and A.~Schneider\orcid{0000-0001-7055-8104}\inst{\ref{aff184}}
\and M.~Schultheis\inst{\ref{aff8}}
\and E.~Sellentin\orcid{0009-0002-2655-3458}\inst{\ref{aff189},\ref{aff42}}
\and S.~Serjeant\orcid{0000-0002-0517-7943}\inst{\ref{aff183}}
\and L.~C.~Smith\orcid{0000-0002-3259-2771}\inst{\ref{aff190}}
\and J.~G.~Sorce\orcid{0000-0002-2307-2432}\inst{\ref{aff191},\ref{aff62}}
\and I.~Szapudi\orcid{0000-0003-2274-0301}\inst{\ref{aff48}}
\and M.~Talia\orcid{0000-0003-4352-2063}\inst{\ref{aff93},\ref{aff3}}
\and R.~Teyssier\orcid{0000-0001-7689-0933}\inst{\ref{aff192}}
\and S.~Tosi\orcid{0000-0002-7275-9193}\inst{\ref{aff32},\ref{aff23},\ref{aff33}}
\and C.~Uhlemann\orcid{0000-0001-7831-1579}\inst{\ref{aff193}}
\and C.~Valieri\inst{\ref{aff29}}
\and D.~Vergani\orcid{0000-0003-0898-2216}\inst{\ref{aff3}}
\and E.~Vilenius\orcid{0000-0002-6184-7681}\inst{\ref{aff59}}
\and S.~Vinciguerra\orcid{0009-0005-4018-3184}\inst{\ref{aff13}}
\and M.~von~Wietersheim-Kramsta\orcid{0000-0003-4986-5091}\inst{\ref{aff94},\ref{aff163}}
\and N.~A.~Walton\orcid{0000-0003-3983-8778}\inst{\ref{aff190}}
\and L.~Wang\orcid{0000-0002-6736-9158}\inst{\ref{aff194},\ref{aff11}}
\and D.~Scott\orcid{0000-0002-6878-9840}\inst{\ref{aff195}}}
										   
\institute{Department of Astronomy, University of Geneva, ch. d'Ecogia 16, 1290 Versoix, Switzerland\label{aff1}
\and
CEA Saclay, DFR/IRFU, Service d'Astrophysique, Bat. 709, 91191 Gif-sur-Yvette, France\label{aff2}
\and
INAF-Osservatorio di Astrofisica e Scienza dello Spazio di Bologna, Via Piero Gobetti 93/3, 40129 Bologna, Italy\label{aff3}
\and
School of Physics, HH Wills Physics Laboratory, University of Bristol, Tyndall Avenue, Bristol, BS8 1TL, UK\label{aff4}
\and
Max Planck Institute for Extraterrestrial Physics, Giessenbachstr. 1, 85748 Garching, Germany\label{aff5}
\and
Universit\"ats-Sternwarte M\"unchen, Fakult\"at f\"ur Physik, Ludwig-Maximilians-Universit\"at M\"unchen, Scheinerstr.~1, 81679 M\"unchen, Germany\label{aff6}
\and
Universit\"at Innsbruck, Institut f\"ur Astro- und Teilchenphysik, Technikerstr. 25/8, 6020 Innsbruck, Austria\label{aff7}
\and
Universit\'e C\^{o}te d'Azur, Observatoire de la C\^{o}te d'Azur, CNRS, Laboratoire Lagrange, Bd de l'Observatoire, CS 34229, 06304 Nice cedex 4, France\label{aff8}
\and
Institute of Space Sciences (ICE, CSIC), Campus UAB, Carrer de Can Magrans, s/n, 08193 Barcelona, Spain\label{aff9}
\and
Institut d'Estudis Espacials de Catalunya (IEEC),  Edifici RDIT, Campus UPC, 08860 Castelldefels, Barcelona, Spain\label{aff10}
\and
Kapteyn Astronomical Institute, University of Groningen, PO Box 800, 9700 AV Groningen, The Netherlands\label{aff11}
\and
Ruhr University Bochum, Faculty of Physics and Astronomy, Astronomical Institute (AIRUB), German Centre for Cosmological Lensing (GCCL), 44780 Bochum, Germany\label{aff12}
\and
Aix-Marseille Universit\'e, CNRS, CNES, LAM, Marseille, France\label{aff13}
\and
Serra H\'unter Fellow, Departament de F\'{\i}sica, Universitat Aut\`onoma de Barcelona, E-08193 Bellaterra, Spain\label{aff14}
\and
Institut de F\'{i}sica d'Altes Energies (IFAE), The Barcelona Institute of Science and Technology, Campus UAB, 08193 Bellaterra (Barcelona), Spain\label{aff15}
\and
Infrared Processing and Analysis Center, California Institute of Technology, Pasadena, CA 91125, USA\label{aff16}
\and
Department of Physics and Astronomy, University of California, Davis, CA 95616, USA\label{aff17}
\and
Kobayashi-Maskawa Institute for the Origin of Particles and the Universe, Nagoya University, Chikusa-ku, Nagoya, 464-8602, Japan\label{aff18}
\and
Institute for Particle Physics and Astrophysics, Dept. of Physics, ETH Zurich, Wolfgang-Pauli-Strasse 27, 8093 Zurich, Switzerland\label{aff19}
\and
Department of Physics and Helsinki Institute of Physics, Gustaf H\"allstr\"omin katu 2, University of Helsinki, 00014 Helsinki, Finland\label{aff20}
\and
INAF-Osservatorio Astronomico di Roma, Via Frascati 33, 00078 Monteporzio Catone, Italy\label{aff21}
\and
ESAC/ESA, Camino Bajo del Castillo, s/n., Urb. Villafranca del Castillo, 28692 Villanueva de la Ca\~nada, Madrid, Spain\label{aff22}
\and
INAF-Osservatorio Astronomico di Brera, Via Brera 28, 20122 Milano, Italy\label{aff23}
\and
IFPU, Institute for Fundamental Physics of the Universe, via Beirut 2, 34151 Trieste, Italy\label{aff24}
\and
INAF-Osservatorio Astronomico di Trieste, Via G. B. Tiepolo 11, 34143 Trieste, Italy\label{aff25}
\and
INFN, Sezione di Trieste, Via Valerio 2, 34127 Trieste TS, Italy\label{aff26}
\and
SISSA, International School for Advanced Studies, Via Bonomea 265, 34136 Trieste TS, Italy\label{aff27}
\and
Dipartimento di Fisica e Astronomia, Universit\`a di Bologna, Via Gobetti 93/2, 40129 Bologna, Italy\label{aff28}
\and
INFN-Sezione di Bologna, Viale Berti Pichat 6/2, 40127 Bologna, Italy\label{aff29}
\and
INAF-Osservatorio Astronomico di Padova, Via dell'Osservatorio 5, 35122 Padova, Italy\label{aff30}
\and
ATG Europe BV, Huygensstraat 34, 2201 DK Noordwijk, The Netherlands\label{aff31}
\and
Dipartimento di Fisica, Universit\`a di Genova, Via Dodecaneso 33, 16146, Genova, Italy\label{aff32}
\and
INFN-Sezione di Genova, Via Dodecaneso 33, 16146, Genova, Italy\label{aff33}
\and
Department of Physics "E. Pancini", University Federico II, Via Cinthia 6, 80126, Napoli, Italy\label{aff34}
\and
INAF-Osservatorio Astronomico di Capodimonte, Via Moiariello 16, 80131 Napoli, Italy\label{aff35}
\and
European Southern Observatory, Karl-Schwarzschild-Str.~2, 85748 Garching, Germany\label{aff36}
\and
Instituto de Astrof\'isica e Ci\^encias do Espa\c{c}o, Universidade do Porto, CAUP, Rua das Estrelas, PT4150-762 Porto, Portugal\label{aff37}
\and
Faculdade de Ci\^encias da Universidade do Porto, Rua do Campo de Alegre, 4150-007 Porto, Portugal\label{aff38}
\and
Dipartimento di Fisica, Universit\`a degli Studi di Torino, Via P. Giuria 1, 10125 Torino, Italy\label{aff39}
\and
INFN-Sezione di Torino, Via P. Giuria 1, 10125 Torino, Italy\label{aff40}
\and
INAF-Osservatorio Astrofisico di Torino, Via Osservatorio 20, 10025 Pino Torinese (TO), Italy\label{aff41}
\and
Leiden Observatory, Leiden University, Einsteinweg 55, 2333 CC Leiden, The Netherlands\label{aff42}
\and
INAF-IASF Milano, Via Alfonso Corti 12, 20133 Milano, Italy\label{aff43}
\and
INFN-Sezione di Roma, Piazzale Aldo Moro, 2 - c/o Dipartimento di Fisica, Edificio G. Marconi, 00185 Roma, Italy\label{aff44}
\and
Centro de Investigaciones Energ\'eticas, Medioambientales y Tecnol\'ogicas (CIEMAT), Avenida Complutense 40, 28040 Madrid, Spain\label{aff45}
\and
Port d'Informaci\'{o} Cient\'{i}fica, Campus UAB, C. Albareda s/n, 08193 Bellaterra (Barcelona), Spain\label{aff46}
\and
INFN -- Sezione di Napoli, Via Cinthia 6, 80126, Napoli, Italy\label{aff47}
\and
Institute for Astronomy, University of Hawaii, 2680 Woodlawn Drive, Honolulu, HI 96822, USA\label{aff48}
\and
Dipartimento di Fisica e Astronomia "Augusto Righi" - Alma Mater Studiorum Universit\`a di Bologna, Viale Berti Pichat 6/2, 40127 Bologna, Italy\label{aff49}
\and
Instituto de Astrof\'{\i}sica de Canarias, E-38205 La Laguna, Tenerife, Spain\label{aff50}
\and
Institute for Astronomy, University of Edinburgh, Royal Observatory, Blackford Hill, Edinburgh EH9 3HJ, UK\label{aff51}
\and
Jodrell Bank Centre for Astrophysics, Department of Physics and Astronomy, University of Manchester, Oxford Road, Manchester M13 9PL, UK\label{aff52}
\and
European Space Agency/ESRIN, Largo Galileo Galilei 1, 00044 Frascati, Roma, Italy\label{aff53}
\and
Universit\'e Lyon 1, CNRS/IN2P3, IP2I Lyon, Villeurbanne, France\label{aff54}
\and
Institut de Ci\`{e}ncies del Cosmos (ICCUB), Universitat de Barcelona (IEEC-UB), Mart\'{i} i Franqu\`{e}s 1, 08028 Barcelona, Spain\label{aff55}
\and
Instituci\'o Catalana de Recerca i Estudis Avan\c{c}ats (ICREA), Passeig de Llu\'{\i}s Companys 23, 08010 Barcelona, Spain\label{aff56}
\and
Institut de Ciencies de l'Espai (IEEC-CSIC), Campus UAB, Carrer de Can Magrans, s/n Cerdanyola del Vall\'es, 08193 Barcelona, Spain\label{aff57}
\and
Universit\'e Lyon 1, CNRS/IN2P3, IUF, IP2I Lyon, Villeurbanne, France \label{aff58}
\and
Mullard Space Science Laboratory, University College London, Holmbury St Mary, Dorking, Surrey RH5 6NT, UK\label{aff59}
\and
Departamento de F\'isica, Faculdade de Ci\^encias, Universidade de Lisboa, Edif\'icio C8, Campo Grande, PT1749-016 Lisboa, Portugal\label{aff60}
\and
Instituto de Astrof\'isica e Ci\^encias do Espa\c{c}o, Faculdade de Ci\^encias, Universidade de Lisboa, Campo Grande, 1749-016 Lisboa, Portugal\label{aff61}
\and
Universit\'e Paris-Saclay, CNRS, Institut d'astrophysique spatiale, 91405, Orsay, France\label{aff62}
\and
Institut universitaire de France (IUF), 1 rue Descartes, 75231 PARIS CEDEX 05, France\label{aff63}
\and
INFN-Padova, Via Marzolo 8, 35131 Padova, Italy\label{aff64}
\and
Aix-Marseille Universit\'e, CNRS/IN2P3, CPPM, Marseille, France\label{aff65}
\and
INAF-Istituto di Astrofisica e Planetologia Spaziali, via del Fosso del Cavaliere, 100, 00100 Roma, Italy\label{aff66}
\and
Universit\'e Paris-Saclay, Universit\'e Paris Cit\'e, CEA, CNRS, AIM, 91191, Gif-sur-Yvette, France\label{aff67}
\and
INFN-Bologna, Via Irnerio 46, 40126 Bologna, Italy\label{aff68}
\and
Department of Physics, University of Oxford, Keble Road, Oxford OX1 3RH, UK\label{aff69}
\and
University Observatory, LMU Faculty of Physics, Scheinerstr.~1, 81679 Munich, Germany\label{aff70}
\and
Dipartimento di Fisica "Aldo Pontremoli", Universit\`a degli Studi di Milano, Via Celoria 16, 20133 Milano, Italy\label{aff71}
\and
INFN-Sezione di Milano, Via Celoria 16, 20133 Milano, Italy\label{aff72}
\and
National Research Council, Herzberg Astronomy and Astrophysics Research Centre, 5071 W. Saanich Rd. Victoria, BC, V9E 2E7, Canada\label{aff73}
\and
Institute of Theoretical Astrophysics, University of Oslo, P.O. Box 1029 Blindern, 0315 Oslo, Norway\label{aff74}
\and
Caltech/IPAC, 1200 E. California Blvd., Pasadena, CA 91125, USA\label{aff75}
\and
Jet Propulsion Laboratory, California Institute of Technology, 4800 Oak Grove Drive, Pasadena, CA, 91109, USA\label{aff76}
\and
School of Physics and Astronomy, Lancaster University, Lancaster, LA1 4YB, UK\label{aff77}
\and
Felix Hormuth Engineering, Goethestr. 17, 69181 Leimen, Germany\label{aff78}
\and
Technical University of Denmark, Elektrovej 327, 2800 Kgs. Lyngby, Denmark\label{aff79}
\and
Cosmic Dawn Center (DAWN), Denmark\label{aff80}
\and
Universit\'e PSL, Observatoire de Paris, Sorbonne Universit\'e, CNRS, LERMA, 75014, Paris, France\label{aff81}
\and
Universit\'e Paris-Cit\'e, 5 Rue Thomas Mann, 75013, Paris, France\label{aff82}
\and
Max-Planck-Institut f\"ur Astronomie, K\"onigstuhl 17, 69117 Heidelberg, Germany\label{aff83}
\and
NASA Goddard Space Flight Center, Greenbelt, MD 20771, USA\label{aff84}
\and
Universit\'e de Gen\`eve, D\'epartement de Physique Th\'eorique and Centre for Astroparticle Physics, 24 quai Ernest-Ansermet, CH-1211 Gen\`eve 4, Switzerland\label{aff85}
\and
Department of Physics, P.O. Box 64, University of Helsinki, 00014 Helsinki, Finland\label{aff86}
\and
Helsinki Institute of Physics, Gustaf H{\"a}llstr{\"o}min katu 2, University of Helsinki, 00014 Helsinki, Finland\label{aff87}
\and
Laboratoire d'etude de l'Univers et des phenomenes eXtremes, Observatoire de Paris, Universit\'e PSL, Sorbonne Universit\'e, CNRS, 92190 Meudon, France\label{aff88}
\and
SKAO, Jodrell Bank, Lower Withington, Macclesfield SK11 9FT, UK\label{aff89}
\and
Centre de Calcul de l'IN2P3/CNRS, 21 avenue Pierre de Coubertin 69627 Villeurbanne Cedex, France\label{aff90}
\and
University of Applied Sciences and Arts of Northwestern Switzerland, School of Computer Science, 5210 Windisch, Switzerland\label{aff91}
\and
Universit\"at Bonn, Argelander-Institut f\"ur Astronomie, Auf dem H\"ugel 71, 53121 Bonn, Germany\label{aff92}
\and
Dipartimento di Fisica e Astronomia "Augusto Righi" - Alma Mater Studiorum Universit\`a di Bologna, via Piero Gobetti 93/2, 40129 Bologna, Italy\label{aff93}
\and
Department of Physics, Institute for Computational Cosmology, Durham University, South Road, Durham, DH1 3LE, UK\label{aff94}
\and
Niels Bohr Institute, University of Copenhagen, Jagtvej 128, 2200 Copenhagen, Denmark\label{aff95}
\and
Universit\'e Paris Cit\'e, CNRS, Astroparticule et Cosmologie, 75013 Paris, France\label{aff96}
\and
CNRS-UCB International Research Laboratory, Centre Pierre Bin\'etruy, IRL2007, CPB-IN2P3, Berkeley, USA\label{aff97}
\and
Institute of Physics, Laboratory of Astrophysics, Ecole Polytechnique F\'ed\'erale de Lausanne (EPFL), Observatoire de Sauverny, 1290 Versoix, Switzerland\label{aff98}
\and
Dipartimento di Fisica - Sezione di Astronomia, Universit\`a di Trieste, Via Tiepolo 11, 34131 Trieste, Italy\label{aff99}
\and
School of Mathematics and Physics, University of Surrey, Guildford, Surrey, GU2 7XH, UK\label{aff100}
\and
European Space Agency/ESTEC, Keplerlaan 1, 2201 AZ Noordwijk, The Netherlands\label{aff101}
\and
Waterloo Centre for Astrophysics, University of Waterloo, Waterloo, Ontario N2L 3G1, Canada\label{aff102}
\and
Department of Physics and Astronomy, University of Waterloo, Waterloo, Ontario N2L 3G1, Canada\label{aff103}
\and
Perimeter Institute for Theoretical Physics, Waterloo, Ontario N2L 2Y5, Canada\label{aff104}
\and
Space Science Data Center, Italian Space Agency, via del Politecnico snc, 00133 Roma, Italy\label{aff105}
\and
Centre National d'Etudes Spatiales -- Centre spatial de Toulouse, 18 avenue Edouard Belin, 31401 Toulouse Cedex 9, France\label{aff106}
\and
Institute of Space Science, Str. Atomistilor, nr. 409 M\u{a}gurele, Ilfov, 077125, Romania\label{aff107}
\and
Dipartimento di Fisica e Astronomia "G. Galilei", Universit\`a di Padova, Via Marzolo 8, 35131 Padova, Italy\label{aff108}
\and
Instituto de F\'isica Te\'orica UAM-CSIC, Campus de Cantoblanco, 28049 Madrid, Spain\label{aff109}
\and
Institut de Recherche en Astrophysique et Plan\'etologie (IRAP), Universit\'e de Toulouse, CNRS, UPS, CNES, 14 Av. Edouard Belin, 31400 Toulouse, France\label{aff110}
\and
Universit\'e St Joseph; Faculty of Sciences, Beirut, Lebanon\label{aff111}
\and
Departamento de F\'isica, FCFM, Universidad de Chile, Blanco Encalada 2008, Santiago, Chile\label{aff112}
\and
Centre for Electronic Imaging, Open University, Walton Hall, Milton Keynes, MK7~6AA, UK\label{aff113}
\and
Department of Physics, Royal Holloway, University of London, Surrey TW20 0EX, UK\label{aff114}
\and
Cosmic Dawn Center (DAWN)\label{aff115}
\and
Universidad Polit\'ecnica de Cartagena, Departamento de Electr\'onica y Tecnolog\'ia de Computadoras,  Plaza del Hospital 1, 30202 Cartagena, Spain\label{aff116}
\and
European University of Technology EUt+, European Union\label{aff117}
\and
Centre for Information Technology, University of Groningen, P.O. Box 11044, 9700 CA Groningen, The Netherlands\label{aff118}
\and
Dipartimento di Fisica e Scienze della Terra, Universit\`a degli Studi di Ferrara, Via Giuseppe Saragat 1, 44122 Ferrara, Italy\label{aff119}
\and
Istituto Nazionale di Fisica Nucleare, Sezione di Ferrara, Via Giuseppe Saragat 1, 44122 Ferrara, Italy\label{aff120}
\and
ICSC - Centro Nazionale di Ricerca in High Performance Computing, Big Data e Quantum Computing, Via Magnanelli 2, Bologna, Italy\label{aff121}
\and
INAF, Istituto di Radioastronomia, Via Piero Gobetti 101, 40129 Bologna, Italy\label{aff122}
\and
Astronomical Observatory of the Autonomous Region of the Aosta Valley (OAVdA), Loc. Lignan 39, I-11020, Nus (Aosta Valley), Italy\label{aff123}
\and
Department of Mathematical Physics, Institute of Physics, University of S\~ao Paulo, R. do Mat\~ao 1371, 05508-090, S\~ao Paulo, SP, Brazil\label{aff124}
\and
Centre de Recherche Astrophysique de Lyon, UMR5574, CNRS, Universit\'e Claude Bernard Lyon 1, ENS de Lyon, 69230, Saint-Genis-Laval, France\label{aff125}
\and
Universidad de La Laguna, Dpto. Astrof\'\i sica, E-38206 La Laguna, Tenerife, Spain\label{aff126}
\and
Oskar Klein Centre for Cosmoparticle Physics, Department of Physics, Stockholm University, Stockholm, SE-106 91, Sweden\label{aff127}
\and
Astrophysics Group, Blackett Laboratory, Imperial College London, London SW7 2AZ, UK\label{aff128}
\and
Univ. Grenoble Alpes, CNRS, Grenoble INP, LPSC-IN2P3, 53, Avenue des Martyrs, 38000 Grenoble, France\label{aff129}
\and
Dipartimento di Fisica, Sapienza Universit\`a di Roma, Piazzale Aldo Moro 2, 00185 Roma, Italy\label{aff130}
\and
Institute of Cosmology and Gravitation, University of Portsmouth, Portsmouth PO1 3FX, UK\label{aff131}
\and
Institut d'Astrophysique de Paris, 98bis Boulevard Arago, 75014, Paris, France\label{aff132}
\and
ICL, Junia, Universit\'e Catholique de Lille, LITL, 59000 Lille, France\label{aff133}
\and
Center for Astrophysics and Cosmology, University of Nova Gorica, Nova Gorica, Slovenia\label{aff134}
\and
MIT Kavli Institute for Astrophysics and Space Research, Massachusetts Institute of Technology, Cambridge, MA 02139, USA\label{aff135}
\and
Department of Astronomy, Indiana University, 727 East Third Street, Bloomington, IN 47405, USA\label{aff136}
\and
CERCA/ISO, Department of Physics, Case Western Reserve University, 10900 Euclid Avenue, Cleveland, OH 44106, USA\label{aff137}
\and
Departamento de F{\'\i}sica Fundamental. Universidad de Salamanca. Plaza de la Merced s/n. 37008 Salamanca, Spain\label{aff138}
\and
IRFU, CEA, Universit\'e Paris-Saclay 91191 Gif-sur-Yvette Cedex, France\label{aff139}
\and
Universiteit Gent, Department of Physics and Astronomy, Proeftuinstraat 86 N3, 9000 Ghent, Belgium
\label{aff140}
\and
Konkoly Observatory, HUN-REN CSFK, MTA Centre of Excellence, Budapest, Konkoly Thege Mikl\'os {\'u}t 15-17. H-1121, Hungary\label{aff141}
\and
MTA-CSFK Lend\"ulet Large-Scale Structure Research Group, Konkoly-Thege Mikl\'os \'ut 15-17, H-1121 Budapest, Hungary\label{aff142}
\and
Universit\'e de Strasbourg, CNRS, Observatoire astronomique de Strasbourg, UMR 7550, 67000 Strasbourg, France\label{aff143}
\and
Max-Planck-Institut f\"ur Physik, Boltzmannstr. 8, 85748 Garching, Germany\label{aff144}
\and
Scuola Superiore Meridionale, Via Mezzocannone 4, 80138, Napoli, Italy\label{aff145}
\and
California Institute of Technology, 1200 E California Blvd, Pasadena, CA 91125, USA\label{aff146}
\and
Departamento de F\'{i}sica Te\'{o}rica, At\'{o}mica y \'{O}ptica, Universidad de Valladolid, 47011 Valladolid, Spain\label{aff147}
\and
Laboratory for Disruptive Interdisciplinary Science (LaDIS), Universidad de Valladolid, 47011 Valladolid, Spain\label{aff148}
\and
Instituto de Astrof\'isica e Ci\^encias do Espa\c{c}o, Faculdade de Ci\^encias, Universidade de Lisboa, Tapada da Ajuda, 1349-018 Lisboa, Portugal\label{aff149}
\and
Department of Physics, Astronomy and Mathematics, University of Hertfordshire, College Lane, Hatfield AL10 9AB, UK\label{aff150}
\and
Departamento F\'isica Aplicada, Universidad Polit\'ecnica de Cartagena, Campus Muralla del Mar, 30202 Cartagena, Murcia, Spain\label{aff151}
\and
Instituto de F\'isica de Cantabria, Edificio Juan Jord\'a, Avenida de los Castros, 39005 Santander, Spain\label{aff152}
\and
Univ. Grenoble Alpes, CNRS, IPAG, 38000 Grenoble, France\label{aff153}
\and
Department of Mathematics and Physics E. De Giorgi, University of Salento, Via per Arnesano, CP-I93, 73100, Lecce, Italy\label{aff154}
\and
INFN, Sezione di Lecce, Via per Arnesano, CP-193, 73100, Lecce, Italy\label{aff155}
\and
INAF-Sezione di Lecce, c/o Dipartimento Matematica e Fisica, Via per Arnesano, 73100, Lecce, Italy\label{aff156}
\and
Dipartimento di Fisica "E. Pancini", Universita degli Studi di Napoli Federico II, Via Cinthia 6, 80126, Napoli, Italy\label{aff157}
\and
INFN Sezione di Napoli, Universita degli Studi di Napoli "Federico II", Compl. Univ. di Monte S. Angelo, Edificio G, Via Cinthia, I-80126, Napoli, Italy\label{aff158}
\and
Institut d'Astrophysique de Paris, UMR 7095, CNRS, and Sorbonne Universit\'e, 98 bis boulevard Arago, 75014 Paris, France\label{aff159}
\and
Department of Computer Science, Aalto University, PO Box 15400, Espoo, FI-00 076, Finland\label{aff160}
\and
School of Physics and Astronomy, University of Nottingham, University Park, Nottingham NG7 2RD, UK\label{aff161}
\and
DARK, Niels Bohr Institute, University of Copenhagen, Jagtvej 155, 2200 Copenhagen, Denmark\label{aff162}
\and
Department of Physics, Centre for Extragalactic Astronomy, Durham University, South Road, Durham, DH1 3LE, UK\label{aff163}
\and
Astrophysics Research Centre, University of KwaZulu-Natal, Westville Campus, Durban 4041, South Africa\label{aff164}
\and
School of Mathematics, Statistics \& Computer Science, University of KwaZulu-Natal, Westville Campus, Durban 4041, South Africa\label{aff165}
\and
Department of Physics and Astronomy, Vesilinnantie 5, University of Turku, 20014 Turku, Finland\label{aff166}
\and
Finnish Centre for Astronomy with ESO (FINCA), Quantum, Vesilinnantie 5, University of Turku, 20014 Turku, Finland\label{aff167}
\and
Serco for European Space Agency (ESA), Camino bajo del Castillo, s/n, Urbanizacion Villafranca del Castillo, Villanueva de la Ca\~nada, 28692 Madrid, Spain\label{aff168}
\and
Van Swinderen Institute, University of Groningen, Nijenborgh 3, 9747 AG Groningen, The Netherlands\label{aff169}
\and
Department of Physics and Astronomy, University of the Western Cape, Bellville, Cape Town, 7535, South Africa\label{aff170}
\and
Ohio University, Physics \& Astronomy Department,1 Ohio University, Athens, OH 45701, USA\label{aff171}
\and
Institute for Theoretical Particle Physics and Cosmology (TTK), RWTH Aachen University, 52056 Aachen, Germany\label{aff172}
\and
INAF-Osservatorio Astrofisico di Arcetri, Largo E. Fermi 5, 50125, Firenze, Italy\label{aff173}
\and
Institute for Theoretical Physics, ETH Zurich, Wolfgang-Pauli-Str. 27, 8093 Zurich, Switzerland\label{aff174}
\and
Centro de Astrof\'{\i}sica da Universidade do Porto, Rua das Estrelas, 4150-762 Porto, Portugal\label{aff175}
\and
Dipartimento di Fisica, Universit\`a di Roma Tor Vergata, Via della Ricerca Scientifica 1, Roma, Italy\label{aff176}
\and
INFN, Sezione di Roma 2, Via della Ricerca Scientifica 1, Roma, Italy\label{aff177}
\and
HE Space for European Space Agency (ESA), Camino bajo del Castillo, s/n, Urbanizacion Villafranca del Castillo, Villanueva de la Ca\~nada, 28692 Madrid, Spain\label{aff178}
\and
STAR Institute, University of Li{\`e}ge, Quartier Agora, All\'ee du six Ao\^ut 19c, 4000 Li\`ege, Belgium\label{aff179}
\and
Center for Frontier Science, Chiba University, 1-33 Yayoi-cho, Inage-ku, Chiba 263-8522, Japan\label{aff180}
\and
Department of Physics, Graduate School of Science, Chiba University, 1-33 Yayoi-Cho, Inage-Ku, Chiba 263-8522, Japan\label{aff181}
\and
Dipartimento di Fisica ``G. Occhialini", Universit\`a degli Studi di Milano Bicocca, Piazza della Scienza 3, 20126 Milano, Italy\label{aff182}
\and
School of Physical Sciences, The Open University, Milton Keynes, MK7 6AA, UK\label{aff183}
\and
Department of Astrophysics, University of Zurich, Winterthurerstrasse 190, 8057 Zurich, Switzerland\label{aff184}
\and
Space physics and astronomy research unit, University of Oulu, Pentti Kaiteran katu 1, FI-90014 Oulu, Finland\label{aff185}
\and
Department of Mathematics and Physics, Roma Tre University, Via della Vasca Navale 84, 00146 Rome, Italy\label{aff186}
\and
Theoretical astrophysics, Department of Physics and Astronomy, Uppsala University, Box 524, 751 20 Uppsala, Sweden\label{aff187}
\and
Minnesota Institute for Astrophysics, University of Minnesota, 116 Church St SE, Minneapolis, MN 55455, USA\label{aff188}
\and
Mathematical Institute, University of Leiden, Einsteinweg 55, 2333 CA Leiden, The Netherlands\label{aff189}
\and
Institute of Astronomy, University of Cambridge, Madingley Road, Cambridge CB3 0HA, UK\label{aff190}
\and
Univ. Lille, CNRS, Centrale Lille, UMR 9189 CRIStAL, 59000 Lille, France\label{aff191}
\and
Department of Astrophysical Sciences, Peyton Hall, Princeton University, Princeton, NJ 08544, USA\label{aff192}
\and
Fakult\"at f\"ur Physik, Universit\"at Bielefeld, Postfach 100131, 33501 Bielefeld, Germany\label{aff193}
\and
SRON Netherlands Institute for Space Research, Landleven 12, 9747 AD, Groningen, The Netherlands\label{aff194}
\and
Department of Physics and Astronomy, University of British Columbia, Vancouver, BC V6T 1Z1, Canada\label{aff195}}     


%
%
 \abstract{We present the pipeline design of the photometric redshift processing function for Data Release 1 (DR1) of the \Euclid mission, and the validation and performance of our cosmology-related products that are released as part of DR1-Foundation. The \Euclid mission aims to constrain the equation-of-state parameter of dark energy, $w$, to percent-level accuracy through the measurement of large-scale structure via weak gravitational lensing and galaxy clustering. As part of DR1-Foundation we release determinations of photometric redshifts (photo-$z$), corrected photometry, and object classes that are used to form the basis of the cosmological samples. The release furthermore includes estimates of the spectral energy distributions of stars and galaxies for the careful modelling of the point-spread function for weak-lensing measurement. We estimate the performance of the photo-$z$ point estimates and probability distribution functions using overlapping spectroscopic redshifts, which are weighted to match different flux-limited target samples. We find that the current performance is somewhat below mission design requirements ($40\%$ in scatter and a factor of 2 in outlier fraction) at the nominal survey depth ($\IE<24.5$), but closely matching requirements at slightly brighter magnitudes ($\IE<24$), consistent with pre-launch analyses and the slightly shallower external data (up to $0.3\,\mathrm{mag}$) with respect to end-of-mission expectations. Data from Rubin-LSST will help alleviate this issue by the time of the final cosmology analyses. The other cosmology-related outputs are tested to the extent that is possible with currently available comparison data. This paper is partnered by an article describing the measurement of the physical properties of galaxies for the purpose of non-cosmological science.}
 
%
%
\keywords{Galaxies: distances and redshifts -- Catalogues -- Methods: statistical -- Surveys -- Cosmology: observations}
%
%
   \titlerunning{Photo-z pipeline and validation for DR1}
   \authorrunning{Euclid Collaboration: Hartley et al.}
   
   \maketitle


\section{\label{sc:intro}Introduction}
Data Release 1 of the \Euclid mission \citep[DR1,][]{DR1cite, DR1-TP001} is the first major release, following the smaller Early Release Observations \citep{EROData} and Quick Data Releases \citep[Q1 and Q2][]{Q1-TP001,Q2-TP002} data sets. It comprises the \gls{EWS} with data of the same depth as the Q1 release; the \glspl{EDF}, which cover the same area as Q1 but to greater depth; and numerous calibration observations, including the \glspl{EAF} for photo-$z$ and colour gradient calibration. 

In this work we present an overview of the photometric redshift processing function (PHZ PF), developed by the organisation unit for photometric redshifts (OU-PHZ) and the Swiss science data centre (SDC-CH), and its output products that are released as part of DR1-Foundation. In particular, we focus on aspects that are new with respect to the Q1 pipeline (which is described in detail in \citealt{Q1-TP005}, EC:T26 hereafter), the most significant difference being the availability of two of the photo-$z$ \glspl{EAF} and the resulting implementation of the pre-launch baseline strategy (see Sect.\,\ref{sc:pipe_photoz} and \citealt{EuclidSkyOverview}). This overview can be found in Sect.\,\ref{sc:pipeline}, along with a description of the two main tools used in the PHZ PF and developed alongside it, \texttt{Phosphoros} and \texttt{NNPZ}. The construction of the reference data sets from the EAF that are required by \texttt{NNPZ} is provided in Sect.\,\ref{sc:ref_samples}. 

DR1 represents the first opportunity to measure constraints for cosmological parameters such as the matter density, $\Omega_{\rm m}$, amplitude of the power spectrum, $\sigma_8$, and the dark energy equation-of-state parameter, $w$ \citep[e.g.][]{Chevallier}, with \Euclid data. It will mark the first weak-lensing analysis from a stage IV experiment, following the stage III weak lensing experiments, such as the Dark Energy Survey (DES, \citealt{DESsurvey}), the Kilo-Degree Survey (KiDS, \citealt{KiDSsurvey}), and Hyper Suprime-Cam Subaru Strategic Programme (HSC-SSP, \citealt{HSCsurvey}). Covering an area of order $1900\,\mathrm{deg}^2$, to an effective magnitude limit of $\IE=24.5$ and with broad-band photometry from the $u$ band to the $\HE$ band, Euclid DR1 occupies a unique position among current weak-lensing surveys that should enable improved constraints with respect to the aforementioned stage III experiments' latest results \citep{DESY6, KiDSlegacy, HSCY3}.

Cosmological parameters from \Euclid's photometric data set will be measured through the combination of $2$-point functions using both galaxy ellipticities and positions. They will form a so-called \threebytwo analysis of cosmic shear, galaxy--galaxy lensing, and galaxy clustering, as has become common in the literature \citep[e.g.,][]{3x2pt}. Among the key ingredients in these analyses are the redshifts of the galaxies used as probes of the large-scale structure, which rely largely on the photometric redshift (photo-$z$) method \citep[e.g.,][]{Hyperz, BPZ, ANNz, Salvato18}. The statistical precision of the anticipated final-mission weak lensing measurements places requirements on the quality of such photo-$z$ and, in turn, requirements on the \Euclid survey design and depths of the photometry \citep{Laureijs11}. 

The role of photo-$z$ in \threebytwo analyses is twofold: to place galaxies in meaningfully-separated bins in redshift (redshift tomography); and to calibrate the distributions in redshift of the galaxies, $n(z)$, in each bin. The first of these tasks implies requirements on the precision and accuracy of single-value best estimates for each galaxy's redshift (`point estimates'). These are expressed via the dispersion in the scaled redshift difference between the point estimate and spectroscopic redshift, $\sigma(\delta z)<0.05$, and the fraction of outlier galaxies with photo-$z$ point estimates very different from their true redshift ($|\delta z| > 0.15$), $f_{\rm outliers}<10\%$ \citep{Amara2007,Laureijs11}; with $\delta z$ the normalised difference, given by $(z_{\rm phot} - z_{\rm spec}) / (1 + z_{\rm spec})$. These requirements can be similarly expressed through their redshift probability distribution functions, $z$PDFs, following \cite{Desprez-EP10}. These requirements are easily met in deep extra-galactic fields at the target magnitude limit of samples for cosmic shear ($\IE<24.5$ in the case of \Euclid), where high \gls{snr} data across many bands are available \citep[e.g.,][]{Hartley22,Weaver22,Taylor23}. However, they remain a considerable challenge for the wide-field areas where the analyses are carried out \citep{Tanaka18,Desprez-EP10,Duncan22}. Moreover, galaxies hosting AGN can remain hidden among the galaxy samples and will require careful treatment for the overall requirements to be met \citep{Shirley26}. 

In the second task, the requirement on the accuracy of the mean redshift for $n(z)$ in each tomographic bin is among the most stringent of all the \Euclid{} design requirements, at $\delta z/(1+z)<0.002$. Multiple complementary strategies will be required to reach this performance, including photometric-space calibration approaches \citep{Masters2015} and inference through angular clustering \citep{Newman08}. Stage III weak lensing experiments have already achieved impressive accuracy through the sophisticated application of these techniques with the support of simulated data sets \citep{Wright25,dAssignies26,Yin26}. However, even the most tightly constrained $n(z)$ produced to date fall far from the required accuracy in stage IV experiments such as \Euclid. In this paper we focus on the photo-$z$ performance required for the first of the tasks (presented in Sect.\,\ref{sc:perf_photoz}), while the calibration of $n(z)$ for cosmological inference will be presented in a separate paper, which will appear alongside the full Euclid DR1 release. 
The photo-$z$ point estimates for a flux-limited $\IE<24.5$ sample are validated against the requirements in Sect.\,\ref{sc:validation}. 

The leading-order source of systematic error in weak lensing analyses is often the measurement and modelling of the \gls{psf}. In the case of a diffraction-limited optical system such as \Euclid, the effective PSF of an astrophysical object is sensitive to its \gls{sed}. Because the visible instrument (VIS) throughput is very broad in wavelength \citep[$\mathrm{FWHM} \sim 370\,\mathrm{nm}$, centred on $731\,\mathrm{nm}$,][]{EuclidSkyVIS}, the objects' \glspl{sed}, be they stars used to model the PSF or galaxies used in the weak lensing analysis, must be extremely well known on average across any given subsample \citep{Cropper13,EH18}. The task of measuring these \glspl{sed} is also part of the PHZ PF and, while such products were released in Q1 and described in \citetalias{Q1-TP005}, they are first used for their intended purpose in DR1 and therefore deserve greater focus in the present paper. The description of how the products are produced can be found in Sects.\,\ref{sc:ref_galaxysed} and \ref{sc:ref_starsed}, and an assessment of their performance in Sect.\,\ref{sc:perf_seds}.   

While \Euclid is primarily designed as a cosmology mission, the combination of image quality, depth, and survey area of DR1 is unlike anything that has been produced by the community previously and will enable an incredible range of non-cosmological science to be carried out, building on the success of the Q1 release. The PHZ PF produces non-cosmological science data products for each of the tiles of photometric data produced by the Merge (MER) PF \citep{DR1-TP009}, containing detailed classification information and a set of outputs for different object classes. These products include catalogues of star model fits, quasi-stellar object (QSO) properties, and a separate photo-$z$ determination that extends to higher redshift for objects taken from the \gls{NIR} detection image. These three specialist catalogues were already described in the PHZ Q1 paper, \citetalias{Q1-TP005}, but a brief description is included in the current work for completeness, found in Sects.\,\ref{sc:qso_templates} to \ref{sc:nir_photoz}. The final product for non-cosmological science applications is the catalogue of galaxy physical properties, such as stellar masses and \glspl{sfr}, assuming that each object is a galaxy (i.e., objects with active nuclei are poorly modelled in this catalogue). Similar to the photo-$z$ determinations, this product makes use of the newly available \glspl{EAF} and warrants a more detailed description than can be included in the present work. Details of the determination of these physical properties and their validation are therefore presented in a separate publication \citep[][hereafter EC:E26]{DR1-TP012}. We conclude this article in Sect.\,\ref{sc:issues} with a brief discussion of two known issues that users of the products should be aware of. 

Throughout the remainder of this paper, we assume as baseline the \citet{Planck15} $\Lambda$CDM cosmology, with $H_0=67.74~\mathrm{km~s^{-1}~Mpc^{-1}}$, $\Omega_{\mathrm{m}}=0.3089$, and $\Omega_{\Lambda}=0.6911$. All magnitudes are given in the AB system \citep{OkeGunn83} and depths quoted are $10\,\sigma$ noise limits for $2\arcsec$ diameter apertures, unless otherwise stated. 

\section{\label{sc:pipeline}Overview of the PHZ pipeline and methods}

\begin{table*}[]
\centering
\begin{threeparttable}
\caption{Summary of methods for the photometric redshift pipeline outputs and the main references for each product.}\label{tab:methods}
\begin{tabular}{m{3cm} lllm{3cm}}
\hline\hline
\noalign{\vskip 2pt}
Data Type & Table Name$^{a}$ & Container Product$^{b}$ & Method & Main Reference \\
\hline
\noalign{\vskip 2pt}
Photometric redshifts & phz\_photoz & dpdPhzPfOutputCatalog & \texttt{NNPZ} & Sect.\,\ref{sc:pipe_photoz} \\
Object classification & phz\_classification & dpdPhzPfOutputForL3$^{c}$ & pRF & Sect.\,\ref{sc:pipe_classification}, \citetalias{Q1-TP005} \\
Galaxy SEDs & phz\_galaxy\_sed & dpdPhzPfOutputCatalog & \texttt{NNPZ} & Sect.\,\ref{sc:pipe_seds} \\
Star SEDs & phz\_star\_sed & dpdPhzPfOutputCatalog & \texttt{NNPZ} & Sect.\,\ref{sc:pipe_seds} \\
Unified photometry & phz\_photoz & dpdPhzPfOutputCatalog & \texttt{NNPZ} & Sect.\,\ref{sc:pipe_unif} \\
Galaxy physical properties & phz\_physical\_parameters & dpdPhzPfOutputForL3 & \texttt{NNPZ} & \citetalias{DR1-TP012} \\
QSO physical properties & phz\_qso\_physical\_parameters & dpdPhzPfOutputForL3 & \texttt{Phosphoros} & \citetalias{Q1-TP005} \\
Star templates & phz\_star\_template & dpdPhzPfOutputForL3 & \texttt{Phosphoros} & \citetalias{Q1-TP005} \\
NIR-detected source photo-$z$ & phz\_nir\_physical\_parameters & dpdPhzPfOutputForL3 & \texttt{Phosphoros} & \citetalias{Q1-TP005} \\
Calibrated $n(z)$ distributions & \dots & DpdBinMeanRedshift & SOM & EC: Wright et al., in prep.\\
\hline
\end{tabular}
 \footnotesize
 \tablefoot{
      $^{(a)}$ For \gls{EWS} catalogues append \_wide; for EDS (including AUX fields and \Euclid medium deep survey) append \_deep; for the \gls{EWS}-like visits of the deep fields append \_deep\_visits\,.
      $^{(b)}$ For \gls{EWS} and \gls{EWS}-like visits of the EDFs. EDS catalogues have a primary container, dpdPhzDeepOutputCatalog, with the galaxy physical properties catalogue in dpdPhzPfOutputForL3 for technical reasons. 
     $^{(c)}$ Classification information is also summarised in dpdPhzPfOutputCatalog.}
\end{threeparttable}
\end{table*}

\begin{figure*}[htbp!]
\centering
\includegraphics[angle=0,width=1.0\hsize]{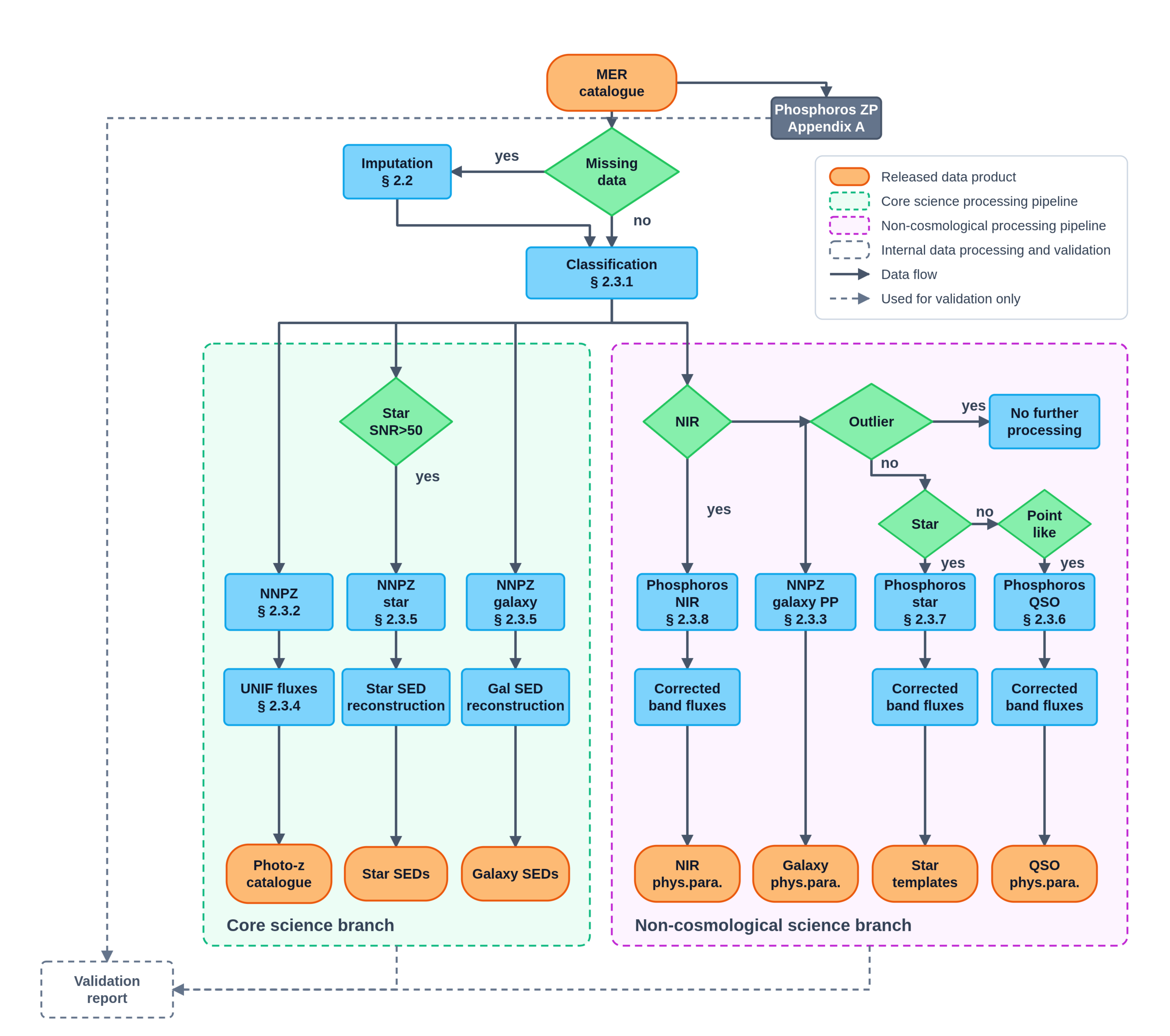}
\caption{Schematic diagram of the PHZ PF main production pipeline for \gls{EWS} tiles, showing the different decision points and subtasks that lead to output products.}
\label{fig:pipeline}
\end{figure*}

The PHZ PF comprises four main pipelines: the main production pipeline; a variant of the production pipeline for \gls{EDF} tiles (Deep pipeline); a sample selection and redshift-tomographic-binning pipeline; and finally a pipeline to produce calibrated redshift distributions, $n(z)$, for cosmological-parameter inference. In addition to these pipelines that are run through the \Euclid{} operations system \citep{DR1-TP033}, OU-PHZ and SDC-CH also develop and execute a number of calibration pipelines that build input products for the main PHZ PF pipelines. In this section we provide a general overview of the main production pipeline and its output products. Only a brief description is provided here for aspects that have not changed since the Q1 release, and we refer the reader to \citetalias{Q1-TP005} for further details.\footnote{Additional information on the PHZ PF outputs is provided in the on-line explanatory supplement, \url{https://euclid.esac.esa.int/dr/dr1/expsup}.}

In Fig.\,\ref{fig:pipeline} we show a schematic overview of the main production pipeline for \gls{EWS} tiles. The pipeline operates on photometric catalogues produced by the MER PF and produces eight catalogues for each tile, separated into two container products depending on whether they are intended for use in core cosmology science or non-cosmology science, as well as validation reports that are used internally within OU-PHZ to assess the success and validity of each pipeline run. Two container products are used in the Data Processing System and Science Archive System to organise our output catalogues (see Euclid Collaboration: Belikov et al. in prep. for a description of the archive system).
The correspondence of the different output catalogues to these containers is provided in Table \ref{tab:methods}, where we also list the main method used for their production. The DpdBinMeanRedshift product will be released as part of the full DR1 and is included here for completeness. 

\subsection{Key pipeline tools}

Before describing the individual catalogue products we begin by introducing two key tools employed to produce them, our Bayesian SED-fitting package, \texttt{Phosphoros}, and our nearest-neighbour algorithm, \texttt{NNPZ}.

\subsubsection{\label{sc:phosphoros}\phosphoros{}}

At its core, \phosphoros{} (Paltani et al., in prep.) is similar to many previous SED template-fitting approaches to measuring photo-$z$, and in particular the \texttt{LePhare} package on which it was initially based \citep{LePhare}. Redshift and galaxy type probabilities are estimated using the $\chi^2$ computed between the measured object fluxes and predicted fluxes of a set of allowed SEDs and redshifts. It is designed with flexibility in mind, allowing for the use of user-defined priors, multiple sets of template SEDs, and a multi-dimensional treatment of the internal properties of galaxies (e.g. dust reddening), within a fully Bayesian framework. The full posterior space for each object can be output, or represented through marginal posteriors, samples from the posterior, or some combination of these. Furthermore, SEDs can be tagged with physical quantities, including those that scale with object luminosity, and these physical quantities can also be output as samples or summary statistics such as median values. As a result, the performance of \phosphoros{} matches or exceeds similar publicly available packages \citep{Desprez-EP10, Desprez23}. Of particular importance for the \Euclid mission is the ability to take into account attenuation from the Milky Way via a forward model and the variation of filter bandpasses across the field of view of a camera \citep{Galametz17,EP-Paltani}.

\subsubsection{\label{sc:nnpz}Nearest-neighbours for photo-$z$: \nnpz{}}

\nnpz{} is a nearest-neighbour machine-learning algorithm designed to emulate and accelerate the behaviour of a template-fitting method such as \phosphoros{}. The principal advantages of a nearest-neighbour method over template fitting are the much lower computational cost and the automatic prior on the galaxy population that is provided by the training data. Throughout this paper we refer to each set of training data as a reference sample because nearest-neighbour approaches are not explicitly trained. It is this nature of the nearest-neighbour algorithm that makes it particularly suited for our purpose, since unlike most machine-learning approaches, we are able to include the impacts of the sky-location dependence of Galactic reddening and variation of filter bandpasses across a camera's field of view without the need to retrain the model. The fluxes of the reference objects are adjusted on the fly using their known \glspl{sed} to forward model the reddening \citep{Galametz17}. Correction for filter variation will be achieved in the same manner as Galactic reddening \citep{EP-Paltani}, but is not accounted for in DR1 as the input information required by the PF are not yet available. 

The reference samples used for the DR1 pipeline are taken from the \gls{EAF} data, and in the case of DR1 exclusively from the COSMOS field. Because these reference samples for \nnpz{} play a central role in the performance and characteristics of our outputs, we devote considerable attention to their construction in Sect.\,\ref{sc:ref_samples}. The use of a machine-learning algorithm with an empirical reference sample confers a further advantage over direct SED template fitting, in that the method is insensitive to systematic errors in the effective band-to-band zero points, provided that any offsets are consistent between survey patches, in particular between the reference data set and the target galaxies in the \gls{EWS} or \gls{EDF}.

The distance measure between a target \gls{EWS} object and the objects in the reference sample plays the role of the likelihood in a traditional template-fitting method and \nnpz{} allows for various definitions of this distance. In the PHZ pipeline, the neighbour distance is defined using the $\chi^2$, or squared Mahalanobis distance,
\begin{equation}
    \mathrm{log~Dist.} = -0.5\,\sum_b (A\,f_{\mathrm{ref},b} - f_{\mathrm{ob},b})^2 / \sigma_{\mathrm{ob},b}^2 ,
\end{equation}
with $f_{\mathrm{ref,b}}$ and $f_{\mathrm{ob,b}}$ the reference and target fluxes through band $b$, $\sigma_{\mathrm{ob},b}$ the associated flux uncertainty and $A$ a free-parameter governing the amplitude offset between a neighbour and target object, controlled via a prior. This prior can be fixed to unity (default behaviour), left completely free (in which case only the SED shape information is used), or one of a number of parametric forms, including a log-normal form that we use in the DR1 pipeline (see Sect.\,\ref{sc:pipe_photoz}). In all cases, the prior probability is clipped to zero where it falls below $p(A)<10^{-10}$, resulting in truncated prior distributions for open-ended distributions such as Gaussian or log-normal. For each reference object the amplitude is chosen such that the final weight, given by the prior times the distance, is maximised. Objects for which no neighbours with weight greater than $10^{-3}$ are found are flagged (\texttt{PHZ\_FLAGS}=1).

Quantities of interest that are known for the reference sample objects, such as their \gls{zpdf}, SEDs, or physical properties are weighted according to the distance and averaged. In this sense, \nnpz{} behaves much like a template-fitting method, with the reference sample objects' measured fluxes taking the place of the synthetic template fluxes, and their known properties replacing the model quantities. However, the use of an empirical reference sample naturally avoids spending computation on combinations of redshift and galaxy type that do not exist in the Universe, and provides greater fidelity for the most commonly encountered galaxies. On the other hand, object types that are unrepresented or poorly represented in the reference data can suffer biases, such as is the case at low redshift in DR1 (see Sect.\,\ref{sc:nnpz_issues}).

Reference sample objects have natural amplitudes to their \glspl{sed}, and this observed brightness plays a key role in the effective galaxy population prior, with the impact of common degeneracies in colour space being reduced as a result (see Sect.\,\ref{sc:perf_photoz}). However, the use of a nearest-neighbour algorithm necessitates some compromises. The neighbourhood in photometric space of any target object is discretely sampled by a finite number of reference objects, which constitutes a source of noise in the output quantities. This sampling noise can be reduced by allowing greater freedom in the amplitude parameter, with the resulting neighbour distances shortened where a target and reference object are close in colour but differ in amplitude. Doing so creates a denser neighbour sampling at the cost of reduced effectiveness in the magnitude aspect of the population prior.\footnote{See \cite{DNF} for related discussions.} 

\subsection{\label{sc:preprocessing}Pre-processing of input catalogues}

The PHZ production pipeline operates on MER final catalogue products \citep{DR1-TP009}. An \gls{EWS} tile will contain external data either from DECam \citep{DECam} or the UNIONS set of surveys \citep{UNIONS}, depending on the sky location of the tile \citep{DR1-TP006}. UNIONS comprises Canada--France--Hawaii Telescope (CFHT) MegaCam data from CFIS in the $u$ and $r$ bands \citep{Ibata2017}, Subaru Hyper Suprime-Cam \citep[HSC][]{Miyazaki2018} data in $g$ and $z$ bands from WHIGS and WISHES respectively, and $i$-band data from Pan-STARRS \citep{Chambers2016}. The pipeline identifies the relevant photometric system and determines which of the expected photometric bands are absent for each object (including the \Euclid bands, \IE, \YE, \JE, and \HE). The number of such bands is indicated in the \texttt{PHOTOMETRIC\_SYSTEM} column of the PHZ output catalogue, and the names of the missing bands are included in the classification catalogue. For the deep tiles, similar sets of bands are defined which accommodate the range of possible combinations that exist in the survey. 

For DR1, the object fluxes used throughout the PHZ pipeline are the PSF-matched aperture fluxes from OU-MER, specifically those in the circular aperture with diameter corresponding to twice the full width at half maximum (FWHM) of the image with the broadest \gls{psf}, \texttt{2FWHM\_APER} columns in the MER catalogues. These fluxes are corrected to pseudo-total fluxes using the Kron-like \texttt{FLUX\_DETECTION\_TOTAL} measurements. 

\subsection{\label{sc:products} PHZ product overview}

\subsubsection{\label{sc:pipe_classification}Object classification}

Object classification is performed by means of a pair of probabilistic random forests (pRFs, citealt{Reis19}), one for star classification and one for galaxy classification. Each pRF provides a probability for a given object to be of that class or not and these probabilities are used to determine which branches of the pipeline an object appears in. These probabilities are also used alongside a pair of threshold values (listed in the classification catalogue) to form the consolidated classification output column, \texttt{PHZ\_CLASSIFICATION}, in which objects could be accepted as one class, both classes or neither. These thresholds were set to return a highly complete galaxy class and a very pure star sample at high \gls{snr}$\,>50$, with core cosmology science in mind. Users are encouraged to set their own class thresholds optimised for their own science goals.

The advantage of a pRF over a standard random forest is that allows the user to take uncertainties in the labels and features into account during training and prediction. In the case of the PHZ pipeline, the features we use are all possible colour combinations from asinh-scaled magnitudes \citep{Lupton99} and the probabilistic nature of the pRF therefore allows us to produce consistently estimated class probabilities across the \gls{EWS} footprint without the need to retrain for varying depths. The training set with labels was constructed in the same way as in the Q1 release, with full details given in \citetalias{Q1-TP005}.

In the Q1 release we used a third classifier to determine objects likely to be QSOs. However, classification of such objects using photometry alone proved to be very challenging. For the DR1 pipeline, objects are included in the QSO catalogue if they are point-like and not classified as stars (using the default consolidation threshold for stars). The morphology information for the classification is provided by the difference between the peak surface brightness of the object minus its \gls{psf} model magnitude (contained in MER final catalogue column, \texttt{MUMAX\_MINUS\_MAG}), with the threshold for an object to be point like set at $\texttt{MUMAX\_MINUS\_MAG} < -2.6$ (see \citealt{Q1-TP004} and \citealt{Q1-SP027} for a validation of this choice). This sample is contaminated at the bright end by stars that do not pass the classification threshold, and will include only AGN where the host light contribution is low or unresolved, consistent with the definition of a QSO. Meanwhile, at faint magnitudes, galaxies and point sources are difficult to distinguish morphologically, even with the \Euclid imaging. It is therefore a starting point for a QSO sample selection, ensuring that photo-$z$ using suitable \glspl{sed} are available for the vast majority of genuine QSOs, and not a ready-made sample of confirmed QSOs. Performance of the classification for DR1 is described in Sect.\,\ref{sc:perf_classification}. 

While most operations in the pipeline are robust to absent photometric information in one or more bands, the pRFs perform better with a consistent set of input features. The fluxes that form these features must also be corrected for Galactic reddening. The method for imputing and correcting the photometry for reddening is unchanged from the Q1 release. Firstly, an \gls{sed}-dependent reddening correction is performed for all objects. Then, a kNN is employed with the objects from the tile with complete data used as the training sample. The closest $15$ neighbours are found for each object that is missing bands, and those bands are imputed using the average values of the neighbours. We invite the interested reader to consult \citetalias{Q1-TP005} for a fuller description of this process.

\subsubsection{\label{sc:pipe_photoz}Photometric redshifts}

Photo-$z$, including the full \gls{zpdf} and associated summary statistics, are computed using \nnpz{} for the DR1 release, in contrast with the Q1 release where \phosphoros{} was used. This switch in methods represents the most significant change to the PHZ production pipeline, made possible by the collection of \Euclid and external data in the \gls{EAF} that are used to construct the reference sample of potential neighbours (see Sect.\,\ref{sc:ref_samples}). For every target, the number of neighbours used by \nnpz{} is set to 30, based on an optimisation performed during pre-launch \Euclid data challenges and a compromise between precision and data volume for the galaxy physical property outputs \citepalias[see][]{DR1-TP012}. In some cases 30 neighbours may not be enough, and thus result in a truncation of the \gls{zpdf}. However, we see no clear evidence for a degradation in the \glspl{zpdf} calibration that would result from such an issue during validation (Sect.\,\ref{sc:validation}). The median effective number\footnote{The effective number of samples, given a set of weights, $w$, is $N_{\mathrm{eff}} = (\sum w)^2 / \sum (w^2)$.} of neighbours for an $\IE<24.5$ sample is $N_{\mathrm{neighb.}}=24.64$ but declines sharply at brighter magnitudes ($\IE<23$).

The second major configuration choice in \nnpz{} is the choice of prior for the allowed amplitude difference between a target object and the reference galaxies. To set this choice we took advantage of the \gls{EWS}-like processing of the EDF-F produced by the MER pipeline. We performed a sweep of the various prior options in NNPZ, changing the associated parameters within sensible limits and optimising for the photo-$z$ performance metrics we describe in Sect.\,\ref{sc:performance}. We find the best results are obtained if we apply a log-normal prior with mean and standard deviation, $\mu = 0.04$ and $\sigma = 0.2$. The prior clipping limit described in Sect.\,\ref{sc:nnpz} restricts the prior amplitude range to $\pm1.5\,\mathrm{mag}$.

In addition to the amplitude prior, we need to allow for any possible differences in the zero-point calibration between the reference sample (COSMOS field) and the target objects from the \gls{EWS}. From an early assessment of the photometric performance of the OU-MER final catalogues we estimate that the tile-to-tile variations in possible zero-point corrections are of the order of $\sim2\%$, and we therefore add a systematic uncertainty at $2\%$ of the object flux in all bands. However, due to a configuration error that could not be fixed ahead of the release, this value was interpreted as the variance, resulting in much smaller distances for most objects. Nevertheless, our photo-$z$ performance is not substantially degraded by this error (see Sect. \ref{sc:perf_photoz}).

Summary statistics are computed from the weighted and averaged \gls{zpdf} of the $30$ neighbours produced by \nnpz{}: the median, central percentile ranges, and a ‘mode’ value, \texttt{PHZ\_MODE\_1} corresponding to the mean of the most probable peak in the \gls{zpdf} following a peak decomposition. The mean redshift of the closest neighbour’s \gls{zpdf} (representing a quantity analogous to a maximum likelihood solution in template-\gls{sed}-fitting photo-$z$) is also output, as the column \texttt{PHZ\_MODE\_2}.\footnote{Despite the column names, neither \texttt{MODE} column refers to the true mode of the \gls{zpdf}, which can be extracted trivially from the \gls{zpdf}.} Finally, the \gls{zpdf}s are smoothed with a narrow Gaussian filter ($\sigma_{z}=0.02$) to remove high-frequency noise, which might otherwise cause artificial clustering in tasks such as \gls{zpdf}-based galaxy cluster detection.

\subsubsection{Galaxy physical properties}

For completeness, we succinctly describe the products containing galaxy physical properties. Full details can be found in \citetalias{DR1-TP012}. The physical properties of galaxies, such as their stellar masses, \gls{sfr}s, and absolute magnitudes are determined in an analogous way to the photo-$z$ PDFs, except that full posterior information cannot be stored and delivered, even for a number of objects the size of the reference sample. Instead, the posteriors of the \nnpz{} reference objects are represented via a set of 50 samples, which between them express the correlations between the various quantities of interest, e.g., between redshift and luminosity or stellar mass; between specific SFR and rest-frame colours etc.

As for photo-$z$ determination, \nnpz{} is run with a neighbour count of 30 and the same freedom in SED amplitude with log-normal prior is used when computing the neighbour distances. The result of the \nnpz{} run is a physical property posterior for each object represented by 1500 samples with relative likelihood weights, and SED amplitude-sensitive quantities such as stellar masses corrected by the appropriate scaling factors. The scale factors are stored in the output catalogue alongside the list of neighbour IDs and distance weights, forming a very compact representation of the full posterior, at the cost of a degree of sampling noise. The output catalogue also includes some simple statistics for each property: the median value; mean of the closest neighbour; and central 68-percentile limits ($15.9\mbox{--}84.1\,\%$). Further details of the methods and validation of the outputs are provided in \citetalias{DR1-TP012}, Sect. \ref{sc:ref_samples} and \ref{sc:performance}.

\subsubsection{\label{sc:pipe_unif}UNIFied photometry}

The \Euclid core cosmology science requires that we are able to calibrate the redshift distribution of a sample of galaxies within a given tomographic redshift bin consistently across the survey footprint. This is done by partitioning the photometric space into discrete volumes, each of which is then calibrated. In practical terms, this implies using a standardised photometric system that is corrected for Galactic reddening and for any differences between the effective filter curves that different objects were observed with. The purpose of the UNIFied (\unif{}) photometry is to achieve this standardisation. For DR1, we assume that each bandpass is invariant across its respective camera's field of view (such variation will be addressed in DR2), and we choose the DECam filter set of the southern DR1 area as the primary filter set. We provide \unif{} fluxes for both north and south filter sets.

Correcting fluxes for reddening requires knowledge of the source SED, as discussed in \cite{Galametz17}. We include this information through the use of \nnpz{} and the reference sample built for photo-$z$ measurement. The high \gls{snr} photometry available in the \glspl{EAF} allows us to obtain reddening-free fluxes for the reference objects through per-object and per-band correction factors via template-SED fitting with \texttt{Phosphoros}. Objects are fit with reddened SEDs, and the ratio of reddened to unreddened model flux for the best-fitting SED is used to correct the fluxes of the reference objects. From the same SED, a set of linear coefficients is computed (one per band) such that when multiplied by a line of sight $E(B-V)$ a close approximation of the galaxy photometry, were it to be viewed in that sky location, are obtained. These coefficients are used during neighbour distance calculation by \nnpz{} to shift each reference object into the (reddened) photometric space of a target object. The predicted \unif{} fluxes for a given band from a single \nnpz{} neighbour are then obtained via the measured \gls{EWS} fluxes ($f_{\mathrm{obs}}$) by
\begin{equation}
f_{\mathrm{UNIF}} = f_{\mathrm{obs}}\, \frac{f_{\mathrm{ref}}}{f_{\mathrm{ref,reddened}}}.
\end{equation}
The output \unif{} photometry is then constructed in an analogous way to the photometric redshifts: taking the weighted average of the predicted photometry in each band from the 30 closest neighbours found by \nnpz{}, after applying their appropriate amplitude scale factors. DECam \unif{} fluxes for objects in the northern hemisphere are similarly obtained through the appropriate reference fluxes, 
\begin{equation}
f_{\mathrm{UNIF, DECam}} = f_{\mathrm{obs,UNIONS}}\, \frac{f_{\mathrm{ref,DECam}}}{f_{\mathrm{ref,reddened, UNIONS}}},
\end{equation}
and vice versa for the UNIONS fluxes in the southern hemishpere.

We do not propagate uncertainty estimates in DR1 as they are not strictly necessary for their purpose of $n(z)$ calibration. Error propagation will be addressed in future data releases. The reference sample used for this task is the same as used for the photo-$z$ and consists only of galaxies, so although \unif{} photometry is computed and output for all objects they should not be used for stars.

\subsubsection{\label{sc:pipe_seds}SEDs for PSF modelling}

The PSF of the \Euclid VIS instrument through its \IE band is diffraction limited and hence the effective PSF of any given source recorded in an image will depend on its SED \citep{EH18}. Since the PSF model is among the leading-order sources of error in cosmic shear analyses \citep{Massey2013} it is crucial that wavelength dependence is taken into account during galaxy shape measurement and the PSF model must therefore be explicitly SED dependent. To allow the construction and application of a chromatic PSF model in the SHE PF (Euclid Collaboration: Miller et al., in prep.),
we produce finely sampled SEDs at intervals of $10\,{\rm nm}$ for both stars and galaxies across the range of wavelengths relevant to the PSF modelling, $450 < \lambda/{\rm nm} < 1000$. Catalogue columns are named for the synthetic narrow-band flux that each interval represents. As for the \unif{} fluxes described earlier, we do not provide uncertainty estimates for the finely sampled SED fluxes in DR1.

For the estimation of SEDs, we again employ \nnpz{}, together with an appropriate reference sample of objects, drawn either from \textit{Gaia} \citep{Gaia} for stars or from the COSMOS \gls{EAF} for galaxies (see Sect.\,\ref{sc:ref_samples}). Stellar SEDs from the dedicated PSF fields (Euclid Collaboration: Duncan et al., in prep.) 
are used during the initial PSF model construction, and also during tile-by-tile processing to refine the PSF model for each \Euclid{} observation (in particular the telescope focus). For this purpose, only high ${\rm S/N}>50$ stars are used, based on the photometric classification (Sect.\,\ref{sc:pipe_classification}) and the MER catalogue \texttt{FLUX\_DETECTION\_TOTAL}. A loose cut removing clearly extended objects (\texttt{MUMAX\_MINUS\_MAG}$\,>-1.5$) was also introduced in DR1 to remove a small number of obvious galaxy contaminants. The \nnpz{} configuration for stellar SEDs uses a uniform scaling prior (i.e., a free SED amplitude) and no correction for Galactic reddening, such that we assume the reference sample contains all relevant combinations of stellar type and reddening. 

Galaxy SEDs are produced for all objects that were not rejected during catalogue ingestion, regardless of their classification. This choice is to avoid introducing any potential biases due to imposing selection criteria at an early stage, since shear measurement can only be made for objects with valid galaxy SEDs. To construct the galaxy SEDs, \nnpz{} is run without scaling and the finely sampled photometry is reddened appropriately \citep{Galametz17} according to the line-of-sight reddening from the {\em Planck} map \citep[\texttt{GAL\_EBV} in the MER final catalogue]{Planckdust}. 

\subsubsection{\label{sc:qso_templates}QSO photo-$z$}

Estimation of the properties of QSO-classified objects is unchanged with respect to the Q1 pipeline, although the method of classifying such objects has changed (see Sect.\,\ref{sc:pipe_classification}). QSO candidates are fit with the set of composite QSO+host galaxy SEDs from \cite{Salvato2009} using \texttt{Phosphoros}. A reasonably restricted and simple model space is used because QSO photo-$z$ are notoriously challenging to measure due to near power-law-like SEDs, a wide range of emission-line strengths, and photometric variability \citep[e.g.,][]{Shirley26}. Variability is particularly problematic when dealing with photometric measurements from images taken several years apart and cannot be included in the PHZ pipeline at present. Redshifts up to $z=6$ are allowed in steps of $\Delta z=0.05$, and intrinsic reddening is assumed to follow the \cite{Prevot1984} SMC law, with colour excess values up to $E(B-V)=0.04$. Outputs include the redshift PDFs, best-fit templates, combined \gls{sed} luminosities, and Milky Way-corrected photometry (again following \citealt{Galametz17}). Some further details of the set-up and general performance can be found in \citetalias{Q1-TP005}. 

\subsubsection{\label{sc:star_templates}Stellar template fits}

Similarly unchanged is the \texttt{Phosphoros} set-up for objects classified as stars. Objects are processed through the star template element of the pipeline based on their probability of being a star during the classification step. No morphology information is used in this decision so as to avoid rejecting objects that might fail such a criterion due, for instance, to minor blending effects. A lower threshold for the probability cut is used at this stage than for the star SED catalogue to boost the completeness of the sample. Stars are fit to a large set of synthetic stellar SEDs from \cite{EP-Serrano}, spanning a wide variety in effective temperature, surface gravity and metallicity. Milky Way reddening correction is applied, using the value provided in the MER catalogue with the assumption that each star lies behind the dust screen. \texttt{Phosphoros} is again used for this task, as described in \citetalias{Q1-TP005}. 

\subsubsection{\label{sc:nir_photoz}Near-infrared object processing}

The final class of objects processed with \texttt{Phosphoros} are those detected in the combined image built from the three \gls{NIR} passbands rather than the \IE band, and this task is again unchanged from Q1 \citepalias[see ][]{Q1-TP005}. Approximately one quarter of the \gls{EWS} objects are \gls{NIR}-detected. A small fraction of such objects are those that have been missed by the deblending algorithm applied to the \IE data, or are residual artefacts from the NISP image processing. The majority, however, are objects with very faint optical SEDs that are bright enough to be identified in the NISP imaging. Such objects are a mix of cool Milky Way stars; heavily dust-reddened galaxies; intermediate- to high-redshift galaxies with very low star formation; and very high-redshift galaxies and QSOs. The SED model set used in \texttt{Phosphoros} therefore includes examples of each of these object classes. Optically bright models are also included to enable the significance of high-redshift solutions to be assessed with respect to lower-redshift object classes. As high-redshift galaxies and QSOs are particularly interesting classes of objects, the redshift grid is extended up to $z=10$, with a coarser step size of $\delta z=0.05$. For the convenience of users interested in high-redshift candidates, we provide the integrated probability of $z>6$ for each source. 

\subsection{\label{sc:pipe_deep}Deep tile pipeline}
The processing function for \gls{EDF} tiles follows very closely the structure and flow of the main \gls{EWS} pipeline described above. The main differences are in the combinations of photometric bands available, the container products used, and of course the depths of the images used to measure the photometry. Deep survey tiles in DR1 may contain DECam ($g$,$r$,$i$,$z$ bands), HSC ($g$,$r$,$r2$,$i$,$i2$,$z$ bands), or both sets, in addition to \Euclid photometry \citep[see][]{DR1-TP006}. Deep tiles without external bands are not processed by the PHZ PF. 

All available bands are used in most of the pipeline stages, with the main exception being the classification step. In the wide-tile pipeline all possible asinh-colour combinations are used for classification, but two adjustments were necessary for deep tiles. Firstly, the different generations of HSC filters, $r,r2$ or $i,i2$ are combined into a single band through inverse-variance weighting. Secondly, it was found that using colours of the same broad-band type (e.g., DECam $i~-$ HSC $i$) degraded classification performance, and so such combinations are not used. The result is a maximum of 62 asinh-colour features, corresponding to the case where all DECam and HSC bands are available.

For the DR1 release, no shear catalogue products were expected in the deep fields, and the SED products from OU-PHZ were therefore not required and not produced. The deep pipeline also has two configuration differences with respect to the wide pipeline: the addition of systematic flux errors was not included in the  photo-$z$ determination; and the redshift grid for object detections coming from the NIR image was extended to $z=12$. The only other difference in the content of the output products is in the determination of galaxy physical properties, where we output an additional table of $50$ samples drawn from the individual galaxies' posterior distributions. While it is not feasible to output and store this vloume of data for the \gls{EWS}, many scientific applications will use such samples to capture the uncertainties in the physical properties. The samples are an ordered set, in the sense that the first value in the redshift vector column corresponds to the same sample as the first value in the stellar mass vector column and so forth.

\subsection{\label{sc:pipe_cosmo}Binning and $n(z)$ pipelines}
The products describing the redshift distributions, $n(z)$, of the cosmology samples will be released alongside the cosmology data-vector products, such as the cosmic shear power spectra, and are not part of the DR1-Foundation release. Here, we provide only a brief overview of the calibration procedure, as implemented in the PHZ PF.

The basis of our calibration strategy was first proposed in \cite{Masters2015}, described in \cite{EuclidSkyOverview}, and will be detailed at the time of the full DR1 release (Euclid Collaboration: Wright et al., in prep.). Its core component is the use of a self-organising map (SOM) to provide a mapping between the observed photometry of galaxies in the cosmology samples and redshift. Since their first introduction, SOM-based approaches to redshift calibration have become commonplace in cosmology analyses \citep[e.g.,][]{Wright25}, and have evolved to take advantage of higher-quality data or additional photometric bands within special deep fields \citep[e.g.,][]{Hartley22}. The \glspl{EAF} are the designated fields to provide redshift information for $n(z)$ calibration for the \Euclid mission. They contain data of much higher \gls{snr} than the target \gls{EWS}, which allows us to add noise realisations that are representative of the target objects. This step is performed through a catalogue-based multi-passband transfer (MPT) function \citep{Kang26}.

Since the set of \gls{EAF} objects with spectroscopic redshifts has a different distribution in magnitude than the target \gls{EWS} data (somewhat brighter on average) and different abundance patterns with respect to a fully representative sample, we use a second SOM based on the deep photometric data. This second, deep, SOM is used to infer $p(z\,|\,\vec{f}/f_{\IE})$ and $p(\vec{f}/f_{\IE})$ for the effectively noiseless case of the deep fields, where $\vec{f}/f_{\IE}$ is the vector of flux ratios with respect to the \IE band for the combined set of \Euclid and DECam bands. The redshift calibration sample that enters the PHZ pipeline is then built by drawing from the deep photometric data set, drawing a realisation from $p(z\,|\,\vec{f}/f_{\IE})$ and adding appropriate noise via MPT. In this sense, the pipeline operates in a similar sense to the 2-SOM approach of \cite{DESY3SOMPZ}. The final output product expresses the $n(z)$ as a binned probability density per cosmology sub-sample in intervals of $\delta z = 0.01$, and contains metadata on the mean, variance, and respective estimated uncertainties of each redshift distribution.

\section{\label{sc:ref_samples}Construction of reference data sets}
The PHZ PF uses \texttt{NNPZ} to produce object-by-object outputs for \Euclid core cosmology and to determine the physical properties of galaxies. Each of these operations requires a sample of reference objects. We describe the construction of the four reference data sets, for photometric redshifts, galaxy physical properties, and the SEDs of stars and galaxies, in the following subsections. Our approach for galaxy physical properties is described in detail in \citetalias{DR1-TP012}, so we only provide here a brief overview of this aspect. 

\subsection{\label{sc:ref_cosmos}Construction of the COSMOS-based reference sample}
Any patch of sky we can observe should contain a set of galaxies that are representative of those found across a wider survey area, subject to the limitations of depth and sample variance. From such objects we can build ideal training or reference sets for machine learning algorithms, provided we can determine suitable ground-truth values for the quantities of interest. 
In practice, these values need not be an exact truth, provided they are much more precise and accurate than could be obtained from the wide-field data alone. This strategy underpinned the selection of the \glspl{EAF} \citep{Scaramella-EP1} in the \Euclid survey design, targeting well-established deep fields that were already rich in ancillary data.

A further requirement of the \glspl{EAF} for the PHZ pipeline is that their photometry is effectively noiseless in comparison to the \gls{EWS}, so that they do not add additional uncertainty in the neighbour matching performed by \nnpz{}, can be degraded and resampled where required \citep{Kang26}, and can be adjusted to account for the variety in reddening and passband throughputs encountered across the \gls{EWS}. For DR1, two of the PHZ \glspl{EAF} were observed: COSMOS ($2\,\mathrm{deg}^2$) and the single pointing covering the Chandra Deep Field south (CDFS, $0.5\,\mathrm{deg}^2$).\footnote{A third \gls{EAF} released in DR1, the AEGIS field, is designated a colour gradient calibration field and no EXT data were prepared in this field.} 
The \Euclid COSMOS data we have available represent approximately one-third of final target depth, at $\sim3$ times \gls{EWS} \gls{snr} depth, while CDFS was observed to full depth ($5$ times \gls{EWS} \gls{snr} depth). To complement the \Euclid data, the organisation unit for external data (OU-EXT) obtained, calibrated, and combined the deepest broad-band data available for the two fields \citep{DR1-TP006}, from HSC \citep{HSCsurvey,EP-McPartland} and DECam \citep[][and references therein]{Hartley22}, and these were subsequently processed through the MER PF deep pipeline alongside the \Euclid data \citep{DR1-TP009}. In order to build a reference sample from the MER catalogues, we require precise redshift information and the means to match the reference galaxies to \gls{EWS} objects appropriately. No suitable external catalogues were available for the CDFS field and so the reference samples are based on COSMOS alone.

\subsubsection{\label{sc:cosmos_photoz}Photometric redshifts in the COSMOS \gls{EAF}}

The COSMOS field is the richest and most heavily observed square-degree-scale field on the sky, containing deep images in more than 30 photometric bands that are able to support photometric redshifts at $1\%$ precision down to the depths required for the \gls{EWS} \citep{Weaver22}. Of particular importance are the medium-band observations taken with the Subaru telescope \citep[passbands with widths $30\mbox{--}60\,\mathrm{nm}$][]{Taniguchi2015}, which enable the tight localisation of strong break features out to $z\simeq1$. The \Euclid COSMOS catalogue does not contain these data in DR1, and so our approach to determining the precise redshifts we need for \texttt{NNPZ} is to use the COSMOS2020 catalogue of \cite{Weaver22} and then cross-match these measurements to the \Euclid objects. While cross-matching carries the risk of misidentification and is impacted by blending in the ground-based COSMOS2020 data, it nevertheless produces higher-quality reference data than could have been achieved with the broad-band-only Euclid DR1 catalogue within the available timescale.

To measure photo-$z$ for the COSMOS galaxies we begin with the classic \texttt{SExtractor} catalogue from COSMOS2020 \citep{Weaver22}, removing all objects with \texttt{FLAG\_COMBINED} set, which reduces the sample to $\num{723896}$ objects. The bands used for this part of the pipeline work are listed in Table \ref{tab:cos2020} (found in Appendix \ref{sc:app_cosmoszp}) with the exception of IRAC channels 1 and 2, which are used for galaxy physical properties but not photo-$z$ as better photo-$z$ performance was achieved without them. Also listed in Table \ref{tab:cos2020} are per-band systematic flux uncertainties that are multiplied by the object flux and added in quadrature with the measured uncertainties. These systematic errors compensate for a multitude of possible defects, including varying zero points across the field, possible background offsets, and residual mis-calibration of photometric zero points following the calibration procedure (detailed in Appendix \ref{sc:app_cosmoszp}). 

Redshift probability distributions for each object are computed with \texttt{Phosphoros}, using a similar template space to that of the Q1 release, but extended to a maximum redshift of $z=10$ and on a finer grid of internal reddening. We use the 33 templates from the COSMOS library of \cite{Ilbert09} including two exponentially declining star-formation history models from \cite{BC03}, with two interpolated SEDs between each neighbouring pair in the type-ordered list and two prescriptions for emission-line strengths for a total of 192 SEDs. Reddening is allowed to take values up to $E(B-V)=0.51$, in steps of $0.03$, with five dust laws: an SMC curve from \cite{Prevot1984} and the \cite{Calzetti00} dust law with three prescriptions of the $\lambda = 217.5\,{\rm nm}$ bump, or no bump (see the \texttt{Phosphoros} documentation\footnote{\url{https://phosphoros.readthedocs.io/en/latest/}} for details). For IGM attenuation, we apply the recipe of \cite{Meiksin06}. The two emission-line prescriptions we use follow empirical line ratios and are based on the \cite{Kennicutt98} relation, between the UV luminosity from the template and \ha line flux, at $100\%$ and $60\%$ strength. The choice for line strengths was determined alongside photometric zero-point correction factors. 

A standard method for testing the calibration of \glspl{zpdf} for a statistical sample is through the use of probability integral transforms (PIT, \citealt{Dawid84}; see also Sect.\,\ref{sc:perf_photoz}). Following the previous steps, the PIT test of the COSMOS \gls{zpdf} still showed evidence of some mis-calibration. This is a common occurrence, caused by underestimated photometric uncertainties, which often neglect the correlations between image pixel values. We find that the photometric errors in the COSMOS2020 catalogue need to be increased by $20\%$, and provide some further details of the determination of this value in Appendix \ref{sc:app_cosmosfluxerr}. Finally, we used the \phosphoros{} SED prior functionality to remove the implicit prior on rest-frame SED type that comes from the relative abundance of different template types and to apply a weak volume prior with effectiveness value of $0.3$ (see the \texttt{Phosphoros} documentation for details of prior effectiveness).

Once the \glspl{zpdf} are determined, we trim the COSMOS reference data to the set required for each task. This includes removing stars and bright AGN based on the COSMOS catalogue information, magnitude cuts in the case of the wide reference samples (allowing the union of $\IE<26$ and $\HE<25$), and the removal of a small number of objects with a high probability of being at $z>6$, since their \gls{zpdf} becomes meaningless once truncated at $z=6$.

\subsubsection{Reddening correction and UNIONS broad bands}

The Milky Way dust reddening in the COSMOS field is low [$E(B-V)\simeq0.017$], but non-zero. Furthermore, the \Euclid COSMOS catalogue contains photometry in the DECam and HSC bands, but not the two CFHT bands ($u$ and $r$) or Pan-STARRS $i$ band that are present in the UNIONS set of northern surveys. In order to use \nnpz{} to transfer the high-quality redshift information from the COSMOS field to the \gls{EWS} we must first determine reddening-free fluxes in all the relevant bands of the \gls{EWS}, \gls{EDF}, and \glspl{EAF}. 

The CFHT $u$-band is present in the COSMOS2020 catalogue, but not in the \Euclid catalogue. To use the flux values from the COSMOS2020 catalogue in the reference sample we therefore first make adjustments for any differences in aperture flux-loss corrections. 
For each object, we perform a $\chi^2$ minimisation between the HSC fluxes in the COSMOS2020 catalogue and those in \Euclid catalogue, and derive the SED amplitude factor to be applied to the CFHT $u$-band flux. We then import these corrected $u$-band fluxes directly into the reference sample catalogue.

For reddening corrections and to produce fluxes for the remaining two UNIONS bands, CFHT $r$-band and Pan-STARRS $i$-band, we use \texttt{Phosphoros} to determine SED-dependent filter transformations for each object. We use the same SED set-up as in the COSMOS2020 photo-$z$ run and compute reddening correction factors based on the best-fitting SED for each of the objects and bands present in the \Euclid catalogue. The coefficients required by \nnpz{} to redden the reference sample photometry during neighbour matching are also determined at this time. Finally, per-object band-transformation factors to convert the HSC$~r2$ to CFHT $r$ and HSC $i$ to Pan-STARRS $i$ are computed from the best-fit \glspl{sed} and applied. 

\subsubsection{Effective redshift prior in COSMOS}

COSMOS is the only field that we use to build the reference data products in DR1, and thus, because the abundance of objects in the reference sample acts as a prior in \nnpz{}, we can expect that the sample variance from the cosmic line-of-sight structure in COSMOS will become subtly imprinted across the whole \gls{EWS} sample. To mitigate the impact of this implicit prior we estimate and apply a function to determine a set of redshift-dependent weights that, when applied to the reference sample objects, should suppress the sample variance of the COSMOS field. In practice, only an approximate solution is possible with a single field, and an improved method will be employed once we have all four \glspl{EAF} in DR2.

We start with the assumption that over- and under-densities are narrow in redshift and therefore that a coarsely binned histogram of redshifts will be largely free of such sample variance issues. In analyses of COSMOS data concerning stellar mass functions or clustering measurements, redshift bin widths of $\Delta z = 0.2$ or greater are often used for this reason \citep[e.g.,][]{Pozzetti10} and so we adopt the same value for the bin spacing. 
Reference sample objects are first binned by their total \HE magnitude as over or under-abundance in galaxy number density follows galaxy bias. At fixed redshift \gls{NIR} brightness correlates well with stellar mass and hence bias and so neglecting the magnitude dependence will lead to over- and under-correction of different sub-populations. Next, we form the coarse-binned redshift histogram for each magnitude sub-sample, and fit a cubic spline through the mid-points of the bins, anchored to zero at $z=0$ and $10$. These smooth curves represent the target prior distribution. 

\begin{figure}[btph!]
\centering
\includegraphics[angle=0,width=1.0\hsize]{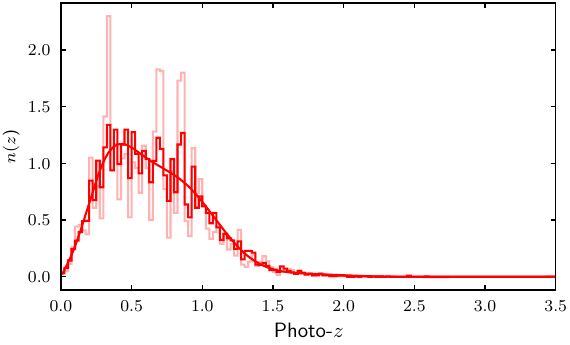}
\includegraphics[angle=0,width=1.0\hsize]{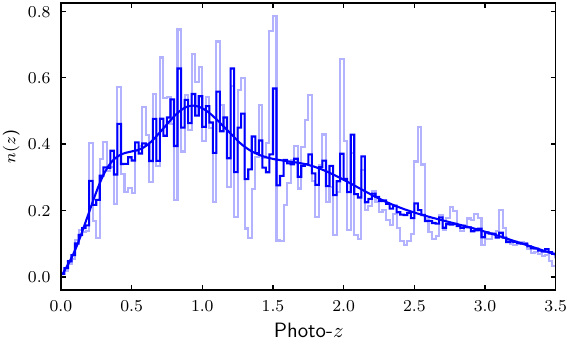}
\caption{Sample variance suppression in the COSMOS reference sample for two \HE magnitude ranges: $17<\HE<21$ (\emph{top}) and $23<\HE<24.5$ (\emph{bottom}). Curved lines show the target distribution, while the heavy and light histograms show the weighted and unweighted redshift distributions of the reference data, respectively.}
\label{fig:ref_prior}
\end{figure}

To remove the fine-scale sample variance, we build fine-binned redshift histograms with bin size, $\Delta z=0.025$, in each magnitude interval. We take the ratio of the fine-binned histogram to the spline curves as the over or under-abundance in a given redshift interval. This 2-dimensional set of ratios (in magnitude and redshift) is then used to build an interpolation function and each reference sample object is given a weight according to its magnitude and maximum posterior redshift value. In practice the weights are truncated at values 0.25 and 4, to avoid amplifying noise. The target prior and impact of the new prior weights can be seen for two of the magnitude intervals in Fig.\,\ref{fig:ref_prior}. Strong, well-known structures in the COSMOS field can be seen in the unweighted redshift histograms (lighter-coloured lines) at redshifts $z$ of 0.35, 0.7 and 0.9. While clearly not perfect, the weights reduce the degree of sample variance present in the reference sample. Nevertheless, we still see some impact of the COSMOS cosmic structure in the pipeline output products (see Sect.\,\ref{sc:issues}).

The COSMOS2020 catalogue lacks sufficient examples of most intrinsically-rare object classes, such as bright objects, low-redshift galaxies and the most highly biased objects in the Universe. These `missing' objects are due to the small size of the field, its selected location on the sky, and the saturation limits of some of the main optical broad bands. Missing such objects from the reference sample will effectively prevent us from being able to determine redshifts and other properties for the brightest galaxies, despite the flexibility introduced by the amplitude prior. To mitigate this issue, we produce a small number of augmented objects, at a rate of less than $1\%$ of the size of the reference sample, to replace the ones that should exist in the parent COSMOS2020 catalogue. These augmented objects are chosen to be at low redshift ($z<0.1$), where the sample has very few objects, and to fill in the magnitude range, $16<i<18$, at a rate that would extend the natural slope of the number counts brighter than 18th magnitude. They are drawn from the existing COSMOS objects at those redshifts and simply boosted in terms of their SED amplitudes. This simplified solution will be replaced in future releases by the use of the \glspl{EDF} to sample rare objects.

\subsubsection{Galaxy physical property reference sample}

Details of the production and validation of the measurements of galaxy physical properties, such as stellar masses and star-formation rates are provided in \citetalias{DR1-TP012}. Here, we provide a very brief overview for completeness. We construct a set of $\num{65538}$ stellar population synthesis (SPS) template SEDs drawn from a 6-dimensional Halton sequence \citep{Halton1960} in the space of redshift, metallicity, dust reddening and star-formation history (SFH) parameters, with delayed-exponential SFH and two dust laws, as used in the Q1 pipeline and described in \citetalias{Q1-TP005}. The SEDs are produced with the \texttt{Bagpipes} package \citep{Carnall18} and output in ASCII format with metadata added for use in \texttt{Phosphoros}. We draw 50 values from the photo-$z$ posterior distribution of each reference sample galaxy. \texttt{Phosphoros} is then run in fixed-redshift mode for each sample, forbidding solutions that imply an onset of star-formation earlier than the Big Bang for the observed redshift. A single draw from the marginal $p({\rm SPS}~|~z)$ is taken and derived parameters, such as stellar mass, are computed. The result of this process is a sparsely sampled joint posterior in redshift and galaxy rest-frame properties described by 50 samples per object. During the post-release analysis of the Q1 data, we found that including a prior on the age of the SFH to disfavour very young galaxy ages improved the measurements \citep{Q1-SP031}. We retain this age prior in the reference data.

\subsection{\label{sc:ref_galaxysed}Galaxy SED reference sample}

The third reference sample of galaxies we build for the DR1 PHZ PF is for the purpose of estimating the detailed SEDs of galaxies through the wavelength interval covered by the \IE passband. The target features in this case are the $10\,\mathrm{nm}$-resolution sampling of the SED in units of \si{\uJy}. Methods to estimate these $10\,\mathrm{nm}$-spaced fluxes were explored in \cite{EU-Tarsitano}, including a purely data-driven method and a template-assisted interpolation between existing photometric data. Neither method was able to produce SEDs at the accuracy required for DR3, motivating additional future data collection and possible combined methods for SED reconstruction. For the DR1 PHZ PF, we use a template-based interpolation approach which is suitable at DR1 precision over most of the redshift range of interest.

For each object we begin with the best-fit \gls{sed} found when building the main photo-$z$ reference sample (Sect.\,\ref{sc:cosmos_photoz}). Next, we identify the pairs of medium-bands from the 12 available in the COSMOS field whose effective wavelengths bracket each of the 55 target wavelengths chosen to represent the SEDs (see Sect.\,\ref{sc:pipe_seds}). For each target wavelength we then have two measured medium-band fluxes, two \gls{sed}-predicted medium-band fluxes, and the flux of the SED at the target wavelength. Similar to how we determine the \unif{} fluxes (see Sect.\,\ref{sc:pipe_unif}), we therefore have two predictions for the target wavelength flux, based on the ratio of medium-band fluxes between the catalogue and template \gls{sed}
\begin{equation}
f_{\lambda} = f_{\lambda,{\rm SED}} \frac{f_{{\rm MB,obs}}}{f_{{\rm MB,SED}}},
\end{equation}
where $f_{\lambda}$ corresponds to the target pseudo-narrow-band flux at wavelength $\lambda$ and the subscript MB refers to one of the medium-bands identified as bracketing the target wavelength. \gls{sed} and obs refer to fluxes from the template \gls{sed} and measured fluxes respectively. Finally, a weighted average of the recovered fluxes from the two medium-band predictions is made, where the weights use the inverse of the distances in wavelength from the medium-bands to the target wavelength.

While this method has been shown to work well on average over a large number of simulated objects, the noisy observed medium-band fluxes can result in unphysical SEDs (and thus unphysical PSF models), especially in the cases of negative measured fluxes. 
To remedy this issue we use the best-fit \gls{sed} template as part of a regularisation. Our desired behaviour is that at high measured \gls{snr} in the medium-bands, the template \gls{sed} has no impact, while at low \gls{snr} it acts to prevent large differences between $f_{\lambda}$ and $f_{\lambda,{\rm SED}}$, and in particular to prevent negative fluxes in the final \gls{sed}. This is implemented by computing the inverse-variance weighted average flux between the interpolated flux described above and the \gls{sed}, using a fixed ${\rm S/N}=3$ for the direct \gls{sed} fluxes. Negative estimated fluxes are set to ${\rm S/N}=0$. This essentially ignores the fluxes measured at less than $3\sigma$ and replaces them with the model fluxes, with a smooth transition between measured and mode fluxes.

\subsection{\label{sc:ref_starsed}Star \gls{sed} reference sample}

The reference sample of stellar \glspl{sed} is built from internal on-the-fly (OTF, \citealt{DR1-TP033}) data covering the dedicated \gls{psf} fields to allow its use in \gls{psf} modelling ahead of the main DR1 production. For the source of truth, the \textit{Gaia} spectrophotometers BP and RP are used, as they cover the full wavelength range required for the SEDs with excellent relative flux calibration \citep{,GAIADR3}. From the PSF fields, all \textit{Gaia} objects classified as stars with a continuous spectrum in the \textit{Gaia} EDR3 archive\footnote{\url{https://gea.esac.esa.int/archive/}} are taken and their BP--RP spectra constructed using the \texttt{\textit{Gaia}XP} package \citep{,GAIAXP}. These objects are then matched to the MER OTF catalogues using the associations determined as part of the MER PF. In addition to the spectrophotometry, the best-fit stellar parameters are collected from the \textit{Gaia} archive, and model \glspl{sed} are built by interpolating between the same set of stellar model \glspl{sed} as used in the pipeline star template fits (Sect.\,\ref{sc:star_templates}).

The \textit{Gaia} spectra and models each provide estimates of the stars' \glspl{sed}, which we evaluated at $10\,\mathrm{nm}$ intervals to match how the galaxy \glspl{sed} are expressed. Model and synthetic broad-band fluxes are also obtained by integrating the appropriate passband over the models and \textit{Gaia} spectra. Naturally, $u$-band and \gls{NIR} fluxes cannot be measured from the \textit{Gaia} spectra and so only model fluxes are available. The reference sample is constructed from the combination of the synthetic broad-band fluxes and sampled \glspl{sed}, giving priority where possible to the latter, meaning that model fluxes are used only for the $u$ and \gls{NIR} bands. 

We make the assumption that the diversity of stellar types and Milky Way reddening in the reference data set is sufficient for the purpose of determining \glspl{sed} for DR1. Without a full 3-dimensional model of the dust distribution there is little other choice we can make. However, given the stringent constraints on the wavelength dependence of the \gls{psf} model \citep{Cropper13} this assumption may not hold at DR3 accuracy levels and investigation into improved solutions is underway.

\section{\label{sc:performance}Performance and validation}
In this section we describe the criteria that we use to determine the scientific validity of the data, the principal reasons for failures in data quality, and the survey area lost as a result. This is followed by an assessment of the performance of the tile-by-tile products that will be used in cosmology analyses, namely the photo-$z$, SED products, and classification. As previously stated, validation of the galaxy physical properties is presented in \citetalias{DR1-TP012} and the other products are essentially unchanged with respect to Q1 \citepalias{Q1-TP005}. 

\subsection{\label{sc:validation}Validation strategy}

\begin{figure}[htbp!]
\centering
\includegraphics[angle=0,width=1.0\hsize]{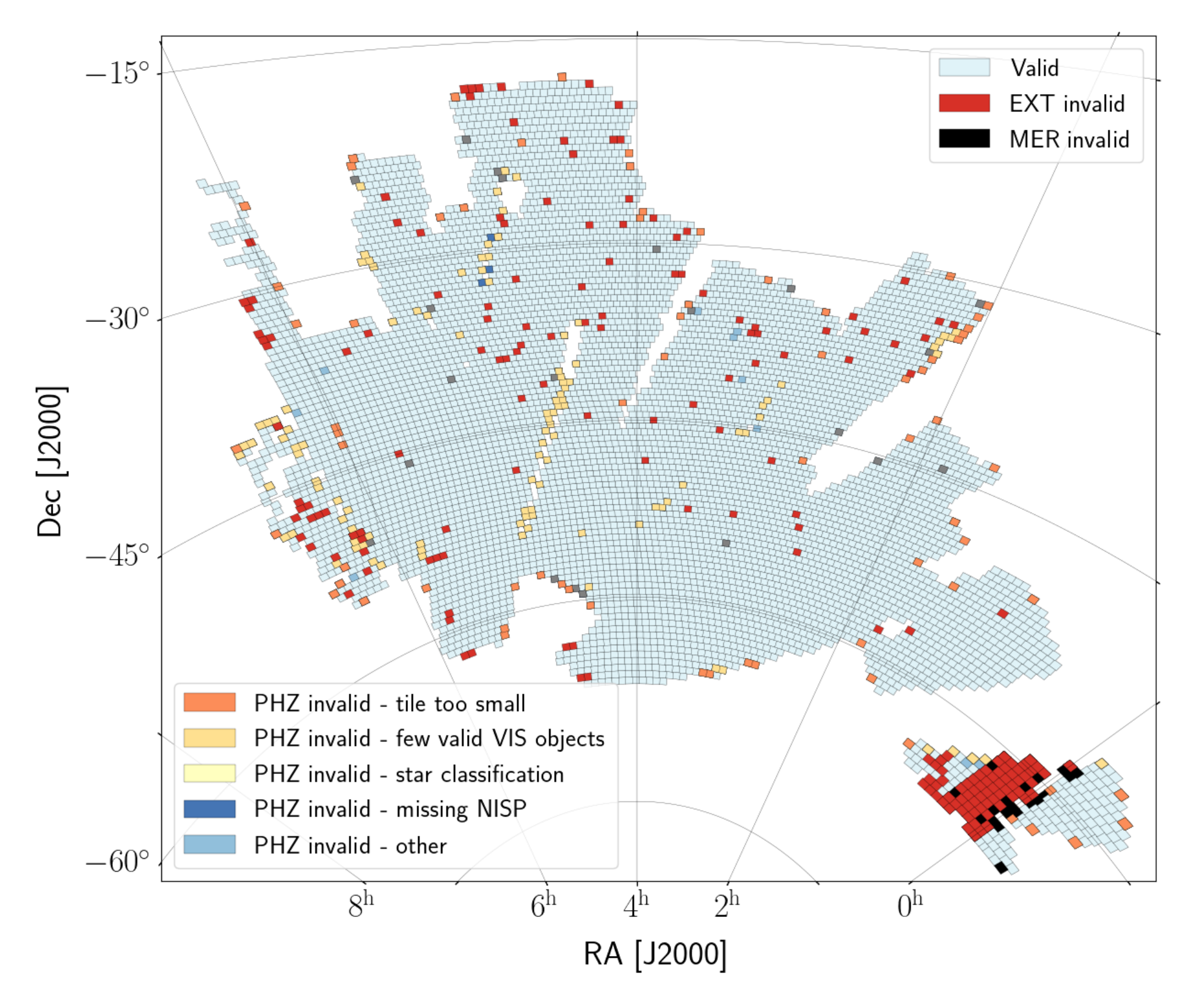}
\raggedleft
\begin{minipage}{0.9\hsize}
\includegraphics[angle=0,width=1.0\hsize]{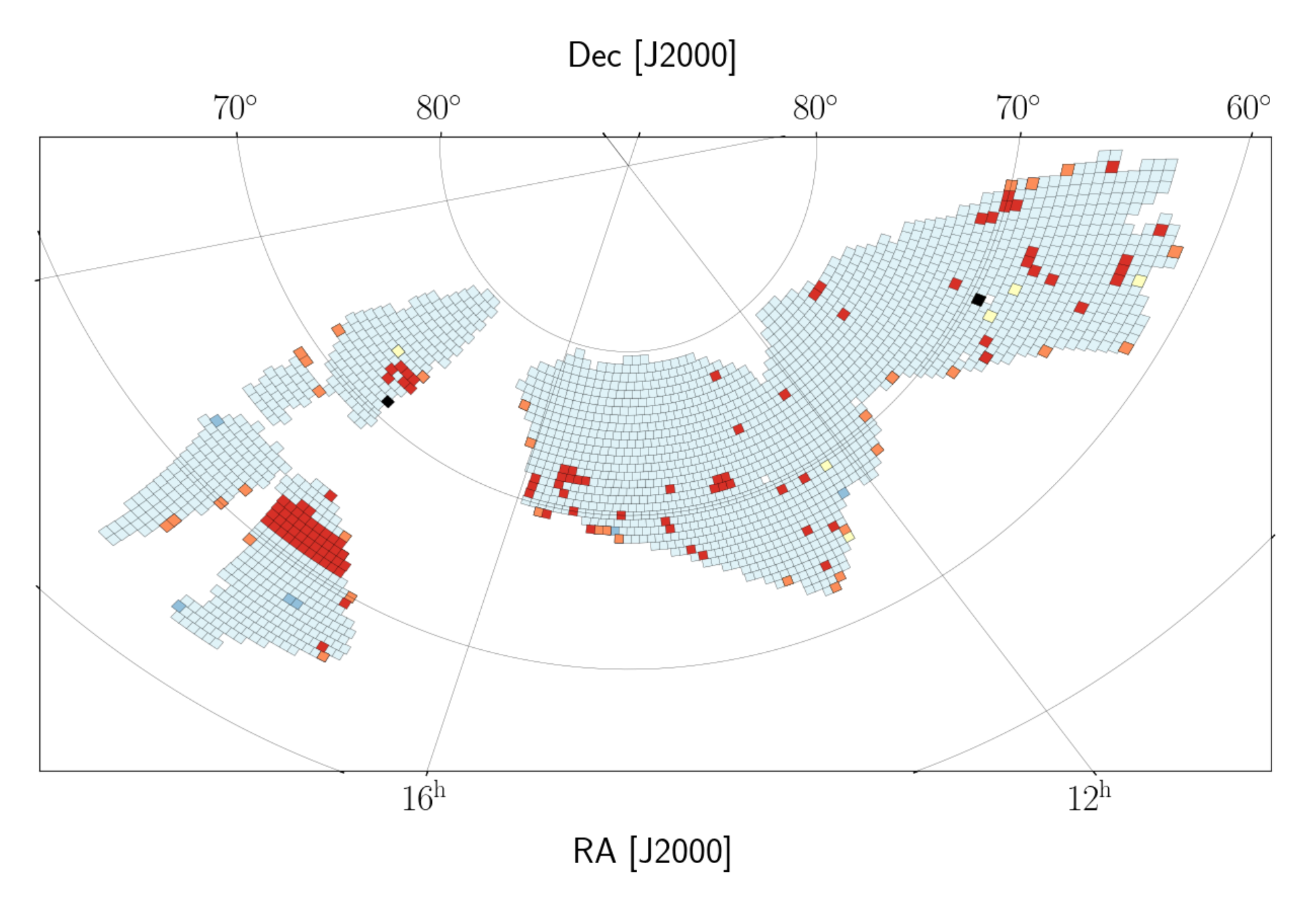}
\end{minipage}
\caption{Maps of the PHZ validation for \gls{EWS} tiles. Each tile is represented by a square, with the colour providing the status and reason for invalidation. See Sect.\,\ref{sc:validation_failures} for further details of the meaning of the PHZ invalidation codes.}
\label{fig:validation_map}
\end{figure}

The final step in the production pipeline is the generation of analysis and validation reports of the inputs and outputs of the pipeline. These form the basis of our tile-by-tile validation process and are broken down into subsections dealing with different technical and scientific aspects of the data. Each subsection consists of a set of pre-defined quality gates that can result in either a pass, an automatic failing condition, or an intermediate label with warnings for near-failing conditions or `dubious' cases requiring a human-made decision. These dubious cases can be considered as soft-fail conditions that might still represent valid output, depending on the specifics of the area of sky they cover, for example regions of higher stellar density. Criteria were set such that data are declared valid if they serve some scientific purpose and do not guarantee that data are valid for all possible science cases.
The major sections of the validation reports deal with the following aspects.
\begin{itemize}
\item The input MER catalogue product, assessing the depth of the \texttt{2FWHM\_APER} measurements, a test of the band-to-band photometric zero-point accuracy, the number and combinations of missing photometric bands, and the area and source density of objects in the tile.
\item Classification performance, assessed via number counts and quantiles of colour distributions with respect to reference data for different magnitude cuts. Since the classification is based purely on photometric information, the morphological information provided in the MER output catalogue is also used to assess the performance of the star classification.  
\item Performance of the photo-$z$, both in terms of point estimates and the dispersion of their \glspl{zpdf}, based on metrics such as: the similarity of the mean of the main peak and median of the \gls{zpdf}; redshift distribution of these point estimates with respect to the COSMOS reference data; $1\sigma$, $2\sigma$, and $3\sigma$ credible interval statistics; and the $\chi^2$ statistic reported by \nnpz. 
\item Technical tests of the contents of all output catalogues and some simple galaxy physical property checks without failing conditions. These two tests were added during re-delivery of the EDFs data products and so correspond only to those tiles.
\end{itemize}

It is worth reminding the reader that during this tile-by-tile validation we do not have access to truth values, such as spectroscopic redshifts, for any of the galaxies and rely on population-level comparisons with reference data. Sample variance on the scale of a single MER tile, especially in the EDFs, is considerable and thus our validation criteria are set to be loose enough to allow for this variance. Assessment of the pipeline performance with respect to spectroscopic redshifts and other labelled data is the subject of the subsections that follow. 

In addition to our own validation criteria, we also take account of the validation performed by both the OU-MER \citep{DR1-TP009} and OU-EXT \citep{DR1-TP006} teams. Because the MER catalogue is used as part of the inputs for the \Euclid spectroscopy products created by OUs for spectral near-infrared (SIR) and spectroscopy (SPE), their catalogues are validated and published regardless of any issues identified in the optical broad-band photometry, to prevent breaking the lineage of the data products. We therefore incorporate the quality assessment performed by those teams into the PHZ validation.
Tiles effectively invalidated by these two OUs are labelled as EXT/MER in Fig.\,\ref{fig:validation_map}. 

\subsection{Validation failure modes}
\label{sc:validation_failures}

The most common failure modes in the validation correspond to cases of absent \Euclid data, such as near the edge of the survey footprint where data from one of the instruments exist but not the other. There are also cases where either the VIS or the NISP data did not pass the initial quality criteria, or were invalidated during Level 2 processing by OU-VIS or OU-NIR \citep{DR1-TP002,DR1-TP003}. These cases correspond to the `few valid VIS objects' and `missing NISP' invalidation criteria shown in Fig.\,\ref{fig:validation_map}. While individual objects are flagged where data are missing, and can thus be removed from scientific analyses, having too large a fraction of such objects in a single tile makes the pipeline unreliable for steps such as the flux imputation and identification of sufficient stars for \gls{psf} modelling. In particular, the classification step in the main production pipeline uses the collection of objects in the tile as an effective training sample during the imputation of missing photometric data. 

For similar reasons we also invalidate tiles that have extremely small areas (typically, $<0.01\,\mathrm{deg}^2$). They often also have incomplete photometry that requires imputation , but also a small number of objects available for this task, leading to poor results. It was determined during the preparation of the Q1 release that losing the small fraction of area that is affected by this issue is preferable to publishing data with different and often peculiar behaviour at the edges of the survey regions. The marginal additional loss in area through excluding heavily incomplete tiles or those with very small area is minor: just $1\%$ of the DR1 \gls{EWS}.

Once cases of incomplete data coverage and the small fragments of tiles on the edge of the footprint have been accounted for, only $0.5\%$ of tiles are invalidated due to anomalies in the PHZ processing. Among these cases, only apparent issues with the star classification are numerous enough to warrant a separate mention in Fig.\,\ref{fig:validation_map}. These tiles have a significant fraction of high-\gls{snr} objects that are confidently classified as stars from their photometry, but also have a low probability to be point like according to morphological measurement from the MER catalogue. These may be blended objects, misclassified galaxies, or problems with either the measured photometry or data imputation. Regardless of the root cause, we consider the PHZ outputs for these tiles to be unreliable. Most of the remaining invalid tiles were rejected because they differed too greatly from one or another of the reference data sets. For example, a few tiles were flagged because the $n(z)$ histogram of their mode photo-$z$ values had far too many high-redshift galaxies, especially corresponding to known degeneracies in the colour space of galaxies; or because the photometric zero-point check suggested that one or more of the zero points was wrong by as much as $0.5\,\mathrm{mag}$.

\subsection{\label{sc:perf_photoz}Photo-$z$ performance}

\begin{figure}[tbp!]
\centering
\includegraphics[angle=0,width=1.0\hsize]{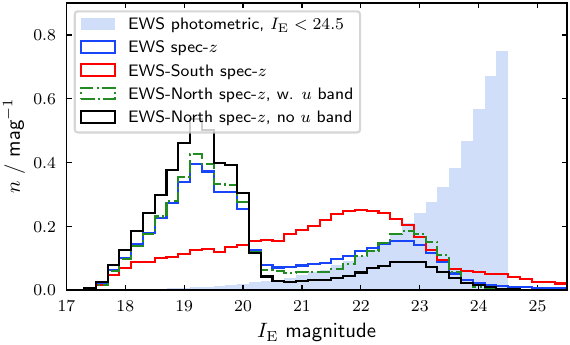}
\includegraphics[angle=0,width=1.0\hsize]{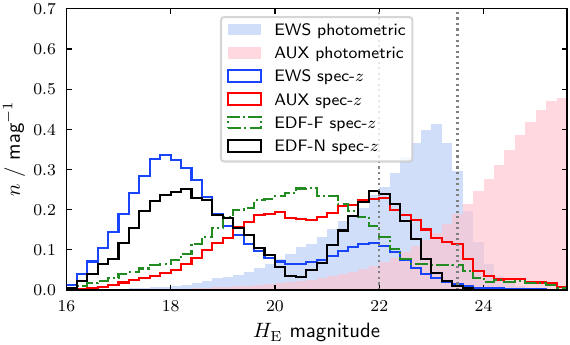}
\caption{Normalised magnitude histograms for the different spectroscopic samples and flux-limited photometric samples considered during the photo-$z$ validation, as listed in the legends. The upper panel shows distributions in \IE magnitude, and the lower panel shows \HE distributions. The vertical dotted lines in the lower figure are the \HE-band flux limits considered for the \gls{EWS}.}
\label{fig:mag_hist}
\end{figure}

\begin{figure*}[htbp!]
\centering
\includegraphics[angle=0,width=1.0\hsize]{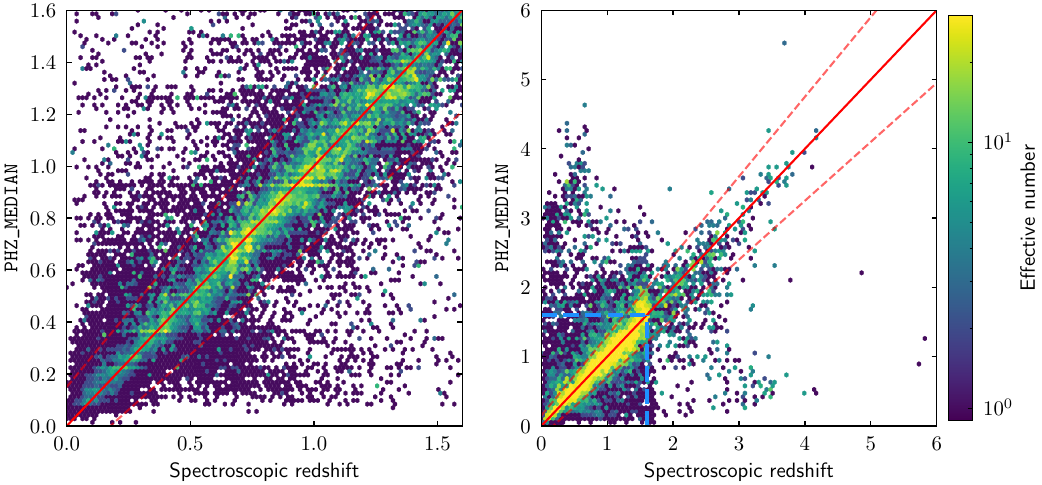}
\caption{Photo-$z$ point estimates (given as the median of the \gls{zpdf}) versus spectroscopic redshifts for galaxies in the \gls{EWS}. The colour bar shows the weighted density of objects, where the weights have been determined to match the magnitude distribution of a flux-limited $17<\IE<24.5$ galaxy sample. The solid red line shows the 1:1 relation, while the dashed lines show the standard definition for outliers, $z\pm 0.15(1+z)$. The left-hand panel focuses-in on the area of the right-hand panel outlined by the dashed blue box.}
\label{fig:photoz_wide}
\end{figure*}

\begin{figure}[htbp!]
\centering
\includegraphics[angle=0,width=1.0\hsize]{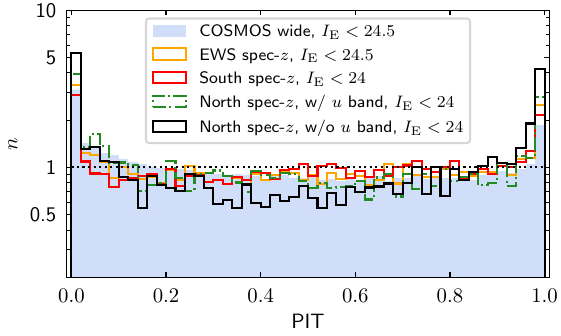}
\includegraphics[angle=0,width=1.0\hsize]{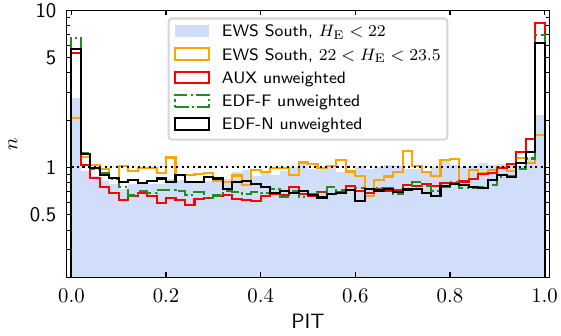}
\caption{Histograms of PIT values for some of the different samples considered in the photo-$z$ validation. Samples with magnitude limits have been weighted to represent flux-limited samples, as in Table \ref{tab:photz}. The horizontal dotted line indicates how a histogram would look if all the \glspl{zpdf} in the sample were perfectly calibrated and the sample large enough to be free of shot noise. Note the change in the $y$-axis scale between the upper and lower panels.}
\label{fig:pit_plots}
\end{figure}

\begin{table*}[]
\centering
\setlength{\tabcolsep}{4.5pt}
\begin{threeparttable}
\caption{Weighted summary statistics by magnitude limit for photometric redshifts in the \gls{EWS}. Samples are further split by survey region (north or south) and, in the north, by the presence or absence of $u$-band data.}\label{tab:photz}
\begin{tabularx}{\textwidth}{l*{11}{c}} 
\hline\hline
\noalign{\vskip 2pt}
 & \multicolumn{2}{c}{\uline{$17<\IE<24.5$}} & \multicolumn{3}{c}{\uline{$17<\IE<24$}} & \multicolumn{3}{c}{\uline{$17<\HE<22$}}&\multicolumn{3}{c}{\uline{$22<\HE<23.5$}} \\
 \noalign{\vskip 2pt}
 & \gls{EWS}$^{a}$ & COSMOS$^{b}$ & N w/ $u$ & N w/o $u$ & S & N w/ $u$ & N w/o $u$ & S & N w/ $u$ & N w/o $u$ & S \\
\hline
\noalign{\vskip 2pt}
$N$ spectra        & $\num{191169}$ & $\num{100593}$ & $\num{77392}$  & $\num{69728}$  & $\num{42656}$ & $\num{67827}$ & $\num{65497}$ & $\num{37859}$ & $\num{10588}$ & $\num{4542}$ & $\num{6382}$ \\
$N_{\mathrm{eff}}$      & $7726$   & $\num{92592}$  & $3405$   & $\num{3844}$ & $7548$  & $\num{13938}$ & $7758$  & $\num{19729}$ & $1943$ & $1106$ & $4106$ \\
Outlier frac.    & $20.16\%$ & $21.75\%$ & $13.10\%$ & $18.53\%$ & $12.50\%$ & $7.12\%$ & $9.07\%$ & $7.61\%$ & $19.15\%$ & $28.55\%$ & $19.13\%$ \\
Mean bias        & $-0.064$ & $-0.097$ & $-0.034$ & $-0.047$ & $-0.047$ & $-0.007$ & $-0.005$ & $-0.014$ & $-0.058$ & $-0.088$ & $-0.072$ \\
Median bias      & $-0.001$ & $-0.006$ & $-0.004$ & $\phantom{+}0.002$ & $\phantom{+}0.002$ & $\phantom{+}0.002$ & $\phantom{+}0.005$ & $\phantom{+}0.003$ & $-0.010$ & $-0.009$ & $\phantom{+}0.001$ \\
$\sigma_{68}$     & $\phantom{+}0.094$ & $\phantom{+}0.106$ & $\phantom{+}0.068$ & $\phantom{+}0.090$ & $\phantom{+}0.065$ & $\phantom{+}0.043$ & $\phantom{+}0.050$ & $\phantom{+}0.044$ & $\phantom{+}0.085$ & $\phantom{+}0.136$ & $\phantom{+}0.090$ \\
NMAD             & $\phantom{+}0.072$ & $\phantom{+}0.080$ & $\phantom{+}0.054$ & $\phantom{+}0.066$ & $\phantom{+}0.054$ & $\phantom{+}0.038$ & $\phantom{+}0.044$ & $\phantom{+}0.039$ & $\phantom{+}0.068$ & $\phantom{+}0.098$ & $\phantom{+}0.076$ \\
\hline
\noalign{\vskip 2pt}
& \multicolumn{9}{c}{Outlier removed stats} \\
\noalign{\vskip 2pt}
Mean bias        & $\phantom{+}0.002$ & $-0.001$ & $-0.003$ & $\phantom{+}0.004$ & $\phantom{+}0.003$ & $\phantom{+}0.001$ & $\phantom{+}0.005$ & $\phantom{+}0.004$ & $-0.008$ & $-0.0012$ & $\phantom{+}0.003$ \\
Median bias      & $\phantom{+}0.004$ & $\phantom{+}0.000$ & $-0.001$ & $\phantom{+}0.006$ & $\phantom{+}0.005$ & $\phantom{+}0.002$ & $\phantom{+}0.006$ & $\phantom{+}0.004$ & $-0.007$ & $\phantom{+}0.000$ & $\phantom{+}0.005$ \\
$\sigma_{68}$     & $\phantom{+}0.056$ & $\phantom{+}0.059$ & $\phantom{+}0.048$ & $\phantom{+}0.052$ & $\phantom{+}0.049$ & $\phantom{+}0.037$ & $\phantom{+}0.042$ & $\phantom{+}0.037$ & $\phantom{+}0.052$ & $\phantom{+}0.061$ & $\phantom{+}0.059$ \\
\hline
\noalign{\vskip 2pt}
& \multicolumn{9}{c}{$z$PDF metrics} \\
\noalign{\vskip 2pt}
PDF outliers     & $5.41\%$ & $3.90\%$ & $5.97\%$ & $9.95\%$ & $5.34\%$ & $6.13\%$ & $8.17\%$ & $5.33\%$ & $6.14\%$ & $9.38\%$ & $3.23\%$\\
KL-div.\ (Nats)& $\phantom{+}0.014$ & $\phantom{+}0.017$ & $\phantom{+}0.036$ & $\phantom{+}0.075$ & $\phantom{+}0.007$ & $\phantom{+}0.009$ & $\phantom{+}0.0421$ & $\phantom{+}0.007$ & $\phantom{+}0.054$ & $\phantom{+}0.107$ & $\phantom{+}0.007$ \\
F-005 (phot)     & $\phantom{+}0.602$ & $\phantom{+}0.571$ & $\phantom{+}0.725$ & $\phantom{+}0.715$ & $\phantom{+}0.671$ & $\phantom{+}0.812$ & $\phantom{+}0.797$ & $\phantom{+}0.777$ & $\phantom{+}0.679$ & $\phantom{+}0.625$ & $\phantom{+}0.547$ \\
F-015 (phot)     & $\phantom{+}0.854$ & $\phantom{+}0.847$ & $\phantom{+}0.927$ & $\phantom{+}0.908$ & $\phantom{+}0.903$ & $\phantom{+}0.968$ & $\phantom{+}0.954$ & $\phantom{+}0.958$ & $\phantom{+}0.893$ & $\phantom{+}0.850$ & $\phantom{+}0.830$ \\
F-005 (spec)     & $\phantom{+}0.404$ & $\phantom{+}0.386$ & $\phantom{+}0.509$ & $\phantom{+}0.460$ & $\phantom{+}0.485$ & $\phantom{+}0.628$ & $\phantom{+}0.580$ & $\phantom{+}0.603$ & $\phantom{+}0.445$ & $\phantom{+}0.359$ & $\phantom{+}0.375$ \\
F-015 (spec)     & $\phantom{+}0.723$ & $\phantom{+}0.710$ & $\phantom{+}0.817$ & $\phantom{+}0.767$ & $\phantom{+}0.808$ & $\phantom{+}0.907$ & $\phantom{+}0.878$ & $\phantom{+}0.889$ & $\phantom{+}0.753$ & $\phantom{+}0.655$ & $\phantom{+}0.707$ \\
\hline
\end{tabularx}
 \footnotesize
 \tablefoot{The point estimate used here is the median. Statistics are computed for the full sample, and in the case of point estimates also after excluding outliers. The different subsurvey regions are shown separately, where possible.
      $^{(a)}$ Due to the magnitude-dependent weighting, the statistics are dominated by a relatively small number of spectroscopic objects in a \gls{EWS}-like pass of the EDF-F. 
      $^{(b)}$ The \Euclid data in the COSMOS \gls{EWS}-like processing are slightly shallower than typical for the \gls{EWS} due to the higher background.}
\end{threeparttable}
\end{table*}

\begin{table}[]
\centering
\caption{Unweighted summary statistics for \gls{EDF}s and \gls{EAF}s, as listed.}\label{tab:photz_deep}
\begin{tabularx}{\hsize}{lcc@{\hspace{-6pt}}c} 
\hline\hline
\noalign{\vskip 2pt}
 &  EDF-N & EDF-F & AUX\\
 &        &       & (COSMOS + CDFS)\\
\hline
\noalign{\vskip 2pt}
No.\ spectra     & $28105$  & $34018$  & $44344$ \\
Outlier fraction & $4.14\%$ & $6.82\%$ & $4.02\%$ \\
Mean bias     & $-0.007$\phantom{$-$} & $-0.005$\phantom{$-$} & $-0.008$\phantom{$-$} \\
Median bias   & $0.001$  & $0.001$  & $0.002$ \\
$\sigma_{68}$ & $0.037$  & $0.040$  & $0.022$ \\
NMAD          & $0.033$  & $0.035$  & $0.018$ \\
\hline
\noalign{\vskip 2pt}
 \multicolumn{4}{c}{Outlier removed stats} \\
 \noalign{\vskip 2pt}
Mean bias     & $-0.003$\phantom{$-$} & $0.001$  & $0.004$ \\
Median bias   & $0.000$  & $0.001$  & $0.002$ \\
$\sigma_{68}$ & $0.034$  & $0.035$  & $0.020$ \\
\hline
\noalign{\vskip 2pt}
 \multicolumn{4}{c}{$z$PDF metrics} \\
 \noalign{\vskip 2pt}
PDF outliers  & $14.62\%$ & $17.51\%$ & $16.40\%$ \\
PIT KL-div.\ (Nats) & $0.050$  & $0.061$  & $0.093$\\
F-005 (phot)  & $0.852$  & $0.850$  & $0.933$\\
F-015 (phot)  & $0.982$  & $0.966$  & $0.986$\\
F-005 (spec)  & $0.706$  & $0.675$  & $0.819$\\
F-015 (spec)  & $0.938$  & $0.906$  & $0.950$\\
\hline
\end{tabularx}
 \footnotesize
 \tablefoot{The point estimate used here is the median. The magnitude limit in each case is $\HE<25.5$.}
\end{table}

The metrics by which a set of photometric redshifts are assessed relative to some set of assumed truth values, that is, spectroscopic redshifts, are now fairly standard in the literature \citep[e.g.,][]{Sanchez14, Salvato18, Desprez-EP10,Schmidt20}. 
To assess the performance of the DR1 photo-$z$ we employ the following set of metrics \citepalias[see also][]{Q1-TP005}, where $\delta z$ is the scaled difference between the photo-$z$ point estimate and spectroscopic redshift, $\delta z = (z_{\rm phot} - z_{\rm spec}) / (1 + z_{\rm spec})$,
\begin{itemize}
\item fraction of outliers, with $|\delta z| > 0.15$;
\item mean scaled bias, $\langle\delta z\rangle$;
\item median scaled bias, ${\rm med}(\delta z)$;
\item $\sigma_{68}$, the central $68$-percentile range of $\delta z$ in the sample;
\item normalised median absolute deviation, \\ ${\rm NMAD}=1.48\,{\rm med}(|~\delta z~-\mathrm{med}(\delta z)~|)$;
\item fraction of objects that are outliers in the sense that the true redshift lies outside of the central $99.8\%$ interval of their $z$PDF;
\item the Kullback--Leibler (KL) divergence of the histogram of probability integral transform (PIT) values with respect to a flat distribution, where the PIT is simply the cumulative redshift distribution function evaluated at the true redshift;
\item and the average integrated probability within either 0.05 (F-005) or 0.15 (F-015) in redshift of some assumed value, either the spectroscopic redshift or photometric redshift point estimate \citep{Desprez-EP10}.
\end{itemize}

We further quote point redshift metric values after excluding outliers, as is common in the deep galaxy survey literature. While this will tend to produce unrealistically optimistic performance, it can nevertheless be useful and informative in understanding the bias and spread around the true value for `good' galaxies. Furthermore, wrong spectroscopic redshifts and mis-classified objects (including strong AGN) often produce outliers in point redshift estimates that cloud the assessment of the performance for normal galaxies. Similarly, the KL divergence of the PIT histogram is computed for objects with values $0.001<{\rm PIT}<0.999$, so as to quantify the distribution of non-outliers. Otherwise, the KL divergence would be driven to first order by the strength of the outlier peaks and provide effectively the same information as the outlier fraction.

\subsubsection{\label{sc:perf_wide}Wide tile performance}

Our aim is to assess the performance of the photo-$z$ for possible galaxy samples to be used in the \threebytwo photometric cosmology analysis, including a baseline $\IE<24.5$ selection, and other magnitude-limited samples. These samples are intended to provide the reader with a sense of how photo-$z$ precision and accuracy depends on flux limit, and thus inform their science sample selection.
Unfortunately, complete spectroscopic samples with pure magnitude limits at the depth of interest for \Euclid do not exist. Instead, we can take two approaches: give spectroscopic objects weights such that their weighted distribution in magnitude matches a target flux-limited distribution (see, e.g., \citealt{Bonnett16}); or use accurate photometric redshifts in place of spectroscopic redshifts as the source of truth, for example using the photo-$z$ from the COSMOS field. We employ both of these methods to assess performance and consider the north and south regions of the DR1 footprint separately, differentiating also between the two combinations of bands in the north part of the survey, with or without $u$-band photometry.

The spectroscopic compilations we use for the purpose of measuring photo-$z$ quality metrics are detailed in Saulder et al., (in prep.). Data sets were included on the basis that they contain internal quality flagging information sufficient to remove most of the poor redshifts, and a median of the acceptable redshifts is used in the cases where multiple solutions exist for a single object. Nevertheless, a small but unknown fraction of the spectroscopic redshifts are likely to be incorrect, especially for the largest data sets containing only automated spectroscopic determinations, namely DESI \citep{DESI} and PRIMUS \citep{Cool2013}. Moreover, the incorrect fraction can be expected to rise at the faintest magnitudes within any given data set, where the \gls{snr} of the data are lowest. We are unable to account for any such incorrect spectroscopic redshifts and so our results should be read as a lower limit to the overall performance of the measurements. 

From the spectroscopic compilation, we take objects with at least one good redshift measurement and use the median of the spectroscopic determinations as the true redshift. We then remove any objects spectroscopically identified as stars or QSOs. Such objects will be removed from the cosmology samples, assuming they can be identified, and QSOs or other AGN are often specifically targeted in spectroscopic campaigns, leading to an over-representation in the aggregated sample (and hence pessimistic performance in galaxy photo-$z$). Some QSOs and strong AGN will remain in the spectroscopic sample, however, since not all spectroscopic surveys provide flagging information to identify them. We also perform object selection based on the \Euclid products, as we would for scientific analyses. These cuts comprise the rejection of spurious sources, objects falling in masked areas \citep{DR1-TP028} and, in the case of \IE flux limits, a compactness criterion to remove point-like objects.

The magnitude distributions of the selected spectroscopic objects for the \gls{EWS} north and south regions are shown in Fig.\,\ref{fig:mag_hist}, alongside those for flux-limited wide samples and two deep fields, EDF-F and EDF-N. Four flux-limited samples are considered when computing performance metrics for the \gls{EWS}: an \IE-selected sample at the nominal \Euclid target depth, $\IE<24.5$; a slightly brighter and highly complete \IE sample, $\IE<24$; a bright \HE-selected sample, suitable for galaxy clustering and galaxy--galaxy lensing measurements, $\HE<22$; and a faint \HE-selected sample reaching approximately the \HE-band completeness limit, $22<\HE<23.5$. Computed values of the performance metrics for these four samples are given in Table\,\ref{tab:photz}, where we have further split three of the samples into three different sub-survey regions, depending on the photometric configuration (DES or UNIONS) and the availability of $u$-band data. 

Comparison of the median photo-$z$ for each spectroscopic object against the assumed true redshift for the nominal-depth sample is shown in Fig.\,\ref{fig:photoz_wide}, where the colour represents the weighted sample density with a logarithmic scale. The larger scatter in Fig.\,\ref{fig:photoz_wide} when compared to the equivalent plot for Q1 in \citetalias{Q1-TP005} is due to the magnitude weights we have applied; in fact, the scatter in DR1 is lower than that in Q1 when calculated in similar conditions. One specific area of improvement worthy of note is the much reduced group of outliers at $z_{\mathrm{spec}}\simeq0.6$ and $z_{\mathrm{phot}}\simeq0.2$. This is due to the implicit prior in galaxy type, magnitude, and redshift that is imposed through the use of \nnpz{} and a representative reference sample.

Figure\,\ref{fig:pit_plots} shows histograms of PIT values for selected subsamples (remaining subsamples omitted for clarity of display). Perfect performance for the $z$PDFs is represented by the horizontal black dotted line, and the fraction of PDF outliers can be readily appreciated by the heights of the first and last histogram bars. In addition to the spectroscopic samples, we also compute metric values for an \gls{EWS}-like processing of the COSMOS field, created specifically for validation purposes. Images emulating the \gls{EWS} were compiled from the available \Euclid{} and DECam data, with the caveat that COSMOS lies in a region of the sky with high zodiacal light and so the delivered \Euclid{} depths are slightly shallower than the DR1 \gls{EWS}. The best-estimated photo-$z$ from the COSMOS2020 CLASSIC/\texttt{LePhare} catalogue \citep{Weaver22} are used as the truth values.

At the nominal \Euclid depth, $\IE<24.5$, both the spectroscopic and COSMOS samples report similar performance for the point estimates of photo-$z$ (taken as the medians of the \glspl{zpdf}). At around $20\%$, the outlier fraction of point estimates is twice the mission design requirement, and the dispersion around the 1:1 line also falls short of the DR3 requirement of $0.05$, at $0.07\mbox{--}0.08$ for NMAD and $0.1$ for $\sigma_{68}$. While each sample may suffer from a low level of catastrophic failures in the truth values, the similarity of the metric values suggests that they are not far from the truth. Indeed, a similar analysis was carried out using data from COSMOS before mission launch in \cite{Desprez-EP10}, with very similar results for the 30-band photo-$z$ validation scores. The DR1 performance, while somewhat below our end-of-mission target, is consistent with pre-launch expectations.

The situation is not as pessimistic as these metrics may first suggest, however. Firstly, a purely Gaussian dispersion with $\sigma=0.1$ would lead to an outlier fraction of $13\%$ under our definition, $|\delta z|>0.15$, meaning that the fraction of genuine catastrophic outliers (such as those resulting from a confusion between strong breaks in their SEDs) is likely to be under reasonable control. Moreover, the dispersion of photo-$z$ point estimates around their true values is driven by several factors, including the number and wavelengths of the photometric bands, the depths of the photometric measurements, and the quality of the photometric calibration. The impact of these factors can be seen in the slightly brighter sample, $\IE<24$, where we are able to separate the sample into the three sub-survey regions.\footnote{The spectroscopic compilation contains virtually no good spectroscopic objects at $24<\IE<24.5$ in the northern hemisphere of DR1.} Object number counts are still quite steep at these depths, and the change in magnitude limit represents close to a $60\%$ increase in average \gls{snr}. This increase in the photometric precision results in a corresponding reduction of around $30\%$ in photo-$z$ dispersion for DR1-South, to the point that the metric values are close to meeting the pre-launch requirements. 

The impact of the inclusion of $u$-band data can be seen from the two sub-regions of DR1-North. The additional band reduces the outlier fraction by five percentage points and the dispersion by around $20\%$. By mission end, the full survey area will be covered by $u$-band data and the \gls{snr} of the photometry in the southern hemisphere will be substantially increased through switching to data from Rubin-LSST \citep{Ivezic19}. The northern area external data will also see an improved homogenisation and may support photo-$z$ at the level of the mission requirements even if the formal depth requirement cannot be met in all bands.

\begin{figure}[t!]
\centering
\includegraphics[angle=0,width=1.0\hsize]{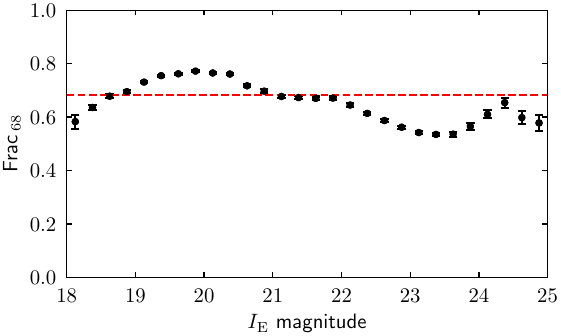}
\caption{Fraction of objects with spectroscopic redshifts falling within the percentile range, $15.9\mbox{--}84.1\,\%$, of their respective \glspl{zpdf} as a function of magnitude. The horizontal red dashed line shows the expected behaviour for perfectly calibrated \glspl{zpdf}.}
\label{fig:frac68}
\end{figure}

Regarding the calibration of the \glspl{zpdf}, the performance we see across the \gls{EWS} samples in terms of the flatness of the PIT distribution is rather good (Fig.\,\ref{fig:mag_hist} and KL-divergence in Table\,\ref{tab:photz}), and compares favourably with the pre-launch data challenge presented in \citet[their figure\,7]{Desprez-EP10}. The main exception is the case of the DR1-North where no $u$-band data are available. The samples discussed thus far provide only limited information on the magnitude dependence of the calibration of the \glspl{zpdf}, and so we complement those metrics with an alternative expression of the calibration quality in Fig.\,\ref{fig:frac68}. Here, we compute the fraction of objects with spectroscopic redshift lying within the central $68$-percentile range ( $15.9\mbox{--}84.1\,\%$) of the respective \gls{zpdf}, $\mathrm{Frac}_{68}$, within narrow intervals in \IE magnitude (see also, \citealt{Saxena24}, their figure.\,7). Uncertainties are computed through bootstrap resampling. This $\mathrm{Frac}_{68}$ value is equivalent to the integral of the PIT distribution over the same percentile range, with a value of 0.682 at all magnitudes representing perfect behaviour. The magnitude dependence is weak, and thus we can conclude that the calibration of our \glspl{zpdf} is of similarly high quality across the broad range in galaxy brightness. 

\subsubsection{\label{sc:perf_deep}Deep tile performance}

\begin{figure*}[htbp!]
\centering
\includegraphics[angle=0,width=1.0\hsize]{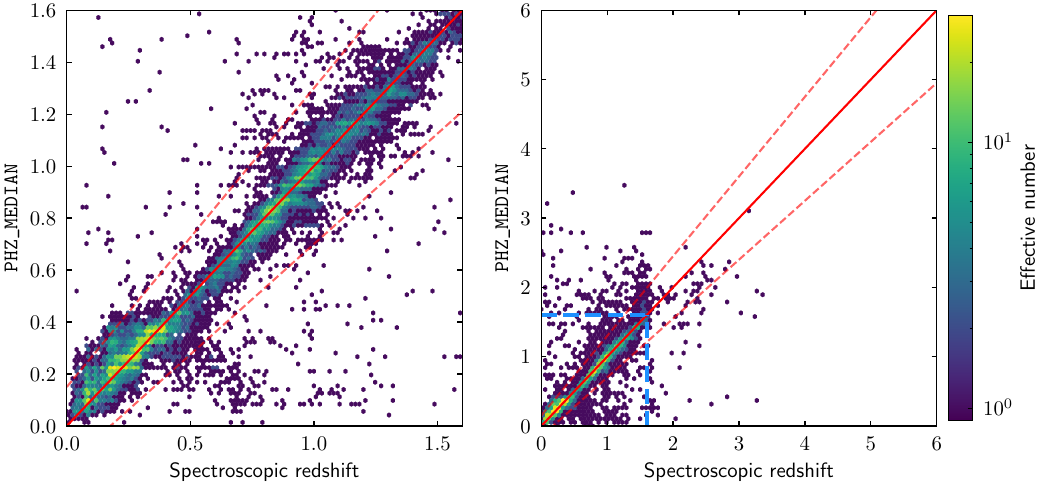}
\caption{Photo-$z$ versus spectroscopic redshift for the EDF-N at $\HE<25.5$. Lines and colours have the same meanings as in Fig.\,\ref{fig:photoz_wide}, except that the objects are not weighted.}
\label{fig:photoz_deep}
\end{figure*}

Forming weighted spectroscopic samples to approximate flux-limited data sets at EDFs depths is impractical and would result in an extremely small effective sample size. Instead, we computed unweighted metrics for the available spec-$z$ samples for the cases of the EDF-N, EDF-F, and the combined set of the two photo-$z$ \glspl{EAF}: CDFS and COSMOS. Relative to a pure flux-limited galaxy sample, the metrics computed in this way are rather optimistic and better reflect quite bright galaxies. Nevertheless, since this approach is common in the literature it allows for simple cross-comparisons. It is also worth reminding the reader that EDF-F has just two layers of \Euclid data in DR1, and part of the field is covered by only wide-depth DECam data \citep{DR1-TP006}, with the result that the performance for this field can be expected to lie between the genuine EDFs area and the \gls{EWS}. Metrics and associated figures are shown in Table \ref{tab:photz_deep} and Figs.\,\ref{fig:photoz_deep} and \ref{fig:pit_plots}, with the same meanings as described for the \gls{EWS} results.

The most directly comparable photo-$z$ results in the literature are from CANDELS \citep{Dahlen13}, the DES Deep Fields \citep[DES DF,][]{Hartley22}, and the UKIDSS Ultra-Deep Survey \citep[UDS,][]{Taylor23}, where each data set contains deep broadband imaging over the optical-to-\gls{NIR} range. The key difference in the Euclid DR1 data is the lack of $u$-band coverage that is present in these other catalogues. Despite the reduced wavelength coverage, the metrics for the performance of point redshift estimates are of a comparable standard to these aforementioned studies, especially in the case of the \glspl{EAF} where the scatter is as low as 0.02 (CANDELS: 0.03, DES DF: 0.02, UDS: 0.02). Meanwhile, the EDFs show only slightly worse performance, at $\mathrm{NMAD}=0.033$ and $0.035$ for EDF-N and EDF-F, respectively. In general, the outlier fractions are slightly worse than the comparison papers', but still matching the DES DF value of $6\%$. We should stress, however, that these quoted NMAD values all depend on the mix of objects available with secure spectroscopic redshifts and should be treated as guideline numbers, even in comparison with one another.

For the \gls{zpdf} metrics listed in Table \ref{tab:photz_deep}, we can compare the performance of the \glspl{EDF} and \glspl{EAF} against those from the DES DF described in \cite{Hartley22}. We have recomputed the KL-divergence values for these samples to be in units of Nats (consistent with the present work) and also corrected an implementation bug in the \cite{Hartley22} calculation. We find new values of 0.020, 0.021, and 0.101 Nats for the unweighted spectroscopic, \cite[; $17<i<23$]{Alarcon20} and \cite[; $17<i<24$]{Laigle16} redshift samples, respectively. In the latter two cases precise photo-$z$ are used as the true redshift, similar to our use of the COSMOS \gls{EWS}-like processing. From the lower panel of Fig. \ref{fig:pit_plots} it is clear that the \glspl{zpdf} in the \glspl{EDF} for DR1 are somewhat under-dispersed with respect to the \gls{EWS}. The values for KL-divergence we report in Table \ref{tab:photz_deep} reflect that weaker calibration and are also larger than the closest equivalent measurement in the DES DF, the unweighted spectroscopic case. However, the more striking difference with respect to the DES DF is in the outlier rate, defined in this case as true redshifts lying entirely outside of the $z$PDF, which at $15\%$ is rather high. As mentioned in Sect.\,\ref{sc:pipe_deep}, a configuration error during the running of the deep pipeline meant that no systematic flux errors were applied during photo-$z$ computation and, given the overall high \gls{snr} of the photometry, any differences in effective zero points between the reference and target data would lead to a greater number of such outliers. This issue will be corrected in the DR2 processing, but any users of the DR1 deep field \glspl{zpdf} will need to account for this increased fraction of outliers.

\subsection{\label{sc:perf_seds}Performance of SED products}
A robust validation of the SED data products (for both stars and galaxies) with respect to the requirements listed in \cite{Cropper13}, and down to their respective target flux limits, is not possible. In neither case do suitable data for this purpose exist and, moreover, it is already known that the data we had available to produce the reference sample for galaxy SEDs were not sufficient to meet our desired performance level \citep{EU-Tarsitano}. Nevertheless, we can perform a limited assessment of their performance using the data that do exist.

\subsubsection{\label{sc:perf_starsseds}Stellar SEDs}
Due to the processing schedule of the data release, the stellar SED reference sample was prepared ahead of the main data processing and used photometric catalogues from an earlier release of the MER processing function. Any changes in the upstream calibration of the photometric data between that time and the main DR1 campaign may then propagate into the stellar SEDs produced by the PHZ PF. Furthermore, the reference stars were drawn from the fields specifically chosen for PSF modelling, which, while distributed across the survey footprint, are small, and sample only specific values of Milky Way reddening. We may therefore be susceptible to reddening-dependent biases, or to differences between the stellar population represented in the reference sample and the main survey area.

\begin{figure}[btp!]
\centering
\includegraphics[angle=0,width=1.0\hsize]{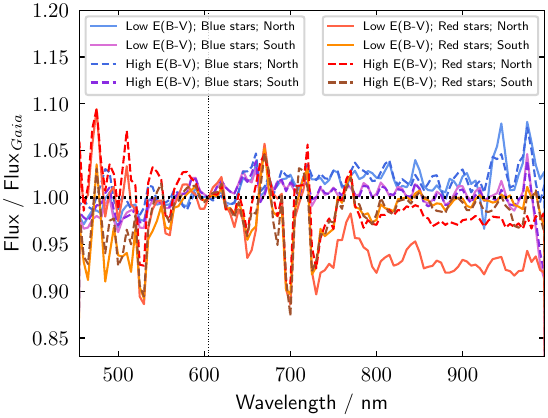}
\caption{Ratio of the average recovered SED from the PHZ pipeline to the average \textit{Gaia} continuous spectrum, for different subsamples in stellar colour and Galactic reddening. The vertical dotted line indicates the wavelength where each individual SED was normalised before averaging.}
\label{fig:star_sed_validation}
\end{figure}

To test our performance in recovering stellar SEDs, we use the \textit{Gaia} XP continuous spectra for stars within two broad regions of the north and south regions of the DR1 footprint, specifically chosen to include high and low Galactic reddening regions. Four subsamples are considered in each hemisphere, of low reddening, $0.02<E(B-V)<0.03$, and high reddening, $0.075<E(B-V)<0.11$, and of blue stars, $0.4<G - G_{\rm RP} <0.5$ and red stars, $1<G - G_{{\rm RP}} < 1.25$. The ranges were chosen to be narrow in order to try to create consistent samples, but also wide enough to allow for a sufficient number of stars for the measurements to be meaningful. Within each subsample, the SEDs from \textit{Gaia} and from the PHZ output product were normalised to unit flux at $605\,{\rm nm}$ (an arbitrarily chosen wavelength, free of significant photospheric absorption lines) and averaged. The resulting ratios of the averaged SEDs are shown in Fig.\,\ref{fig:star_sed_validation}.  

\begin{table}[]
\centering
\caption{Validation of the \gls{sed}s of stars of different colours and locations in the \gls{EWS}, expressed through the fractional bias in mean wavelength of the average \gls{sed} with respect to {\em Gaia}.}\label{tab:star_SED}
\begin{tabular}{l@{\hspace{3pt}}c@{\hspace{6pt}}cc}
\hline\hline\noalign{\vskip 2pt}
Sample  &  Hemisphere & No.\ stars & $\delta\lambda / \lambda_{{\rm true}}$\\
\noalign{\vskip 1pt}
\hline
\noalign{\vskip 2pt}
Low $E(B-V)$; Blue & N & \phantom{0}681 & $\phantom{+}8.02\times10^{-4}$ \\
Low $E(B-V)$; Blue & S & 2123 & $-1.51\times10^{-4}$ \\
High $E(B-V)$; Blue & N & \phantom{0}181 & $\phantom{+}8.03\times10^{-4}$ \\
High $E(B-V)$; Blue & S & \phantom{0}946 & $-2.71\times10^{-4}$ \\
Low $E(B-V)$; Red & N & \phantom{0}\phantom{0}85 & $-2.37\times10^{-3}$ \\
Low $E(B-V)$; Red & S & \phantom{0}180 & $\phantom{+}5.66\times10^{-4}$ \\
High $E(B-V)$; Red & N & \phantom{0}127 & $-8.21\times10^{-4}$ \\
High $E(B-V)$; Red & S & \phantom{0}\phantom{0}90 & $\phantom{+}3.09\times10^{-4}$ \\
\hline
\end{tabular}
\end{table}

The systematic uncertainty budget for the PSF model is broken down in \cite{Cropper13}, with $3.5\times10^{-4}$ allocated to any size bias related to the wavelength dependence of the PSF (expressed in terms of the size squared, $R^2$). This budget encompasses contributions from both the stellar SEDs used for the PSF model construction and galaxy SEDs for its application to each object. We take a simplistic approach and consider an error budget split equally into each of these constituent contributions. As described in \cite{EH18}, the PSF size-squared scales approximately as the square root of wavelength, and we can therefore relate the allocated error budget to an effective requirement on the recovery of the mean wavelength of the photons captured by the VIS camera for any given object \citep{EU-Tarsitano}. Propagating a $\lambda^{0.5}$ dependence leads to a requirement of $\delta\lambda / \lambda < 5\times10^{-4}$. In Table \ref{tab:star_SED} we list the values of these fractional mean wavelength biases for the eight subsamples of \textit{Gaia} stars considered. In general, the stellar SEDs in the South perform rather well, meeting or only slightly exceeding the \Euclid requirement. Meanwhile, the SED recovery in the north is substantially poorer at roughly twice the requirement or greater. This situation should improve in DR2, with improved processing of the external data and a richer reference data set of northern fields.

\subsubsection{\label{sc:perf_galaxyseds}Galaxy SEDs}

\begin{figure}[btp!]
\centering
\includegraphics[angle=0,width=1.0\hsize]{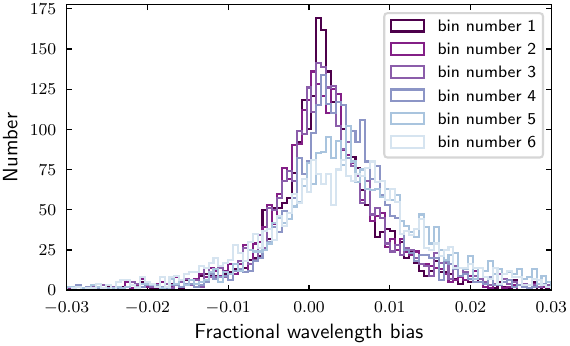}
\caption{Histograms of the fractional difference in SED-weighted \IE mean wavelength, between the PHZ pipeline output galaxy SEDs for a \gls{EWS}-like pass of the EDF-F and a linear interpolation between observed medium-band fluxes from C10. Six equi-populated bins in redshift are shown, covering the range $0.2<z_{\rm phot}<2.5$.}
\label{fig:gal_sed_validation}
\end{figure}

To test the galaxy SEDs we use the medium-band fluxes contained in the MUSYC catalogue produced by \cite{Cardamone10} in the CDFS field. This filter set comprises 18 bands with mean width of approximately $30\,{\rm nm}$, of which 12 are also present in the COSMOS2020 catalogue that was used to build the galaxy SED reference sample. Because the reddest medium-band filter has mean wavelength $856.3\,{\rm nm}$, we extend the wavelength range to cover the whole \IE passband by including the $z^{\prime}$- and $J$-band fluxes from the same catalogue. While many of the photometric bands used to produce the reference SEDs are in common with this CDFS-based test set, the data reduction and calibration differ, and any metric of performance will also be impacted by sample variance from the small CDFS area. Nevertheless, since these are the only observational data available we proceed to compute fractional wavelength differences, similar to those of the stellar SEDs above.

The MUSYC catalogue is cross-matched by sky position to the same \gls{EWS}-like pass of the EDF-F that was used for the photo-$z$ validation. Following the match, point sources are identified using a compactness criterion, \texttt{MUMAX\_MINUS\_MAG}$\,<-2.6$, and photometric zero-point corrections are derived by fitting a set of stellar SEDs and computing median differences between the catalogue and predicted SED fluxes. Using stars to derive the necessary corrections avoids the risk of tuning the zero points to match external galaxy templates, which might produce somewhat optimistic results.

The test sample is prepared in a similar manner to the spectroscopic samples used for the photo-$z$ validation. Namely, point sources, near-IR-detected objects, objects falling in masked regions, and galaxies fainter than $\IE=24.5$ are removed. The SEDs of the test sample galaxies are produced by a linear interpolation of the zero-point-corrected fluxes between the medium-band filter mean wavelengths, and the red extension from the broad bands. This interpolation between bands is an approximation to the true \glspl{sed} that contributes a systematic uncertainty to the present analysis, which itself is of a similar magnitude to the requirement \citep{EU-Tarsitano}.

Six equally populated redshift bins are constructed in the range $0.2<z<2.5$, mimicking the process that will be followed for the DR1 cosmology analysis. Histograms of the per-object fractional wavelength differences between the pipeline SED estimates and the MUSYC-derived SEDs are shown in Fig.\,\ref{fig:gal_sed_validation}. Not shown in the figure is a fraction of outliers ($\sim7\%$), which are located almost exclusively on the edges of the area covered by the \cite{Cardamone10} catalogue. The histograms show a clear redshift dependence to the peak of the distributions, becoming increasingly poor in higher redshift bins. The $3\,\sigma$-clipped mean bias values similarly show poorer performance in the higher redshift bins, with values $2.02\times10^{-3}, 1.73\times10^{-3}, 2.21\times10^{-3}, 3.53\times10^{-3}, 4.99\times10^{-3}$, and $3.58\times10^{-3}$ for bins 1--6 respectively. While these values are approaching an order of magnitude short of the end-of-mission requirement of $5\times10^{-4}$, they are not entirely unexpected, given the results of \cite{EU-Tarsitano} and the limited data we have at hand to perform this analysis. Further data collection to improve both the measurement and validation of the galaxy \glspl{sed} in underway.

\subsection{\label{sc:perf_classification}Classification performance}

\begin{figure}[htbp!]
\centering
\hspace*{-0.5cm}\includegraphics[angle=0,width=1.06\hsize]{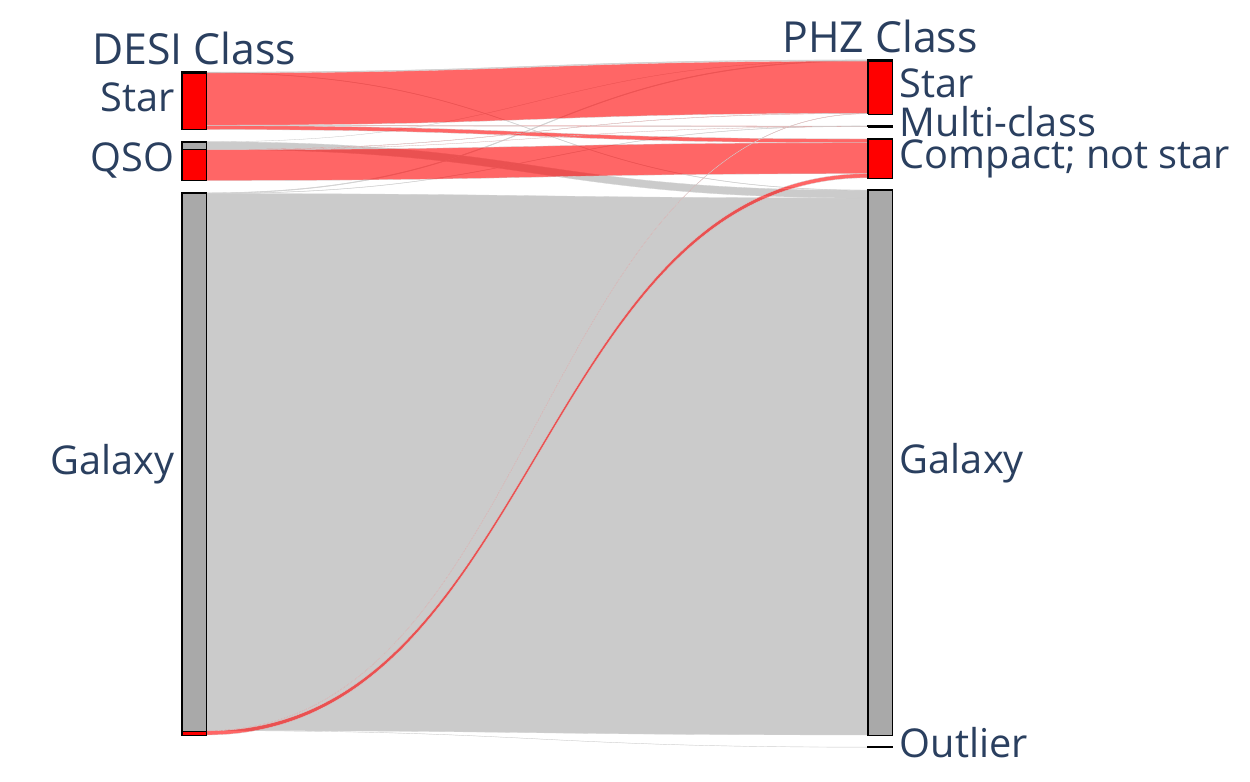}
\caption{Correspondence between spectroscopically determined object types from DESI and the PHZ pipeline classification, from a \gls{EWS}-like pass of the EDF-N, where the fraction of objects is shown by the thickness of the connecting lines. The red colour represents objects that are deemed compact by a cut on the MER catalogue column, $\texttt{MUMAX\_MINUS\_MAG}<-2.6$.}
\label{fig:class_validation}
\end{figure}

As described in Sect.\,\ref{sc:pipe_classification}, the DR1 star and galaxy classifiers are largely unchanged from the Q1 release, and the detailed characterisation of these two classifiers presented in \citetalias{Q1-TP005} is therefore applicable also to the DR1 products. The major change in the classification scheme for DR1 is the removal of the QSO classifier, with the decision point for objects to enter the QSO catalogue based instead on a combination of the probability to be a star from the photometric classifier and reported as unresolved in the MER catalogue (see Sect.\,\ref{sc:pipe_classification}). The result of this new strategy is a largely complete, but heavily stellar-contaminated base sample of QSOs from which the user can further down select to form their science sample; for example, by applying a tighter star probability criterion, colour cuts, or multiwavelength information. In the remainder of this section we show the correspondence between spectroscopic class labels from DESI and the output classes using the default probability acceptance thresholds.   

The EDF-N was observed intensively by DESI during the programme's science verification period and was one of the densely sampled fields in the one-percent data release \citep{DESIEDR}. After cross-matching this release to a \Euclid single-visit EDF-N catalogue we have $\num{29436}$ galaxies, 3121 stars, and 2103 QSOs as labelled by the DESI \texttt{redrock} pipeline \citep{redrock}, which is sufficient for the present purpose. In Fig.\,\ref{fig:class_validation} we show a Sankey diagram of the class correspondence between the DESI labels and the output classes (contained in the \texttt{PHZ\_CLASSIFICATION} column, or trivially derived from the \Euclid catalogues in the case of QSOs). The red colour in the figure represents objects that satisfy the compactness criterion, \texttt{MUMAX\_MINUS\_MAG}$\,<-2.6$.

Of the DESI stars, 2894 ($93\%$) pass the strict threshold for stars to be used in PSF modelling, with 211 of the remainder ending up in the compact and `not star' class. The final 16 are included among the 254 stars that pass the deliberately permissive probability threshold to be classified as galaxies. All but 49 ($0.2\%$) of the DESI galaxies are classified as galaxies by the pipeline, with 58 accepted by the star classifier and three unclassified. These numbers might suggest that the star class purity is falling short of the requirement ($\sim98\%$ purity for DESI objects, versus $99\%$ requirement). However, for the 46 DESI galaxies classified exclusively as stars by the pipeline, their morphologies and colours suggest that most are likely misclassified by the DESI spectroscopic pipeline. Finally, for the DESI QSO sample, 1678 of the 1689 ($99\%$) that pass the compactness criterion are classified as not being stars and hence form the basis for a QSO candidate sample. Meanwhile, $20\%$ of the DESI QSOs are not compact in the \Euclid imaging, and therefore do not pass the definition for QSOs in the Euclid DR1 catalogues. All such objects are accepted by the galaxy classifier, which is expected based on the emphasis for completeness in the galaxy class.

The numbers quoted here are presented to aid a user of the PHZ catalogues to build some intuition of how the classification behaves; they should not be read as formal completeness or purity estimates, since the DESI samples are in general much brighter than the full flux-limited populations of each class. 

\section{\label{sc:issues}Known issues and variant production runs}
Our validation and post-production analyses, including those of science teams within the Euclid Consortium, have revealed a few issues with the data products that are worth noting here. In addition, two supplementary runs of the main pipeline for \gls{EWS} data were prepared and run as part of cosmology pipeline tests and mitigation of issues with some of the input data products.

\subsection{\label{sc:nnpz_issues}Known issues}

As described in Sect. \ref{sc:ref_cosmos}, the relative abundance of different galaxy types, redshifts, and luminosities in the reference sample acts as a prior in the photo-$z$ determination. Having only the COSMOS field available and suitable for building the reference samples creates severe challenges, in particular in producing good measurements for the brightest galaxies. Firstly, due to the small size of the COSMOS field and its choice to avoid low-redshift structures, there are extremely few very low-redshift galaxies. Moreover, to preserve the positive impact of the effective magnitude prior that is implicit in the reference sample, the allowed range in SED amplitude difference between \gls{EWS} and reference objects was restricted to $\pm1.5\,\mathrm{mag}$. The result of these two factors is that the reference sample is under-represented by rare galaxies, including low-redshift objects ($z<0.2$), very high mass galaxies at moderate redshifts ($z<0.4$) and some of the most extreme galaxies in the Universe. These galaxy classes are therefore strongly disfavoured by \nnpz{}. Our strategy of augmentation and imposing an explicit redshift prior to overcome these limitations was unfortunately not as effective as we would have liked. In the majority of cases, the redshifts are still quite reasonable, but the stellar masses and other amplitude-dependent physical properties are biased low.

Secondly, the redshift point estimates tend to cluster around specific values. Such behaviour is quite typical in template-based photo-$z$ estimates, and usually referred to as `redshift focussing' or aliasing. Machine-learning methods tend to be less susceptible to redshift focussing, particularly in cases where the photometric noise is matched between the training and target samples (although see \citealt{Bonnett16} for a counter-example). On the other hand, machine learning methods are vulnerable to sample variance in the training data, or sampling noise if the training data are not sufficiently numerous to smoothly represent the feature space. In the case of \nnpz{} we use deep photometry for the training data and so noise aliasing is certainly a possibility. However, some of the favoured and disfavoured redshift values correspond to known over- and under-densities in the COSMOS field (e.g., the strong structure at $z\simeq0.35$).

\subsection{\label{sc:nounoi}Variant production runs}
We performed two variant runs of the main production pipeline for the \gls{EWS}, both for the north tiles only. Firstly, the EXT processing was only able to use $u$-band data across half of the northern sky footprint. As the photo-$z$ characteristics clearly depend on the combination of bands available (see Sect.\,\ref{sc:perf_photoz}) we needed to be able to provide a more homogeneous northern data set to supplement the main run. This aim was achieved through excluding the $u$-band photometry. These products themselves are not published as part of DR1-Foundation. However, the OU-Level 3 (LE3) products that require them will publish the necessary information where they are used.

Secondly, as described in the paper detailing the external data \citep{DR1-TP006}, the individual images were combined into the coadded mosaics with flags identifying known bad pixels but with no further outlier rejection. The resulting Pan-STARRS $i$-band images therefore carried a high rate of residual artefacts, which would have required outlier rejection in order to be masked. Following the formal validation of DR1-Foundation, a \texttt{Healpix} map of the affected regions was produced \citep[see][]{DR1-TP028}. Nevertheless, it posed a particular difficulty for the task of optical cluster detection, and so again we produced an additional set of products, using neither the $u$ band nor the $i$ band from the northern external data. As in the case of the no $u$-band variant, derived values from this supplemental run will be made available in the OU-LE3 products where relevant. 

\section{\label{sc:summary}Summary}
We have presented the processing function for photometric redshifts for the Euclid DR1-Foundation release and the validation and performance of the cosmology-oriented tile-by-tile output products. A description of the derivation of galaxy physical properties, such as stellar masses and star-formation rates, and associated output products can be found in \citetalias{DR1-TP012}. Meanwhile, calibrated sample-level redshift distributions for cosmological inference will be presented in a future publication at the time of the full DR1 release.

With respect to the Q1 release, the object-by-object \gls{zpdf}s are now produced with our nearest-neighbour package, \texttt{NNPZ}, in line with the pre-launch strategy outlined in \cite{EuclidSkyOverview}. The reference sample of redshifts for \texttt{NNPZ} was built from data in the COSMOS field, using an association between the \Euclid catalogue and the classic photometric catalogue from \cite{Weaver22}. From the latter we derive photometric redshifts, paying particular attention to the calibration of the $z$PDF. The performance of the photo-$z$ are much improved compared with Q1, as expected given that we are now able to implement our full pipeline vision. Nevertheless, the outlier fraction and scatter of the point estimates do not yet meet the mission design requirements, which is at least in part due to shallower depth in the ground-based photometry with respect to design assumptions. In the southern hemisphere, the eventual switch to data from Rubin-LSST will alleviate this issue by mission end.

Other important outputs from the pipeline for the DR1 cosmology analyses include star and galaxy SEDs for PSF modelling and object classes that are used in sample selections. A formal validation of these aspects cannot be performed due to the lack of suitable truth information to the flux limit of the target samples. Nonetheless, we have presented estimates of the performance based on available data. 

The correspondence between spectroscopic object class labels from DESI and the object classes in the pipeline (including taking a compactness criterion into account from the MER catalogue) is very much in line with expectations. The classification for stars has a deliberately strict threshold to ensure high purity, while the galaxy class is highly complete at the cost of some contamination. 

The SEDs of stars in the southern hemisphere of the DR1 \gls{EWS} outputs recover the continuous spectra from \textit{Gaia} very well. In the north, we identify deviations exceeding the nominal mission requirement, particularly for redder stars. The accuracy of galaxy SEDs falls some way short of the mission target, with the bias in mean effective wavelength increasing with redshift within the six bins defined. Additional data are being taken so that we are able to meet the end-of-mission requirements.

Tile-by-tile validation of the PHZ products reveals that the main reasons for invalidating data at the PHZ PF level are incomplete \Euclid photometric data and tiles with very small areas. Of the $\num{8960}$ MER tiles covering the DR1 \gls{EWS} area, $93.6\%$ are valid across the whole photometric pipeline chain, with approximately $95\%$ of the nominal survey area available for scientific exploitation.

\begin{acknowledgements}
  \AckDRone
  \AckDatalabs
  \AckEC
  This project used public archival data from the Dark Energy Survey (DES). Funding for the DES Projects has been provided by the U.S. Department of Energy, the U.S. National Science Foundation, the Ministry of Science and Education of Spain, the Science and Technology FacilitiesCouncil of the United Kingdom, the Higher Education Funding Council for England, the National Center for Supercomputing Applications at the University of Illinois at Urbana-Champaign, the Kavli Institute of Cosmological Physics at the University of Chicago, the Center for Cosmology and Astro-Particle Physics at the Ohio State University, the Mitchell Institute for Fundamental Physics and Astronomy at Texas A\&M University, Financiadora de Estudos e Projetos, Funda{\c c}{\~a}o Carlos Chagas Filho de Amparo {\`a} Pesquisa do Estado do Rio de Janeiro, Conselho Nacional de Desenvolvimento Cient{\'i}fico e Tecnol{\'o}gico and the Minist{\'e}rio da Ci{\^e}ncia, Tecnologia e Inova{\c c}{\~a}o, the Deutsche Forschungsgemeinschaft, and the Collaborating Institutions in the Dark Energy Survey.

The Collaborating Institutions are Argonne National Laboratory, the University of California at Santa Cruz, the University of Cambridge, Centro de Investigaciones Energ{\'e}ticas, Medioambientales y Tecnol{\'o}gicas-Madrid, the University of Chicago, University College London, the DES-Brazil Consortium, the University of Edinburgh, the Eidgen{\"o}ssische Technische Hochschule (ETH) Z{\"u}rich,  Fermi National Accelerator Laboratory, the University of Illinois at Urbana-Champaign, the Institut de Ci{\`e}ncies de l'Espai (IEEC/CSIC), the Institut de F{\'i}sica d'Altes Energies, Lawrence Berkeley National Laboratory, the Ludwig-Maximilians Universit{\"a}t M{\"u}nchen and the associated Excellence Cluster Universe, the University of Michigan, the National Optical Astronomy Observatory, the University of Nottingham, The Ohio State University, the OzDES Membership Consortium, the University of Pennsylvania, the University of Portsmouth, SLAC National Accelerator Laboratory, Stanford University, the University of Sussex, and Texas A\&M University.

Based in part on observations at Cerro Tololo Inter-American Observatory, National Optical Astronomy Observatory, which is operated by the Association of Universities for Research in Astronomy (AURA) under a cooperative agreement with the National Science Foundation.

This work was enabled by observations made from the CFHT and Subaru telescopes. The scientific community is honored to have the opportunity to conduct astronomical research on Maunakea in Hawai‘i. We recognize and acknowledge the very significant cultural role and reverence of Maunakea to the Kanaka Maoli (Native Hawaiians) community.
\end{acknowledgements}

%
%

\bibliography{Euclid, DR1, phz}
%
%

%
%

\begin{appendix}

\section{\label{sc:app_zeropoints}Photometric zero-point assessment}

It is common in the literature to estimate effective zero-point shifts for each photometric band in a catalogue that, when used during photometric redshift measurement, maximise some metric of quality. These shifts may represent genuine corrections to the photometric zero points of the images from which the fluxes were measured, or they may compensate for systematic errors such as model mis-specification in template-based photo-$z$ methods, wrong filter transmission curves, or even issues in background subtraction. In practice, such zero-point shifts should be treated with caution as correcting for some specific subsample (such as galaxies with spectroscopic redshifts) may leave us blind to systematic errors that are introduced in other subpopulations. In the Q1 release we computed such corrections for each band across the data set, using a combination of stars and galaxies with spectroscopic redshifts from a representative set of tiles \citepalias[see][]{Q1-TP005}. The corrections in this case largely compensated for differences in the PSF normalisations from one band to another that were present in the catalogue photometry \citep{Q1-TP004}. 

\begin{figure}[htbp!]
\centering
\includegraphics[angle=0,width=0.9\hsize]{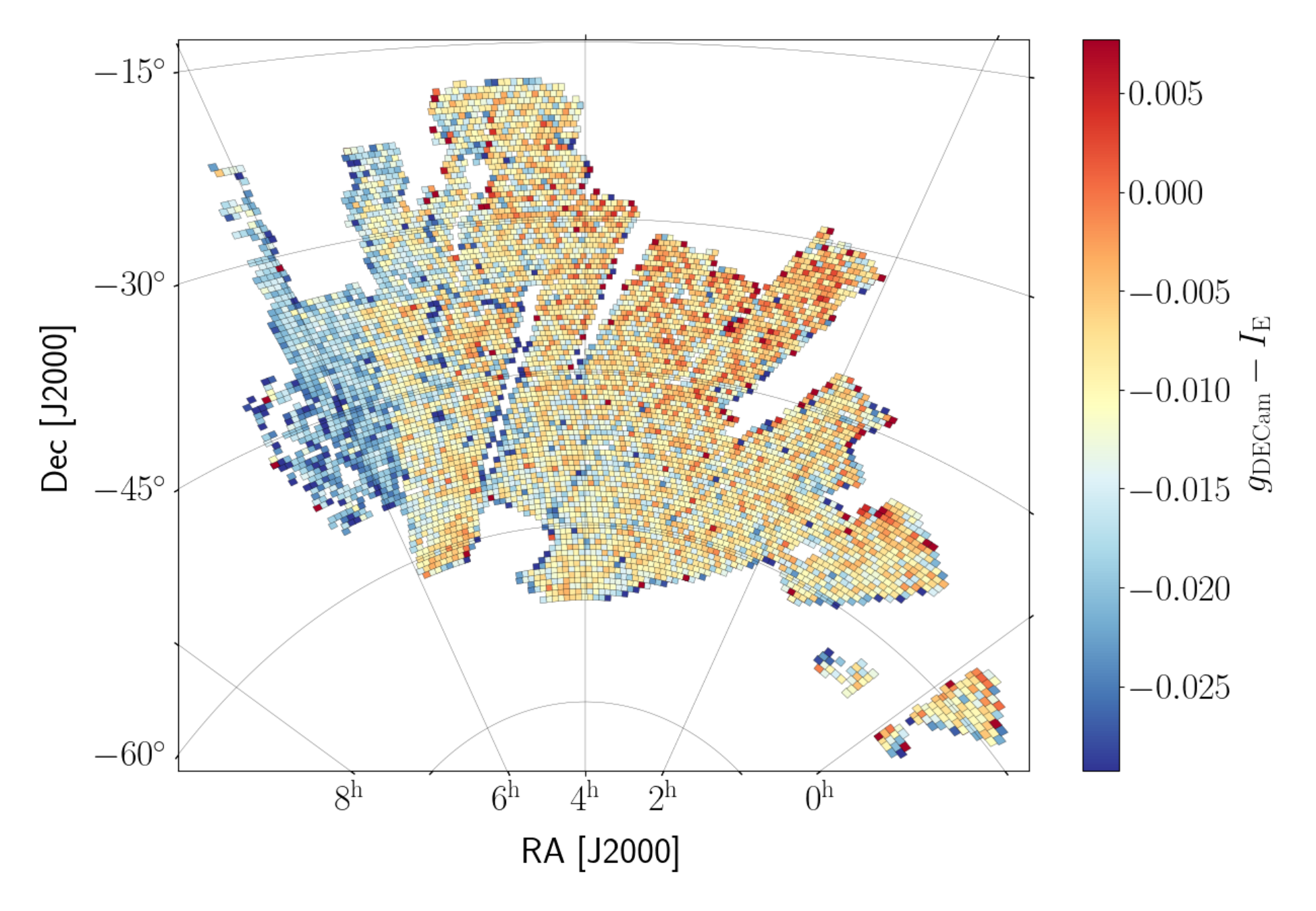}
\includegraphics[angle=0,width=0.9\hsize]{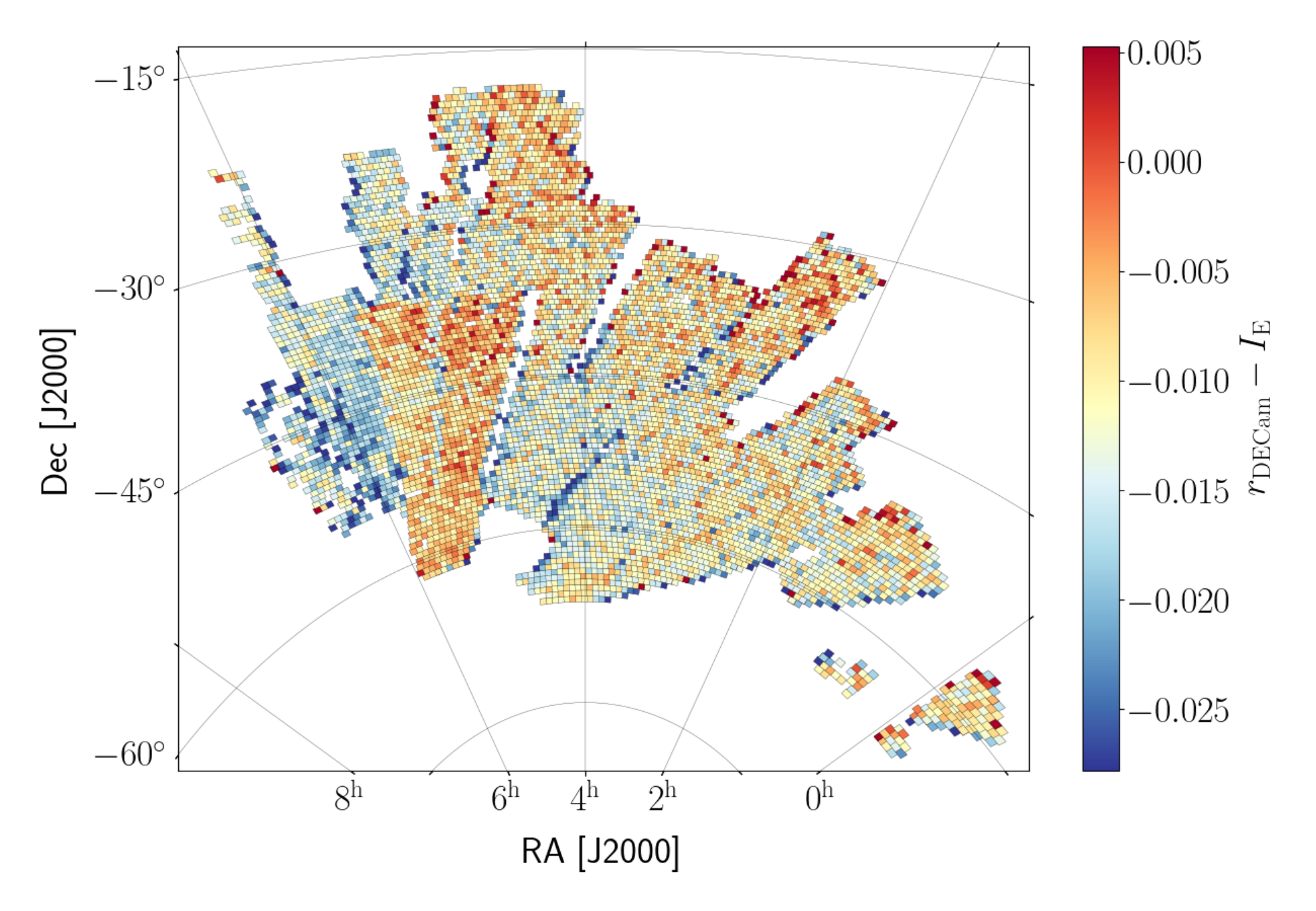}
\includegraphics[angle=0,width=0.9\hsize]{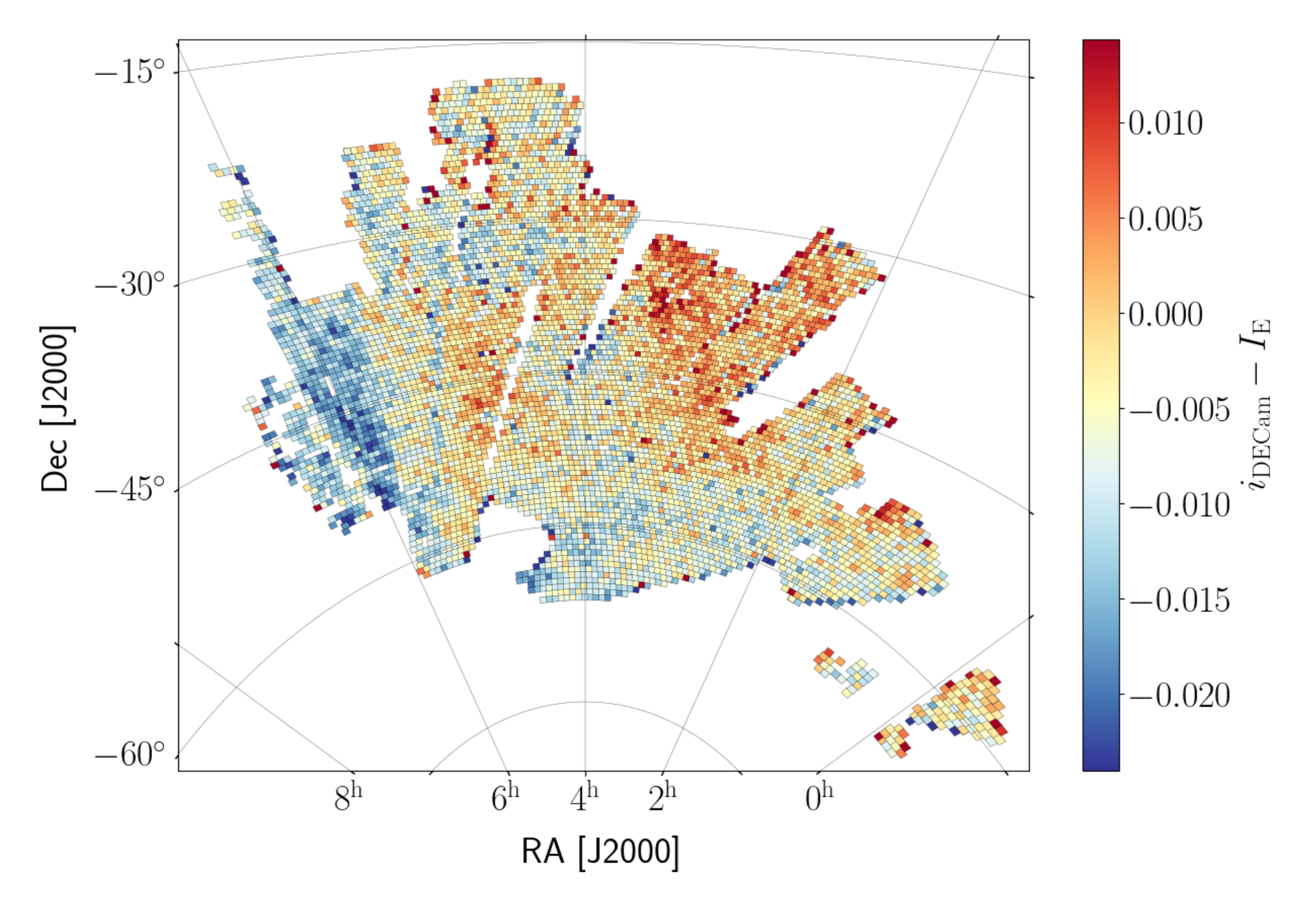}
\includegraphics[angle=0,width=0.9\hsize]{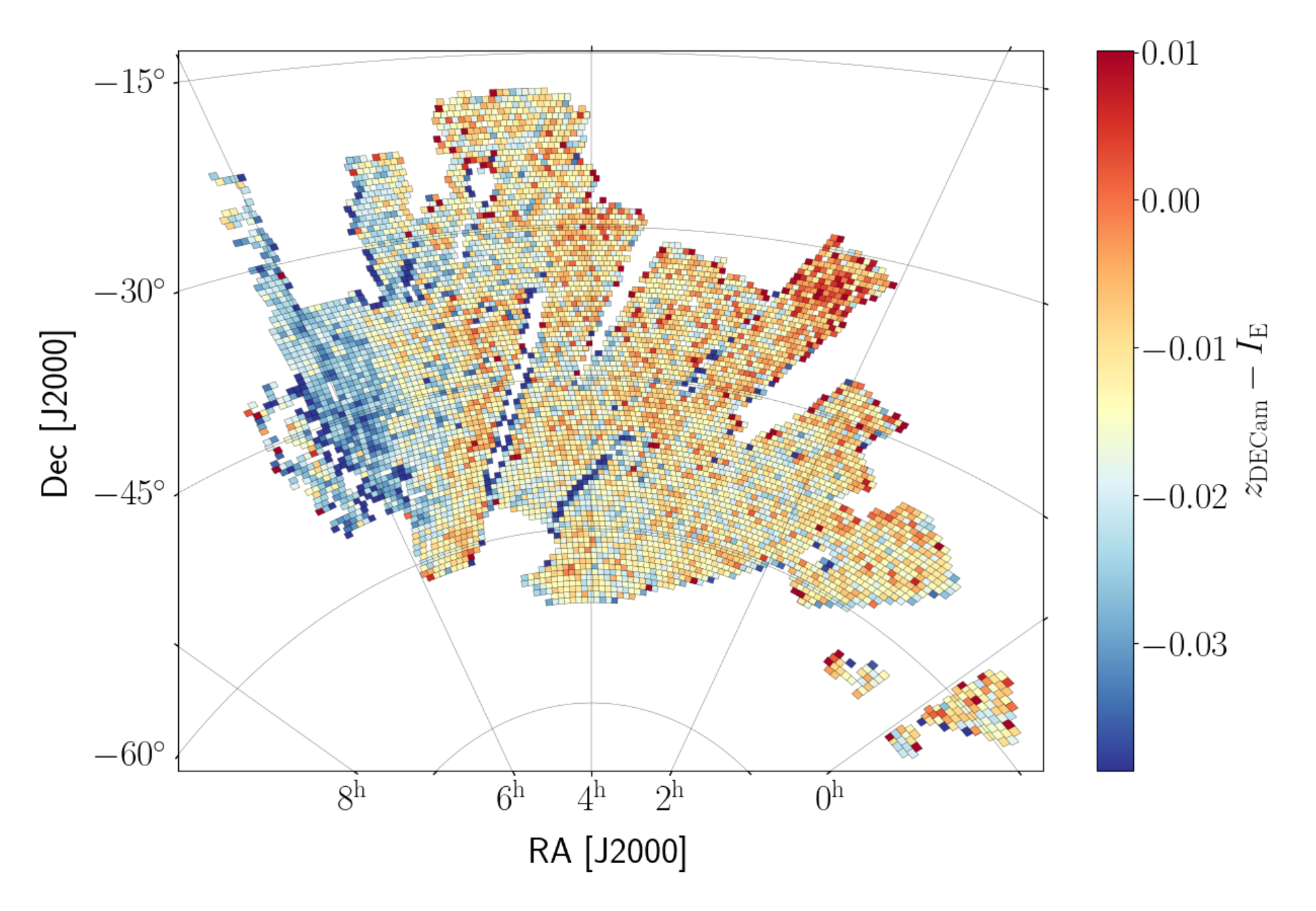}
\caption{Per-tile relative zero-point difference with respect to \IE for the external ground-based photometry, computed using \gls{sed} fits to stellar objects.}
\label{fig:south_zp}
\end{figure}

\begin{figure}[htbp!]
\centering
\includegraphics[angle=0,width=0.9\hsize]{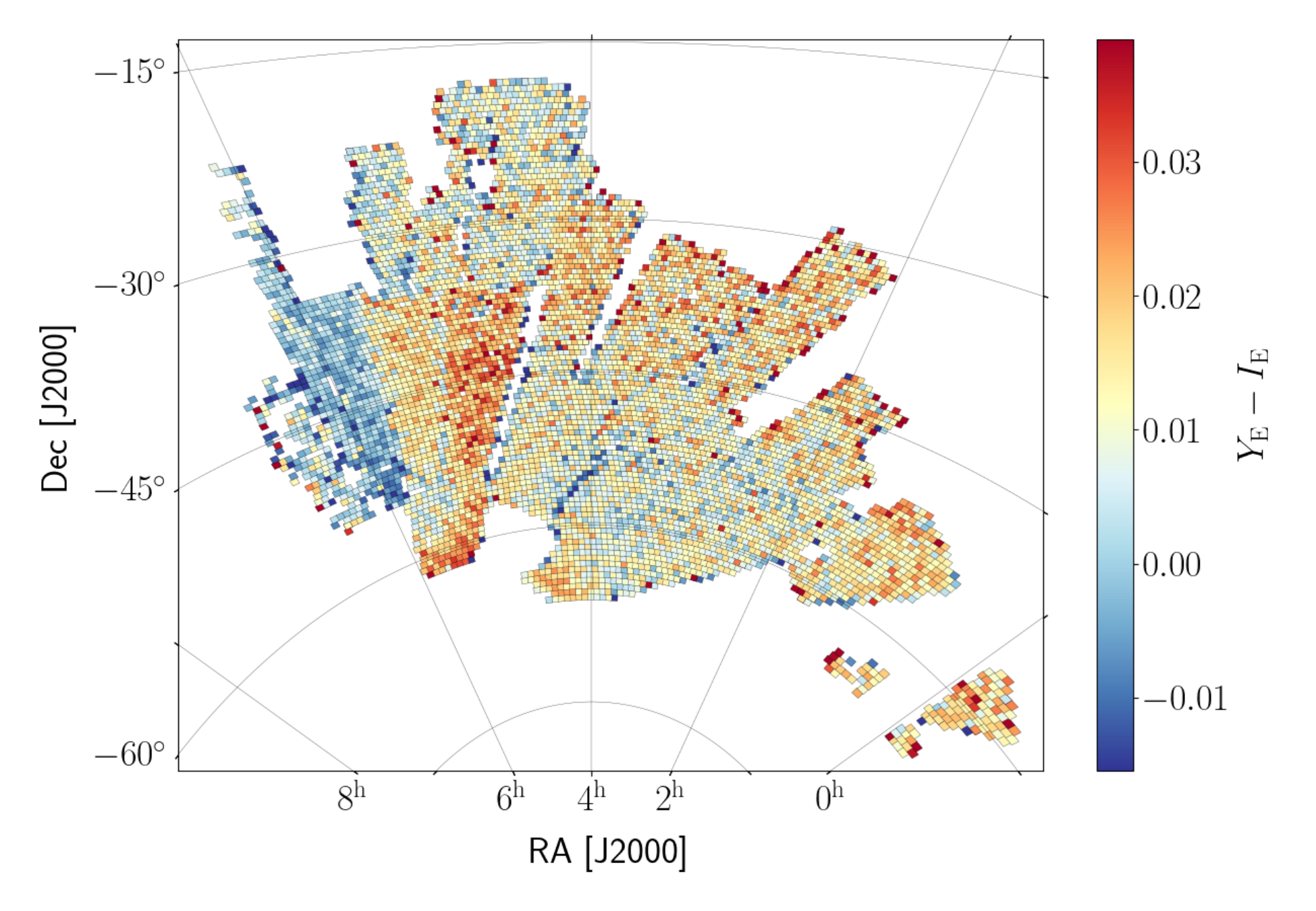}
\includegraphics[angle=0,width=0.9\hsize]{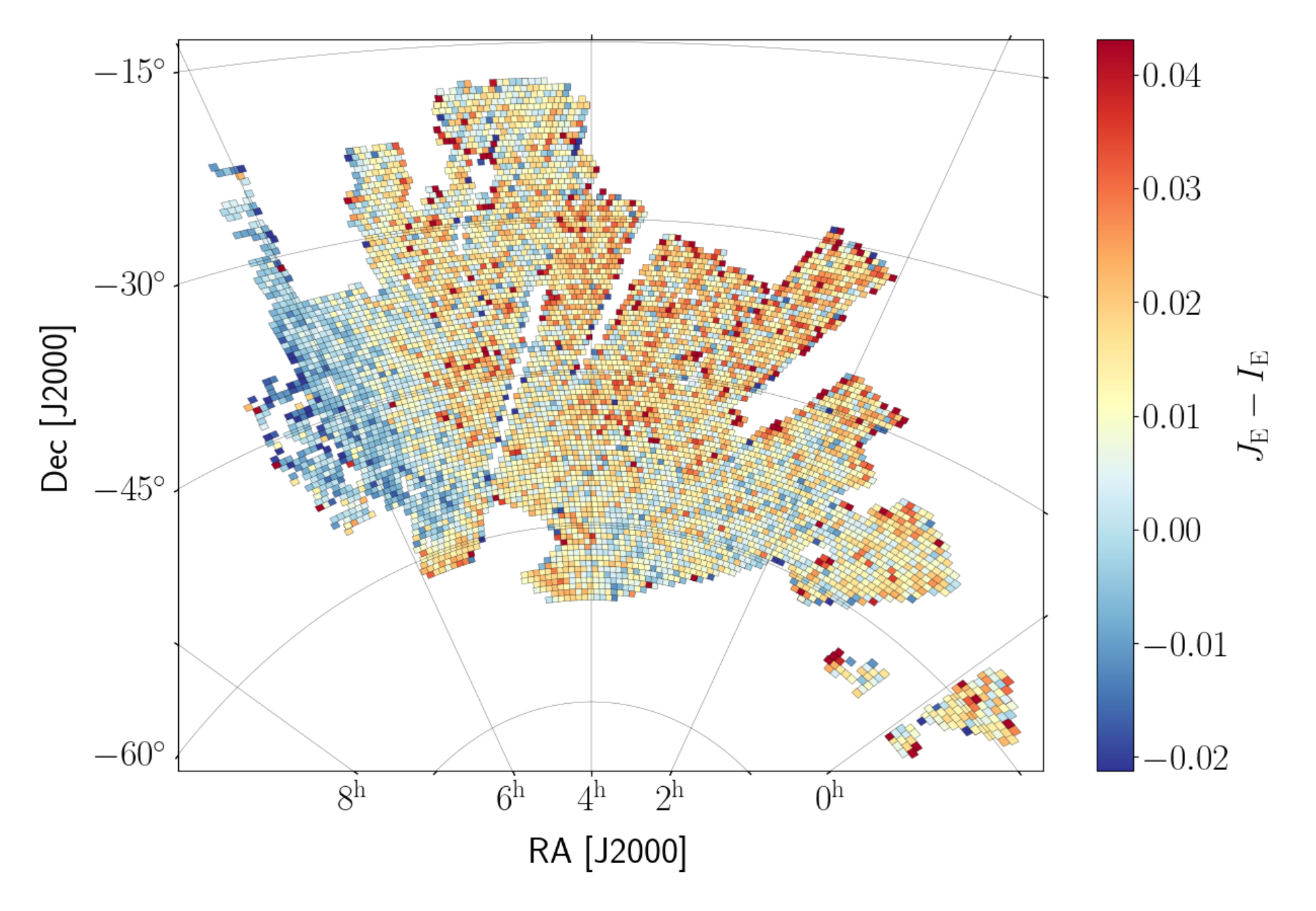}
\includegraphics[angle=0,width=0.9\hsize]{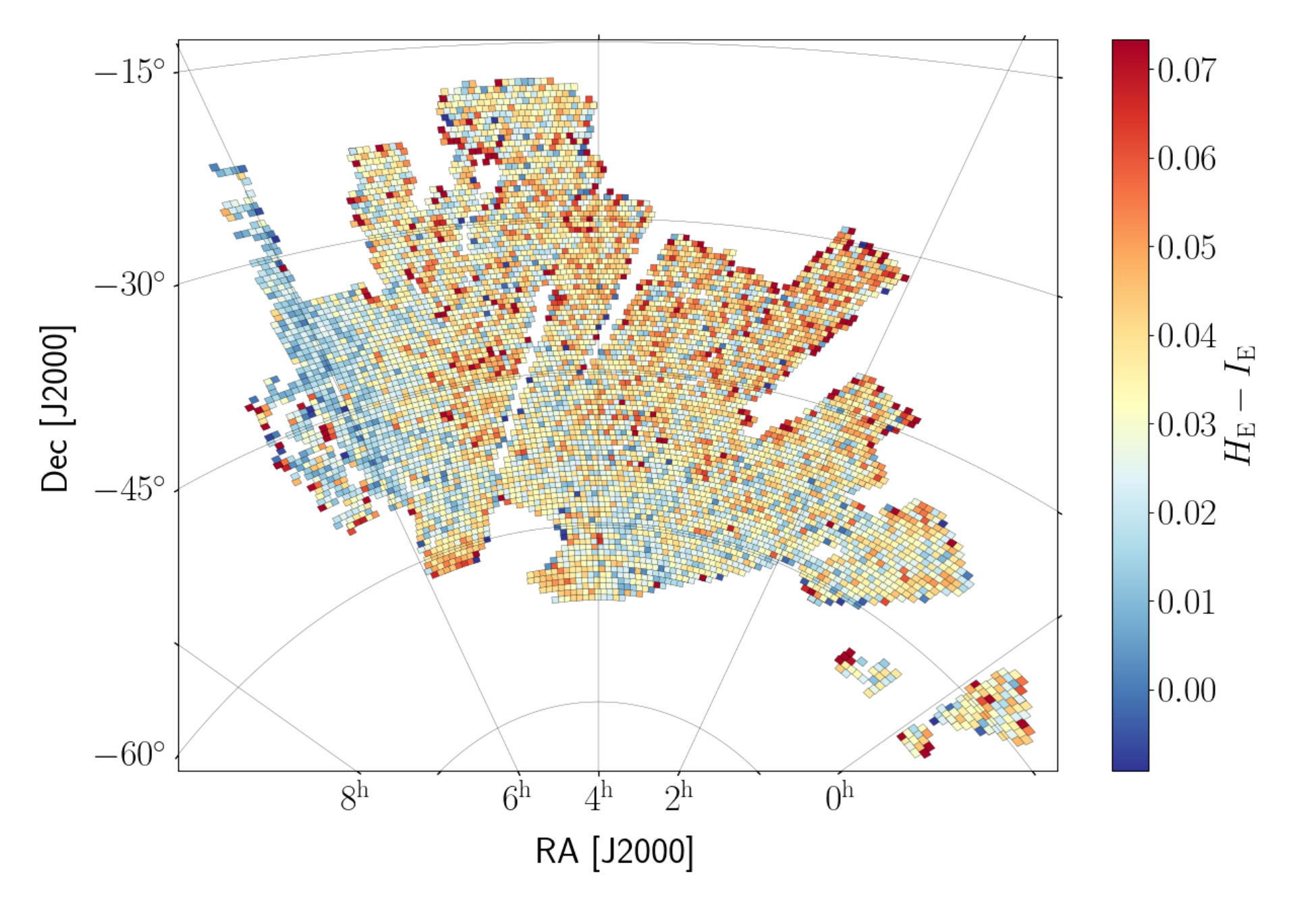}
\caption{Same as Fig.\,\ref{fig:south_zp}, but for the \Euclid NIR bands.}
\label{fig:south_zp_cont}
\end{figure}

\begin{figure}[htbp!]
\centering
\includegraphics[angle=0,width=0.9\hsize]{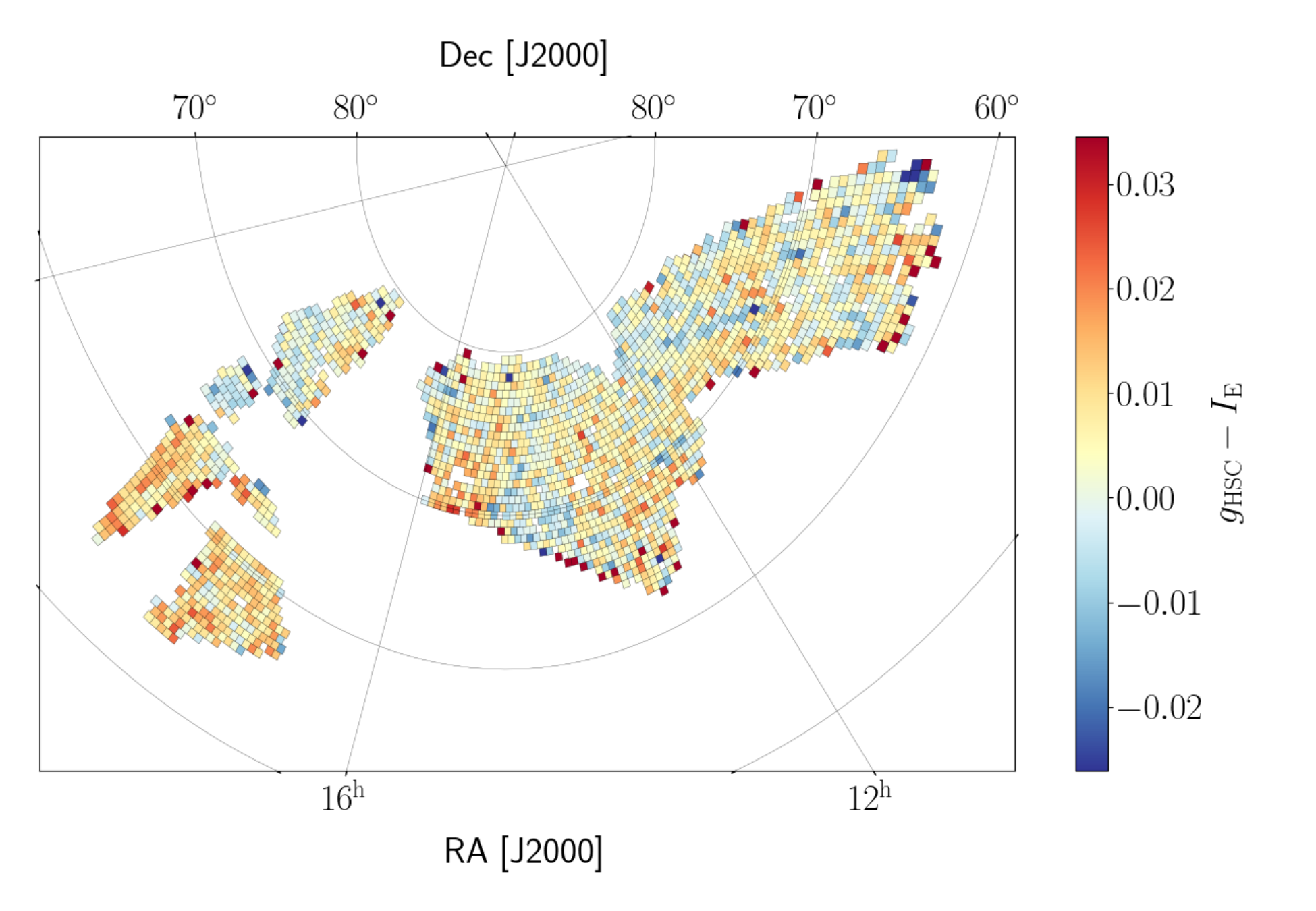}
\includegraphics[angle=0,width=0.9\hsize]{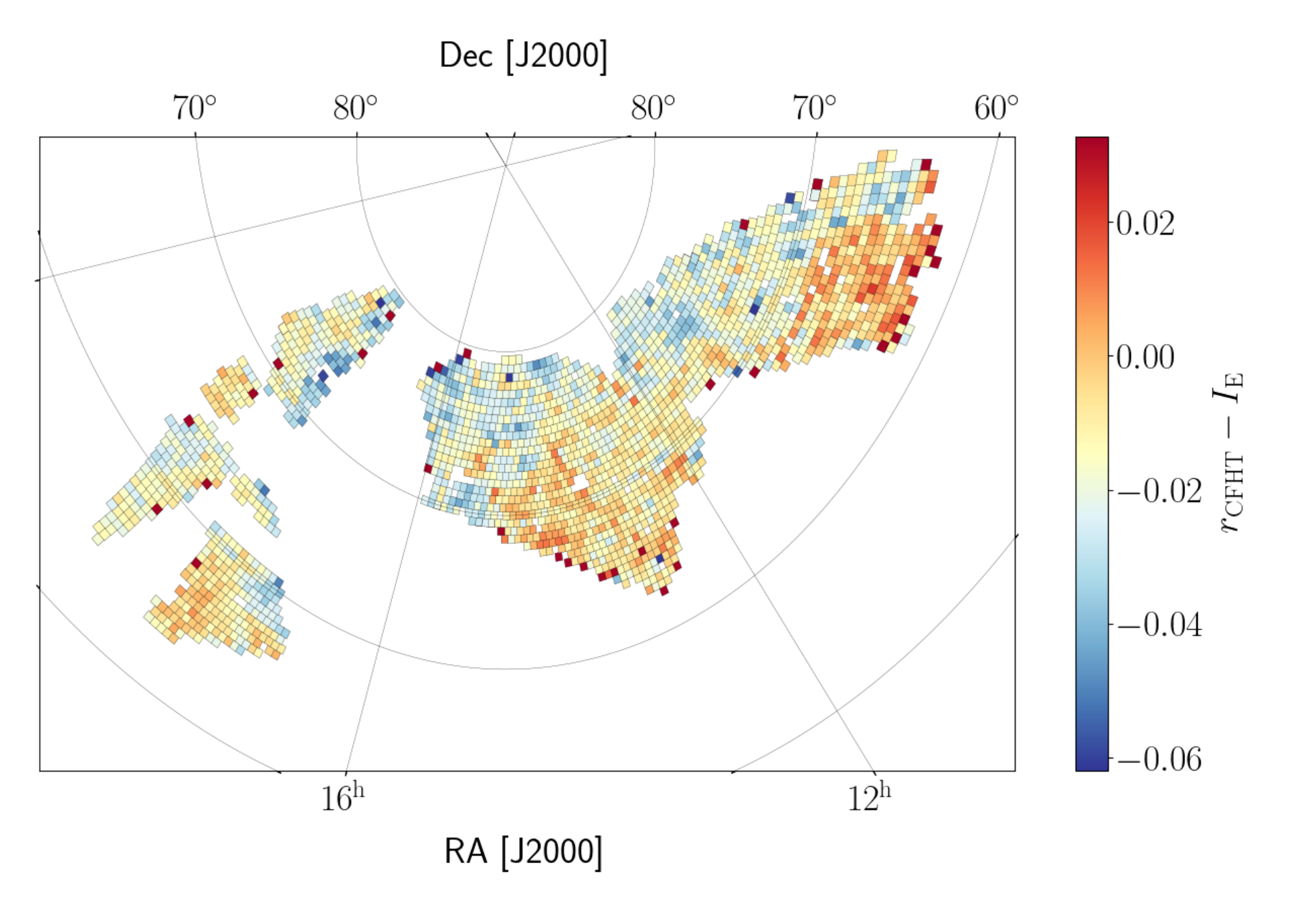}
\includegraphics[angle=0,width=0.9\hsize]{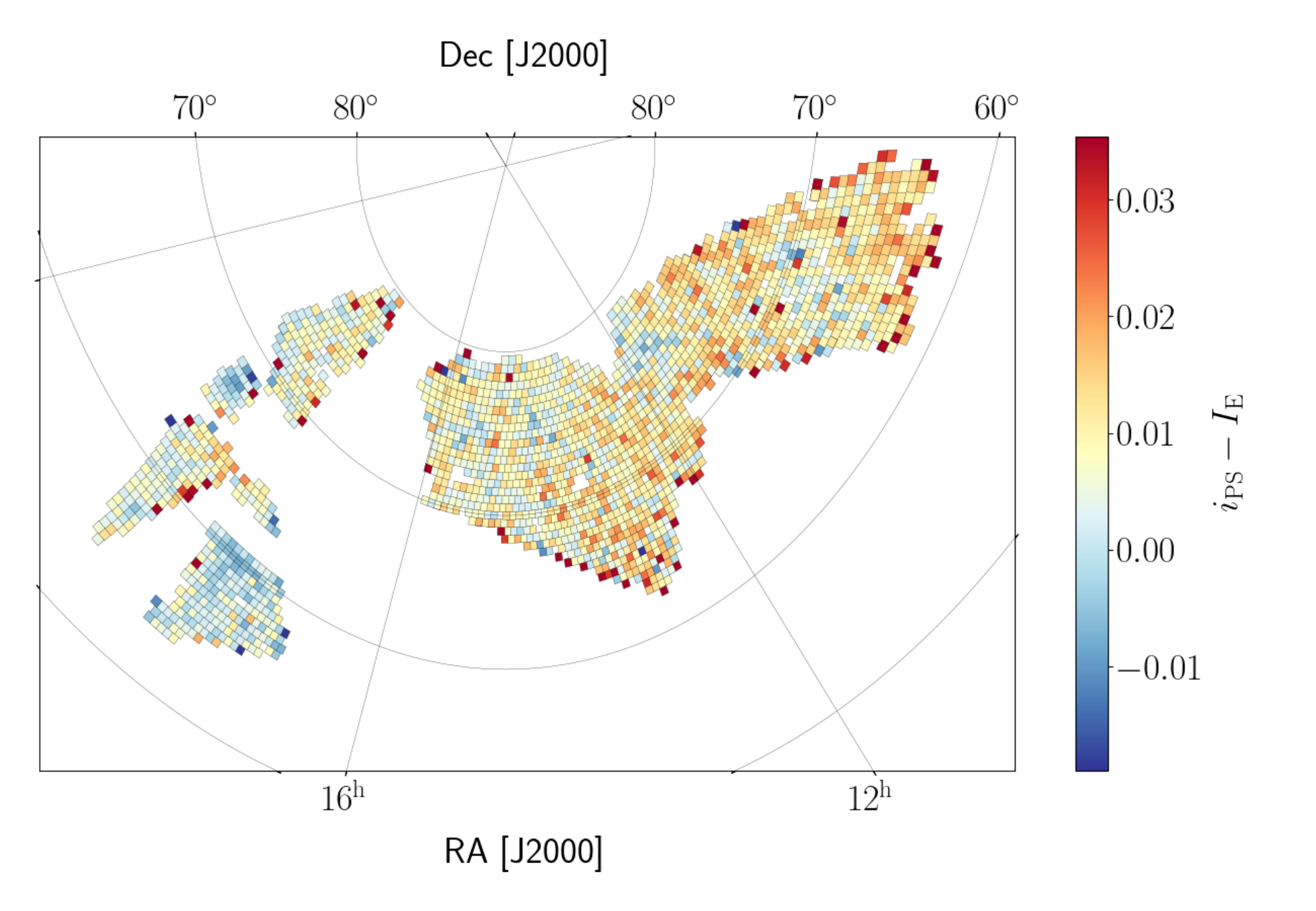}
\includegraphics[angle=0,width=0.9\hsize]{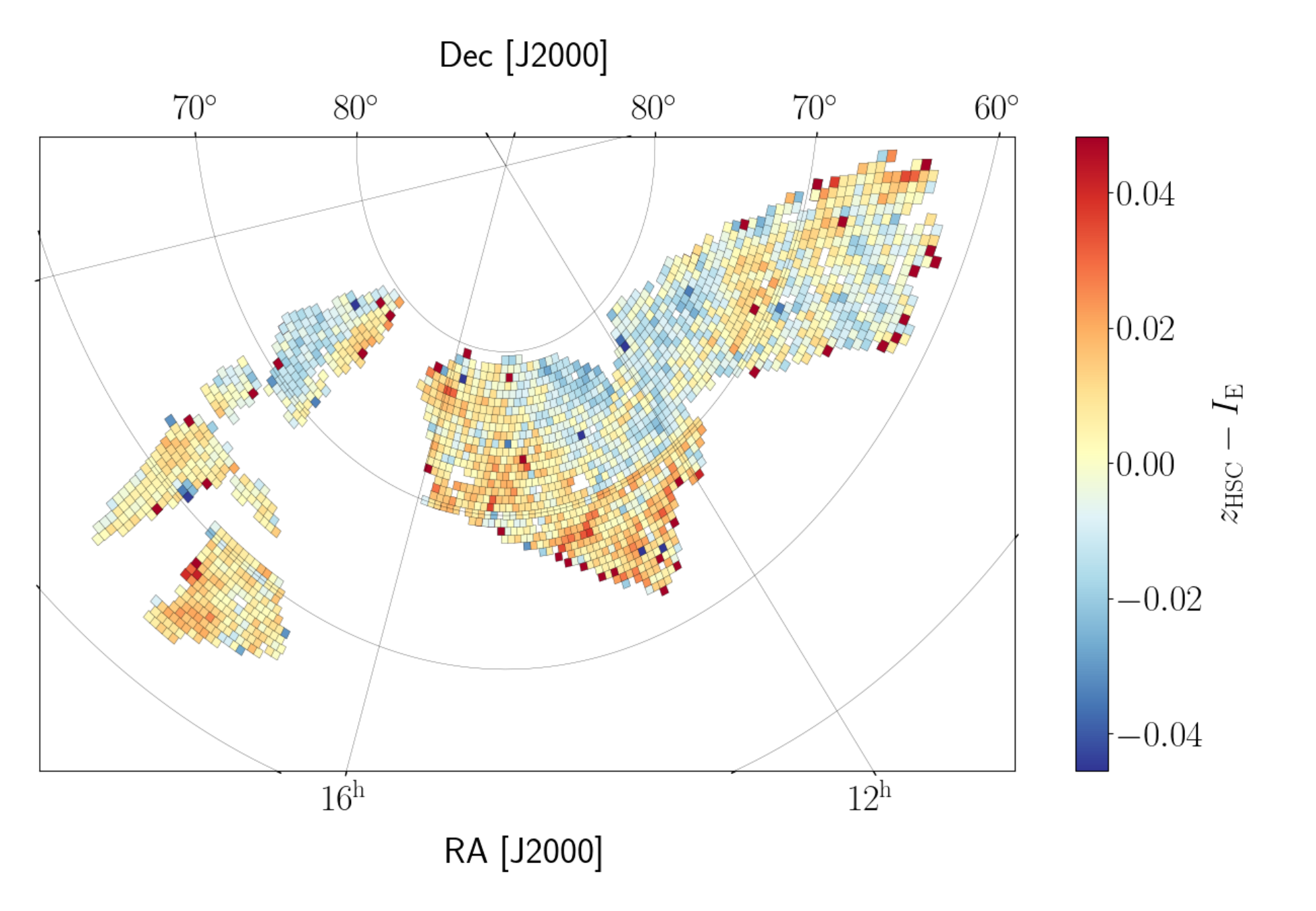}
\caption{Same as Fig.\,\ref{fig:south_zp}, but for the northern part of the \gls{EWS}.}
\label{fig:north_zp}
\end{figure}

\begin{figure}[htbp!]
\centering
\includegraphics[angle=0,width=0.9\hsize]{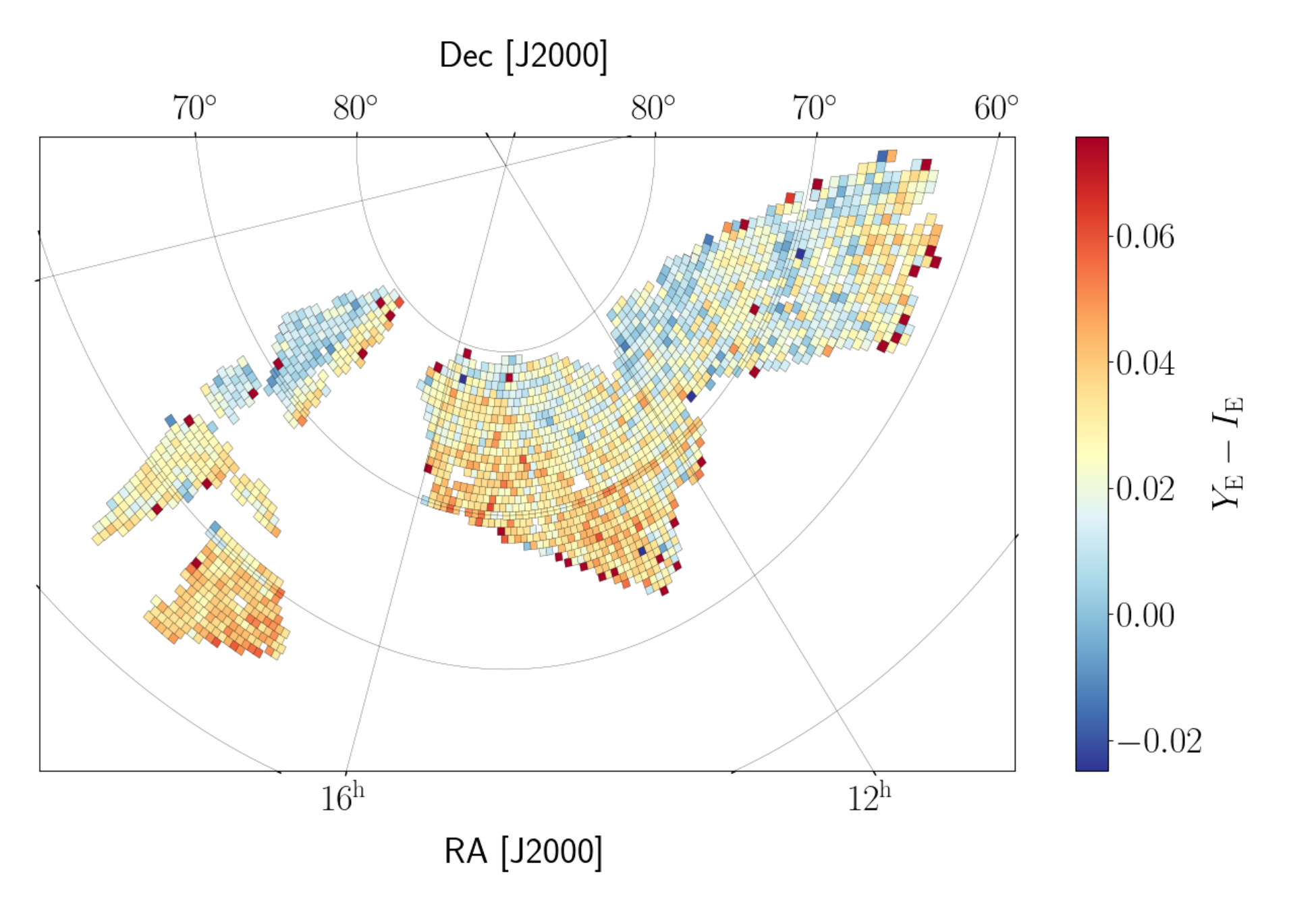}
\includegraphics[angle=0,width=0.9\hsize]{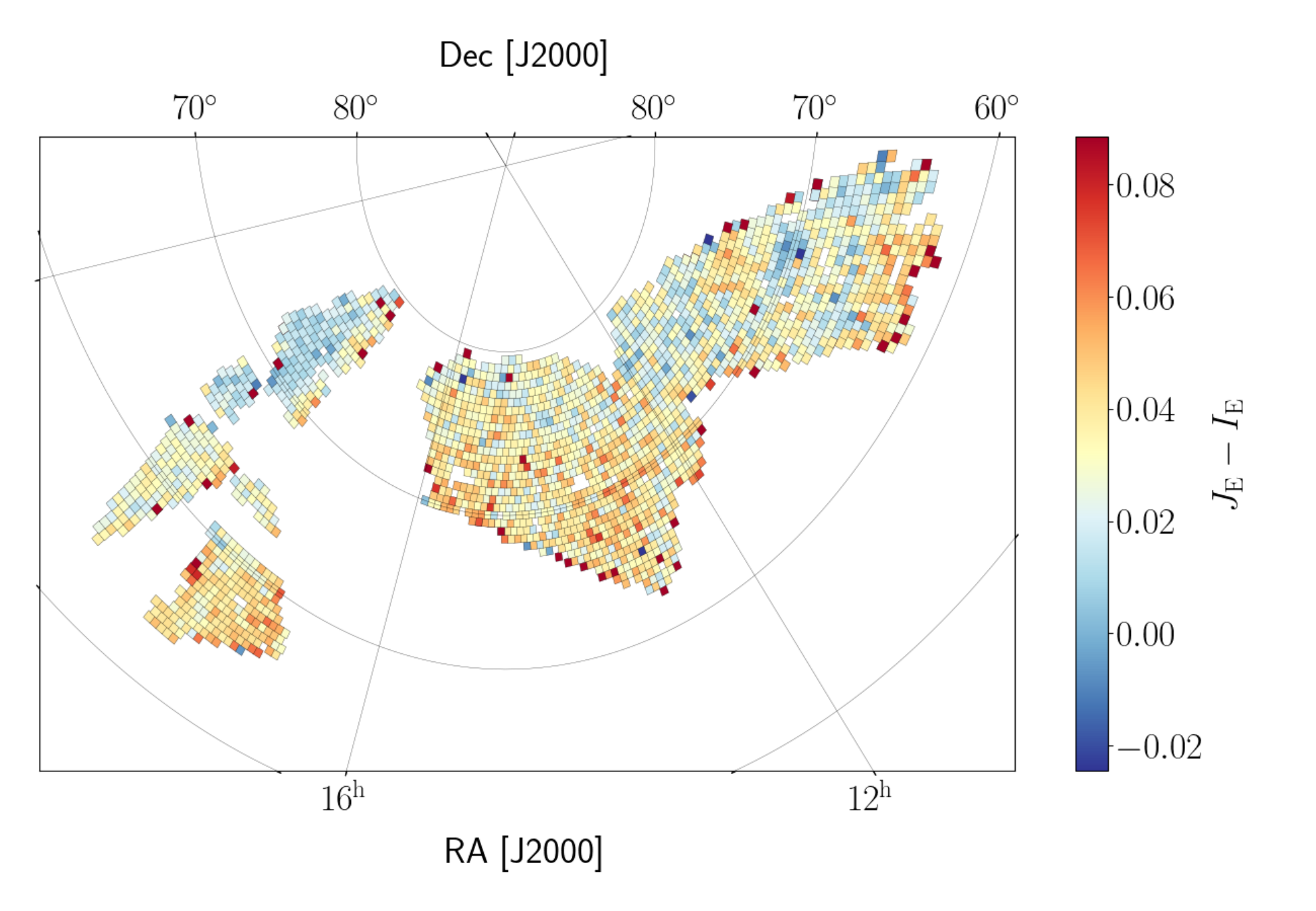}
\includegraphics[angle=0,width=0.9\hsize]{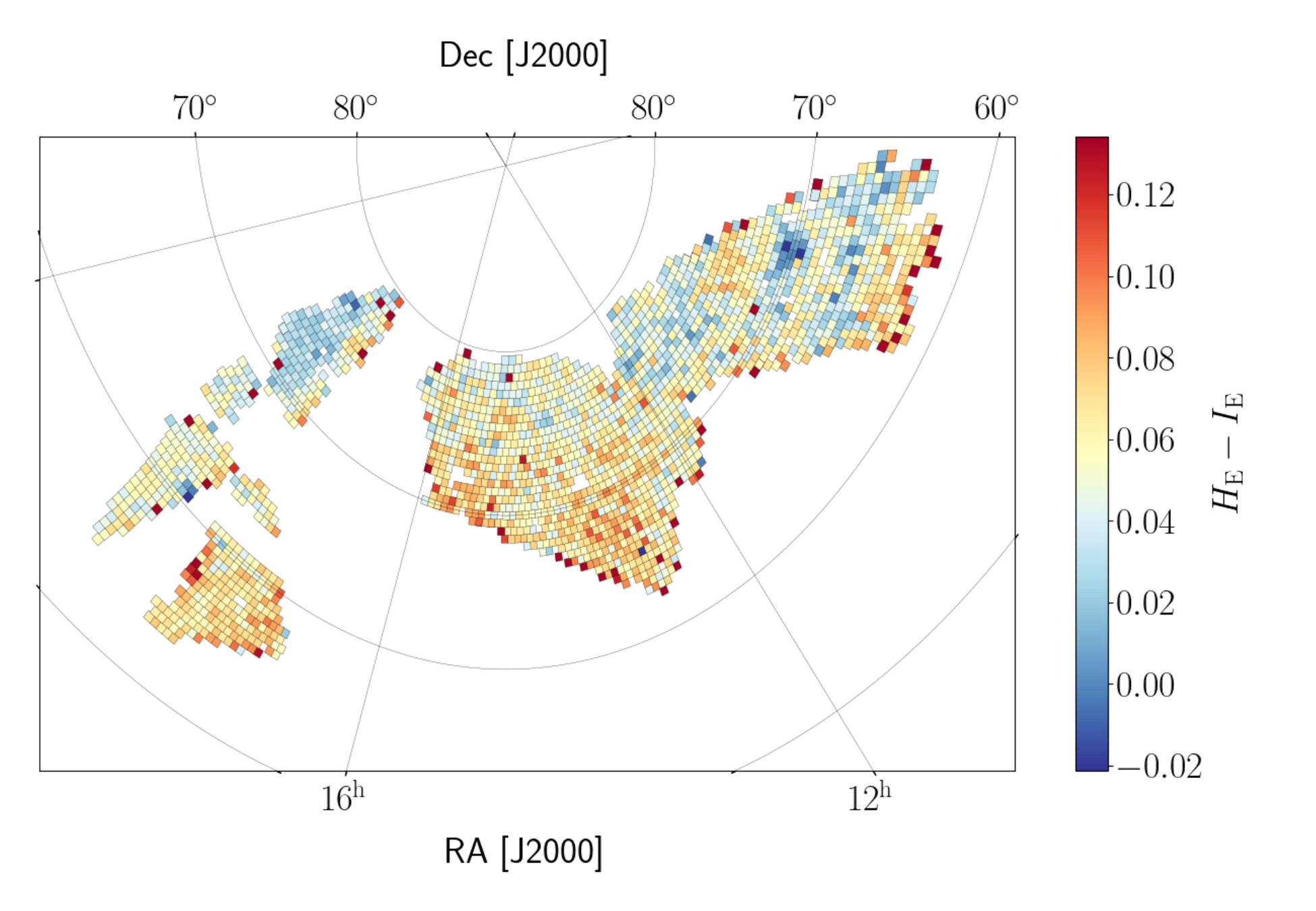}
\includegraphics[angle=0,width=0.9\hsize]{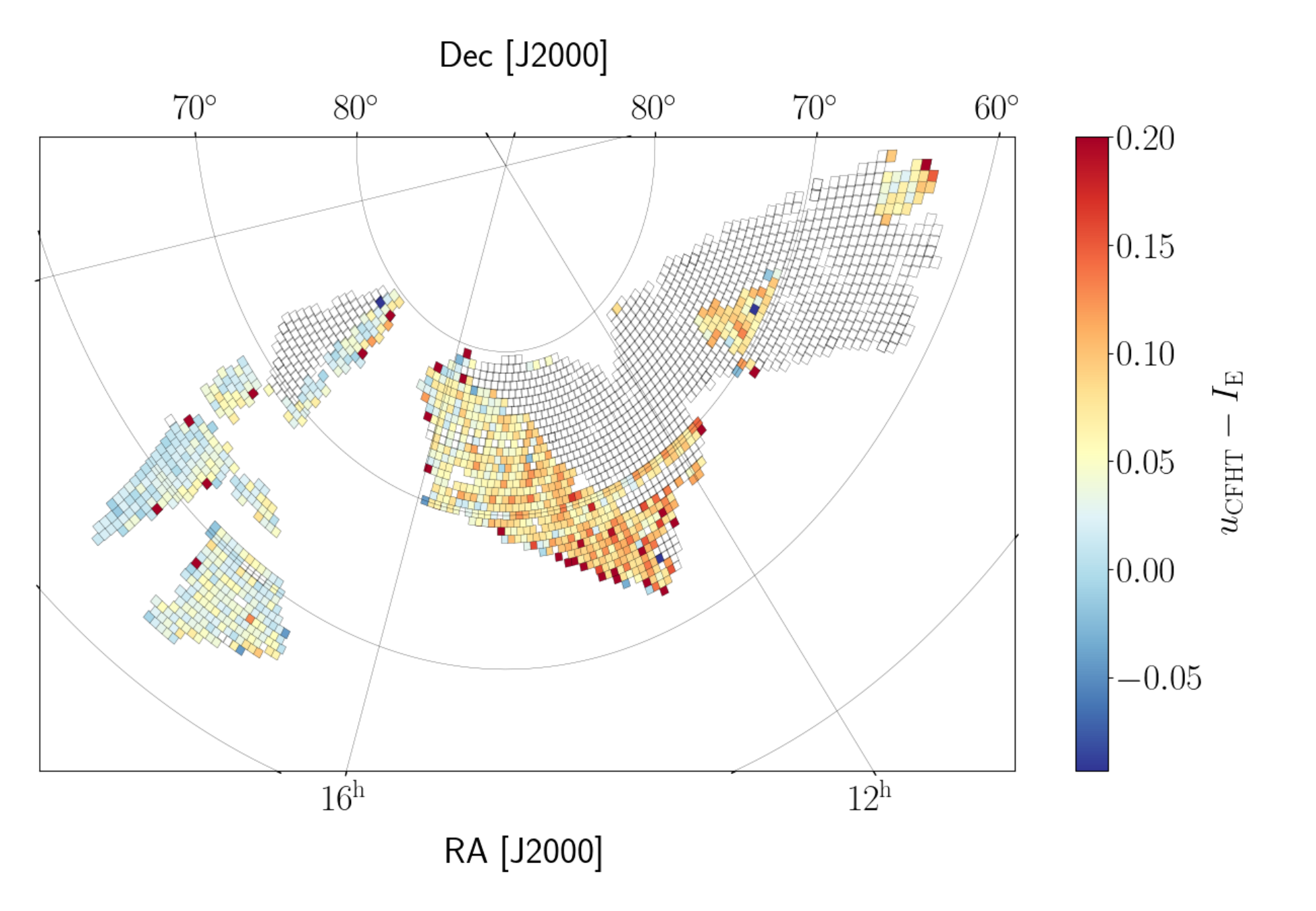}
\caption{Same as Fig.\,\ref{fig:south_zp_cont}, but for the northern part of the \gls{EWS}.}
\label{fig:north_zp_cont}
\end{figure}

For DR1, the methods chiefly depend on the internal consistency of the data processing and not so much on the absolute calibration of each photometric band. Furthermore, we did not have access to the final \gls{EWS} catalogues while preparing the calibrations. Applying zero-point shifts would therefore be both risky and in principle unnecessary, and thus we do not use them in DR1. Nevertheless, we retain the step in the pipeline to compute offsets based on stars in order to assist our validation and detect possible issues. These shifts are included in the output catalogues as a \texttt{FITS} table in the second extension of the file. The method for computing these shifts follows the same approach as we used in Q1, with objects identified as matches to \textit{Gaia} objects by MER, with compact source morphologies ($\texttt{MUMAX\_MINUS\_MAG}<-2.6$ in the MER catalogue) and satisfying the colour criterion $0<\IE - \YE<0.9$. These typically number around 100 objects per \gls{EWS} tile. We fit these selected objects with a large set of stellar SEDs using \texttt{Phosphoros} and for each band compute the median value of the ratios between the best-fit SED fluxes and the catalogue fluxes across the sample. These median ratios are then applied as correction factors to the catalogue photometry and the whole process repeated four further times (there is no test for convergence). The five factors per band are multiplied together and then scaled such that the relative corrections between bands are preserved, but the correction for the \IE band is arbitrarily set to unity. These correction factors are output in the catalogues, but we stress that they are not applied for the remainder of the pipeline stages.

Maps of the band-to-band zero-point offsets derived this way can be found in Figs.\,\ref{fig:south_zp} to \ref{fig:north_zp_cont}. Note the change in colour scale from one map to another.  The zero-point test confirms the generally good level of stability in the photometry across the \gls{EWS}, as reported by \citet{DR1-TP002}, \citet{DR1-TP006}, and \cite{DR1-TP009}, particularly in the DR1-South. However, we see a gradient in zero points across the southern region that is repeated in each band and so is either a gradient in the zero-point of the \IE data or, more likely, stellar population gradients affecting the accuracy of the zero-point determinations. We also see the clear outline of the early DR1 data that were taken before the first attempted ice decontamination, when the \IE passband was subtly different from later exposures \citep{DR1-TP026}. That passband difference is not currently accounted for in the pipeline.

\section{\label{sc:app_cosmoszp}COSMOS zero points}

The photometry of the COSMOS2020 catalogue is known to suffer from significant flux zero-point miscalibrations, which are typically corrected for as part of the photo-$z$ measurement process \citep{Weaver22}. 
As can be seen from table\,3 of \cite{Weaver22}, the estimated zero-point corrections required to optimise the photo-$z$ outputs vary substantially depending on the photo-$z$ algorithm and \gls{sed} template set used. We therefore determine zero-point corrections for the COSMOS2020 catalogue specific to our \texttt{Phosphoros} set-up. Corrections are determined by optimising the mean bias and scatter for a sample of objects with spectroscopic redshifts from \cite{Khostovan26} and are reported in Table \ref{tab:cos2020}. The corrections follow a similar pattern to those of the Classic/\texttt{LePhare} combination of COSMOS2020, which is not surprising considering that the base SED templates we use are in common with those in the \texttt{LePhare} run of COSMOS2020. However, the corrections are slightly smaller in magnitude on average and favour slightly bluer SEDs overall.

\begin{table}[]
\centering
\caption{List of the bands used to build the reference data set for \texttt{NNPZ}, their flux zero-point correction factors, and systematic flux uncertainty contributions. No correction factor is included for the IRAC channels as they were not used during photo-$z$ measurement, only for physical property estimation.}
\label{tab:cos2020}
\begin{tabular}{lcc}
\hline\hline
\noalign{\vskip 2pt}
Band & Correction factor & Systematic error \\
  \hline
  \noalign{\vskip 2pt}
  GALEX NUV & 0.98494 & 0.05 \\
  CFHT $u$ & 0.93997 & 0.02 \\
  CFHT $u^*$ & 0.99185 & 0.02 \\
  HSC $g$ & 0.91888 & 0.02 \\
  HSC $r$ & 0.93190 & 0.02 \\
  HSC $i$ & 0.96249 & 0.02 \\
  HSC $z$ & 0.95230 & 0.02 \\
  HSC $y$ & 0.94030 & 0.02 \\
  SuprimeCam $B$ & 1.08539 & 0.02 \\
  SuprimeCam $V$ & 0.92390 & 0.02 \\
  SuprimeCam $r^{+}$ & 1.00014 & 0.02 \\
  SuprimeCam $i^{+}$ & 0.98712 & 0.02 \\
  SuprimeCam $z^{++}$ & 0.94256 & 0.02 \\
  SuprimeCam IB427 & 1.01037 & 0.02 \\
  SuprimeCam IB464 & 1.00842 & 0.02 \\
  SuprimeCam IA484 & 1.00241 & 0.02 \\
  SuprimeCam IB505 & 0.99986 & 0.02 \\
  SuprimeCam IA527 & 1.02185 & 0.02 \\
  SuprimeCam IB574 & 1.05799 & 0.02 \\
  SuprimeCam IA624 & 0.99528 & 0.02 \\
  SuprimeCam IA679 & 0.84868 & 0.02 \\
  SuprimeCam IB709 & 1.01720 & 0.02 \\
  SuprimeCam IA738 & 1.02716 & 0.02 \\
  SuprimeCam IA767 & 1.04391 & 0.02 \\
  SuprimeCam IB827 & 1.02713 & 0.02 \\
  SuprimeCam NB711 & 0.99772 & 0.05 \\
  SuprimeCam NB816 & 1.04134 & 0.05 \\
  VISTA $Y$ & 0.98596 & 0.05 \\
  VISTA $J$ & 1.00349 & 0.05 \\
  VISTA $H$ & 1.06828 & 0.05 \\
  VISTA $K_{\rm s}$ & 1.01515 & 0.05 \\
  VISTA NB118 & 1.03005 & 0.05 \\
  IRAC Ch1 & \dots & 0.1 \\
  IRAC Ch2 & \dots & 0.1 \\
  \hline
\end{tabular}
\end{table}

\section{\label{sc:app_cosmosfluxerr}COSMOS flux error recalibration}

Poorly estimated flux uncertainties, for example through the neglect of correlated pixel noise in the images or source shot noise, result in miscalibration of \glspl{zpdf}. In the case of under-dispersed \glspl{zpdf} in the PIT test, the calibration can be improved by applying factors to increase the flux uncertainties. The PIT test requires true redshifts to be known, for which we must rely on spectroscopic redshifts. However, the subsample of COSMOS objects with spectroscopic redshift is biased, particularly in magnitude distribution. We wish to test the calibration of the posteriors for a sample that is appropriate to the one that is the main target case, namely the \gls{EWS}. We therefore degrade the photometry of the spectroscopic subset of COSMOS2020 galaxies to better match the distribution of the expected \gls{EWS} sample, re-run the photo-$z$ measurement for these objects, and check the calibration of the resulting \gls{zpdf}. We find that in order to obtain well-calibrated \gls{zpdf} for the target magnitudes we need to inflate the photometric uncertainties in the COSMOS2020 catalogue by $20\%$. With this additional uncertainty factor, we then proceed to produce the full set of COSMOS2020 redshift probability distributions that we use for the reference sample in \texttt{NNPZ}.

\end{appendix}

\end{document}